\documentclass[11pt]{article}

\usepackage{amsmath,amsfonts,amssymb,amsthm,stmaryrd}
\usepackage{dsfont}
\usepackage[dvips]{graphicx}
\usepackage{latexsym}
\usepackage{psfrag}
\usepackage[normalem]{ulem}
\usepackage[latin1]{inputenc}
\usepackage{hyperref}
\usepackage{color}
\usepackage{enumerate}
\numberwithin{equation}{section}
\usepackage{float}
\usepackage[T1]{fontenc}
\usepackage{comment}
\usepackage{mathtools}

\definecolor{figframe}{rgb}{0.35,0.50,0.62}
\definecolor{figback}{rgb}{0.97,0.98,0.99}

\usepackage{framed}

\newcommand{\be}{\begin{equation}}
\newcommand{\ee}{\end{equation}}
\newcommand{\barray}{\begin{array}}
\newcommand{\earray}{\end{array}}
\newcommand{\bea}{\begin{eqnarray}}
\newcommand{\eea}{\end{eqnarray}}
\newcommand{\bs}{\begin{subequations}}
\newcommand{\es}{\end{subequations}}

\newcommand{\bit}{\begin{itemize}}
\newcommand{\eit}{\end{itemize}}
\newcommand{\bd}{\begin{description}}
\newcommand{\ed}{\end{description}}

\def\nn{\nonumber}

\newcommand{\Id}{\mathds{1}}

\def\la{\langle}
\def\ra{\rangle}
\newcommand{\bra}[1]{\left\la {#1}\right|}
\newcommand{\ket}[1]{\left|{#1}\right\ra}
\newcommand{\mean}[1]{\left\la{#1}\right\ra}

\newcommand{\p}{\partial}

\newcommand{\tr}{{\rm Tr}}
\newcommand{\f}{\frac}

\newcommand{\ii}{\mathrm{i}} 
\newcommand{\ff}{\mathrm{f}} 
\renewcommand{\a}{\alpha} \renewcommand{\b}{\beta} \newcommand{\g}{\gamma}  
\renewcommand{\d}{\delta}   
    
    \renewcommand{\l}{\lambda}
        \let\om=\omega
 \newcommand{\s}{\sigma}  \renewcommand{\t}{\tau}

\DeclareMathOperator{\Tr}{Tr}

\newtheorem{remark}{Remark}

\DeclareMathOperator{\sinc}{sinc}

\DeclareMathOperator\arctanh{arctanh}

\usepackage[
    backend=biber,
    style=numeric-comp,
    sorting=none
]{biblatex}

\begin{document}

\title{
Quantum Stochastic Thermodynamics of Macroscopic Systems:\\ An Algebraic Approach}

\author{
\Large Antoine Rignon-Bret\thanks{\texttt{arignonbret@gmail.com}}\,, Cyril Elouard\thanks{\texttt{cyril.elouard@univ-lorraine.fr}}
\\[0.5em]
\small\textit{Universit\'e de Lorraine, CNRS, LPCT, F-54000 Nancy, France}
}

\maketitle

\begin{abstract}
  We build a framework to describe the thermodynamic behavior of possibly macroscopic quantum systems. In contrast with established quantum thermodynamic approaches requiring access to the full density matrix of the system, our framework is based on a coarse-grained description, that is, the statistics of measurements of a few coarse observables. When the chosen observables commute, the measurement outcomes define a classical macrostate whose entropy can be quantified by the so-called observational entropy, which treats on an equal footing the uncertainty about which macrostate the system lies in, and the uncertainty about which microstate within a macrostate. We extend this notion to a more general measurement set, possibly non-commuting, forming a subalgebra of the total system operator space, and use Jaynes principle to define an algebra-dependent entropy which interpolates between the von Neumann and observational entropies. By considering the data of initial and final sets of measurements, between which the system can evolve due to an internal and/or environment-induced dynamics, we derive the second law fulfilled by the coarse-grained dynamics. Unlike versions of the second law based on the von Neumann entropy, our inequality captures both irreversibility coming from non unitary environment-induced dynamics, and from internal equilibration. It takes the usual form of a positive entropy production when the system is initially in internal equilibrium -- i.e. when the variables one chose to ignore in the coarse-graining reach a micro-canonical distribution -- and otherwise feature correcting terms capturing the influence of the ignored nonequilibrium resources. We also derive fluctuation theorems constraining the fluctuations of thermodynamic functions at the coarse-grained level.
By considering a quasi-static path of measurement schemes, we identify emerging quantum macroscopic notions of work and heat fulfilling both the first and second law. We identify an additional work contribution coming from the manipulation of the algebra the system is confined to, either by externally imposed constraints or by quantum measurement backactions. We finally apply our framework on paradigmatic examples illustrating its main features, such as the impact of varying the coarse-graining scheme. Our approach unifies paradigms from macroscopic and stochastic thermodynamics in a genuinely quantum framework, laying the basis of a versatile, experimental-friendly, thermodynamic toolbox to analyze complex quantum dynamics.
    
\end{abstract}

\newpage

\tableofcontents

\section{Introduction}
The tremendous success of macroscopic thermodynamics stems from its ability to infer powerful constraints on the evolution of many-body systems, without needing to solve their full dynamics \cite{Callen91}. In contrast, a reduced (coarse-grained) description only involving a few, macroscopically accessible, state variables, is sufficient.
This possibility relies on a notion of internal equilibrium: the degrees of freedom neglected by the coarse-grained description are assumed to relax rapidly, relative to the macroscopic timescales of interest, toward a state determined by the retained macroscopic constraints.
Then irreversibility is an emerging phenomenon, closely associated with the information lost in reducing the description to only state variables, as quantified by the Boltzmann entropy and the second law \cite{Lebowitz93}.

More recently, it has been shown that the laws of thermodynamics are equally useful for characterizing elementary systems coupled to reservoirs, such as Brownian particles  \cite{Sekimoto,Seifert12} or quantum open systems \cite{Alicki79}. In this context, irreversibility has a completely different origin, namely the apparent stochasticity of the dynamics induced by the environment, leading to a lack of predictability of the system state even if the degrees of freedom composing the latter are fully described. Remarkably, the field of classical (resp. quantum) stochastic thermodynamics demonstrated that the environment-noise-averaged work and heat obey the same form of first and second laws, when entropy is defined as the Shannon entropy of the state distribution \cite{Seifert05} (resp. the von Neumann entropy of the density operator \cite{Alicki79,Esposito10}). Non negligible fluctuations of thermodynamic variables contain additional physics, captured by fluctuation theorems yielding constraints beyond the average second law \cite{Gallavotti95,Jarzynski97,Crooks99,Cohen99} (resp. \cite{Kurchan00,Esposito09,Campisi11,Elouard17,Manzano18}). 

Strikingly, the resulting laws of thermodynamics hold arbitrarily far from equilibrium, a significant advance with respect to the historical formulation of macroscopic thermodynamics. More precisely, all the degrees of freedom of the system being kept in the description, there is no need to restrict the set of admissible states for the theory. The other side of the coin is that, as the system is described without any coarse-graining, there is no notion of internal entropy production in that description. Any closed dynamics is treated as reversible, and it is not possible to attribute entropy production to the internal dynamics of a large quantum system, as it would be needed to investigate topics such that the quantum-to-classical transition with a thermodynamic standpoint.
This is at odds with the exponential complexity of reversing many-body unitary dynamics, and the  well-observed and documented phenomenology of equilibration and thermalization of closed quantum systems, shown to emerge for either local or coarse (finite-resolution) variables of a nonintegrable quantum system \cite{Deutsch91,Linden09,Short11,Goldstein15,Gogolin16}.

Incidentally, an operational limitation of quantum open system formulations of thermodynamics is that computing thermodynamic functions requires the knowledge of the complete quantum state -- the density operator or the wave-function -- of the quantum system of interest, and resolve its full dynamics.
A prominent example is the von Neumann entropy, commonly used as the thermodynamic entropy in this context \cite{Alicki79,Esposito10}, which is a functional of the full system density operator.
Similarly, in measurement-based formulations, the system and the environment have to be measured in a full rank basis for stochastic thermodynamic quantities to be accessed. Such requirement is relatively mild for elementary quantum systems, and can be accommodated for specific cases of engineered reservoirs  -- leading to numerous experimental implementation exploring average \cite{Cottet17,Yan18,Peterson19,vonLindenfels19,Klatzow19,Dassonneville26} and stochastic \cite{Batalhao14,An15,Naghiloo18,Hernandez-Gomez21,Onishchenko24} quantum thermodynamic laws. However, it quickly becomes prohibitive for ensembles of several degrees of freedom, let alone many-body quantum systems, coupled to natural reservoirs.
These observations motivate recent investigations of coarse-grained quantum entropy functions based on a reduced description of the system. An example is the observational entropy \cite{vsafranek2019quantum, vsafranek2019quantum2, Safranek2020Classical, vsafranek2021brief, strasberg2021first, Strasberg:2020uxk, BuscemiSchindlerSafranek2023, Strasberg:2024drm, Schindler:2025zfu} based on the measurement of a single coarse observable of the system, which was involved in pioneer attempts to formulate general fluctuation theorems and quantum versions of laws of thermodynamics at a coarse-grained level \cite{strasberg2021first, RubinoBruknerManzano2026}.

In this article, our goal is to build a quantum thermodynamic framework interpolating between fine-grained (stochastic) quantum thermodynamics and macroscopic equilibrium thermodynamics. That is, given a chosen coarse-grained description of the system, singling out a set of external, accessible variables (i.e. properties of the macrostates) from other internal, inaccessible properties (describing the microstates), we wish to formulate the apparent laws of thermodynamics obeyed by the external variables. We do not expect that any such splitting, i.e. any possible choice of coarse-graining, will give rise to usual thermodynamic laws under arbitrary conditions -- conversely, the microscopic properties of the system and its dynamics will determine whether a reduced thermodynamic description is possible and useful. Therefore, we seek a framework whose level of coarse-graining is freely tunable to find a compromise between the simplicity of the macrostate description (the freedom to use only a few, easily measurable/simulable variables) and a sufficient characterization of the nonequilibrium properties of the system so that an equivalent of thermodynamic laws holds. To enable adjusting this level of coarse-graining in a controlled way, deviations from the usual thermodynamic laws associated to a given coarse-graining must be characterized in the theory.

We achieve this program by proposing a notion of coarse-graining formally based upon defining a subalgebra of accessible observables of the system. This formulation incorporates the paradigmatic cases where the reduced description simply results from the a coarse measurement of an observable (i.e. unable to resolve all the eigenstates individually), and where it results from a partial trace, but also more general cases where the statistics of measurements of several non-commuting observables are combined into a reduced quantum description. We then employ Jaynes principle to define a coarse-grained entropy, function of the reduced description only, which interpolates between the von Neumann entropy (in the absence of coarse-graining), the Shannon entropy (for a fully classical macrostate distribution) and the Boltzmann entropy (when the macrostate dynamics is deterministic); this entropy reduces to the observational entropy when the coarse-graining is based on a single measurement basis and otherwise extends the notion to a more general type of coarse-graining. 
By considering measurements of the system at two different times, possibly involving distinct measurement schemes and levels of coarse-graining -- i.e., different choices of accessible subalgebras -- we derive the second law and fluctuation theorems obeyed by our coarse-grained entropy.
We consider both an open system situation, where only observables of a system of interest are accessible to build the reduced description and allowing the dynamics of the system to be nonunitary due to the action of an inaccessible environment, and an autonomous bi-partite situation where two interacting systems isolated from any environment can be each measured in a coarse-grained manner. We then introduce a notion of quasi-static transformation, along which the macrostate of the system can be tracked, in order to define emerging notions of work and heat increments which bridge the gap between quantum and standard macroscopic definitions of those quantities. Finally, we treat extended examples illustrating some key features of the framework, and in particular, how different choices of the accessible subalgebra defining the reduced description affect the apparent laws of thermodynamics, reflecting the resources available to different observers with different measurement/control capabilities.\\

The article is organized as follows. In Section \ref{sec: kinematics}, we  review notions of macrostates, coarse-grained states, and observational entropy introduced in the literature. We show how these concepts
can be reformulated in terms of a commutative subalgebra of accessible observables before extending them to the more general setting of a non-commutative algebras suitable to describe quantum macrostates inferred by combining several measurements in different bases.
Focusing in a first place on the case of a commutative accessible subalgebras generated by a single measurement basis, we develop in Section \ref{sec: two point thermodynamics}  
the two-point-measurement formulation of the second law for the observational entropy. We derive its average form from both the open-system and autonomous bi-partite system perspectives, and then formulate the corresponding stochastic description and associated fluctuation theorems. Still in the commutative case, we consider in Section \ref{sec: heat and work} quasi-static trajectories from which we introduce notions of work and heat increments for both open-system and autonomous viewpoints, and derive the integral fluctuation theorems they obey. 
Finally, in Section \ref{sec: non commutative subalgebras}, we extend the full formalism to the general algebraic setting and show that the definitions and concepts introduced in the commutative case naturally carry over to general coarse-graining associated to noncommutative algebras, i.e. gathering information from several measurements in incompatible bases.

\begin{figure}[th!]
\begin{framed}
\centering
\includegraphics[width=0.85\textwidth]{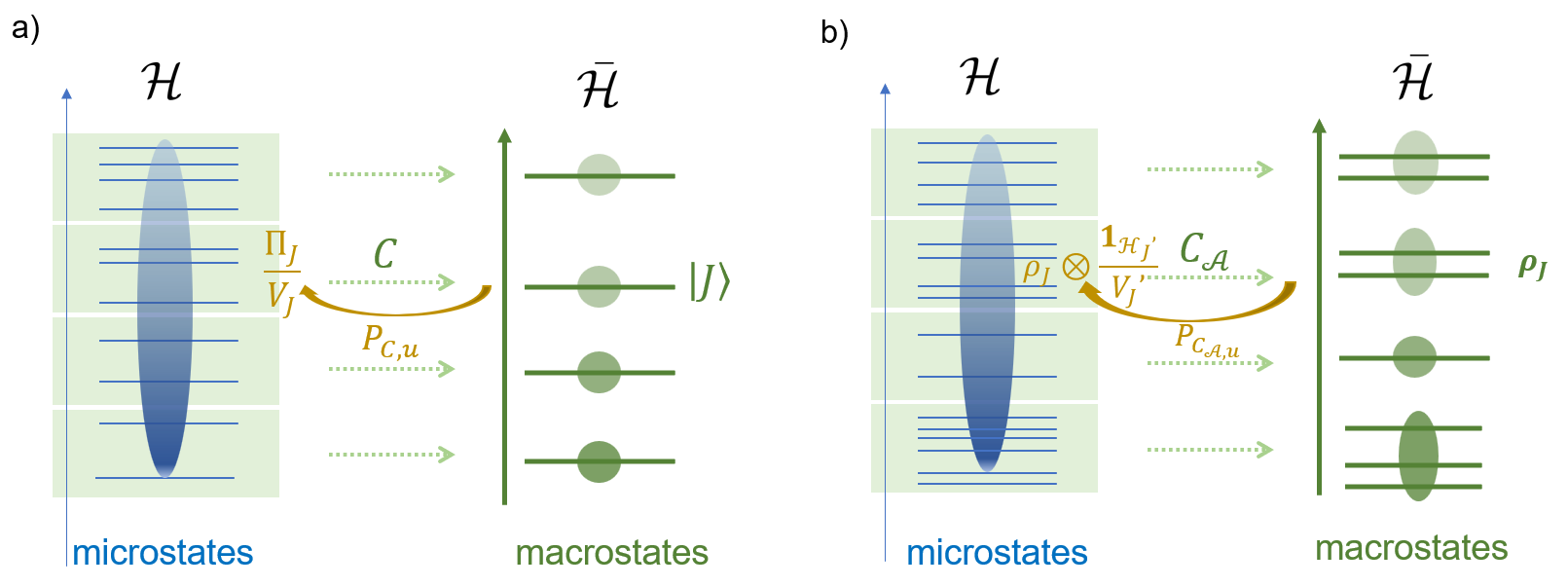}
\caption{{\bf Coarse-graining and macrostates}. Let us consider a quantum system of Hilbert space ${\cal H}$ and let us choose a basis of ${\cal H}$ (depicted by the levels on the figure) to represent the density operator of the system. This basis is not necessarily the energy eigenbasis.
a) \textbf{Commutative algebra}: Accessing the system through a coarse measurement in the chosen basis leads to a statistical mixture of macrostates $\ket{J} \in \bar{\cal H}$. Each macrostate is associated with a subspace of the original Hilbert space ${\cal H}$ of dimension $V_J$, with corresponding projector $\Pi_J$ (green shaded rectangle). The micro--macro map $C$ maps a microstate in ${\cal H}$ to the corresponding macrostate in $\bar{\cal H}$. The Petz recovery map $P_{C,u}$, or equivalently Jaynes' maximum-entropy principle, can then be used to reconstruct the most probable microstate compatible with the coarse-grained information, namely the coarse-grained state. In particular, it maps $\ket{J}$ to $\Pi_J/V_J$.
b) \textbf{Non-commutative algebra}: In the non-commutative case, the statistics of several measurement schemes are combined to assign, within each sector $J$, a density operator $\rho_J$ supported on a subspace ${\cal H}_J$ of $\bar{\cal H}$. The algebra of accessible operators is then non-commutative. The micro--macro channel $C_{\mathcal A}$ maps an arbitrary microstate $\rho$ on ${\cal H}$ to a classical mixture of density operators $\rho_J$ supported on the corresponding subspaces ${\cal H}_J$. The Petz recovery map $P_{C_{\mathcal A},u}$, or equivalently Jaynes' maximum-entropy principle, reconstructs the most probable microstate compatible with the statistics of the accessible measurements, namely the coarse-grained state. In particular, it maps $\rho_J$ to $\rho_J \otimes \Id_{{\cal H}'_J}/V'_J$.
}
\label{fig1}
\end{framed}
\end{figure}

\section{Kinematics: Accessible subalgebra, coarse-grained state and entropy}
\label{sec: kinematics}

In this section, we introduce the notion of coarse-graining of a system based on the specification of an accessible subalgebra describing the measurement capabilities of a specific observer. In particular, we introduce the notions of macrostates, coarse-grained states, and coarse-grained entropies used in this article, generalizing the notion of observational entropy which has recently attracted considerable attention. We first review this framework in the case of a coarse-graining generated by the measurement of a single (coarse) observable, recovering the formalism underlying seminal works formulating thermodynamic laws based on the observational entropy \cite{BuscemiSchindlerSafranek2023, RubinoBruknerManzano2026}. We then use the algebraic interpretation of these concepts to extend them in a natural way to the case where the observer can combine the statistics of several measurements, in possibly incompatible eigenbasis to build macrostates.
In that case, the accessible observables are represented by a type-I von Neumann algebras. The reader unfamiliar with the notion of von Neumann algebra and the algebraic approach to quantum mechanics is referred to Appendix \ref{app: algerba review} for a very brief and intuitive introduction, and to the references therein.

\subsection{Commutative case: coarse-graining from a single measurement basis} \label{sec: coarse graining commutative}

\subsubsection{Macrostate and coarse-grained state}\label{sec:cgdef}

We consider a quantum system described by a Hilbert space ${\cal H}$. We denote by ${\cal B}({\cal H})$ the set of all bounded operators acting on ${\cal H}$ and by $\rho$ the (a priori unknown) density operator of the system. In this section, we consider the case where the observer can only access the system via a projective measurement associated to a set of orthogonal projectors $\{ \Pi_{J} \}$ generating the outcomes $J$ with probability distribution 
\be \label{eq: probdist}
    p_{J} := \tr{(\rho \Pi_{J})}.
\ee

Although the framework is fully general, we will be particularly interested in the case where the rank $\text{Tr}(\Pi_J)$ of the projectors is larger than $1$ to describe a \emph{coarse} measurement. 
This can represent a measurement with finite precision, therefore unable to resolve the fine structure of the observables being measured.
When considering larger and larger systems, the spectrum of observables typically become denser, such that any realistic measurement scheme will become coarser and coarser. We then assume that the only information available to the observer about the state of the system is the probability distribution \eqref{eq: probdist}.
This information can be encoded in a \emph{macrostate} belonging to a smaller Hilbert space $\bar{\mathcal{H}}$ of dimension $\bar{D}$ given by the cardinal of the set of projectors $\{\Pi_J \}$, i.e. the number of distinct measurement outcomes (See Fig.~\ref{fig1}a). For a many-body system, we expect $\bar{D} \ll D= \dim \mathcal{H}$.

Therefore, a coarse-graining procedure can be defined in which a microstate $\rho \in \mathcal{B}(\mathcal{H})$ is mapped to a macrostate $\bar{\rho} \in \mathcal{B}(\bar{\mathcal{H}})$ via the micro-macro channel of Complete Positive Trace Preserving (CPTP). \emph{micro-macro channel}
\begin{align} \label{eq: macromap}
    \mathcal{B}(\mathcal{H}) &\longrightarrow \mathcal{B}(\bar{\mathcal{H}}) \nn \\
    \rho &\longrightarrow \bar{\rho} = C(\rho) = \sum_{J} \tr{(\rho \Pi_{J})} \ket{J} \bra{J} = \sum_{J} p_{J} \ket{J} \bra{J},
\end{align}
where $\{\ket{J}\}$ denotes a basis of macrostates each associated with one of the measurement outcomes $J$.
Note that for a projective measurement, the image of map \eqref{eq: macromap} is always diagonal in the macrostate Hilbert space. Indeed, $\bar{\rho} = C(\rho)$ can be interpreted as the state of the memory storing the measurement results, i.e. a classical macrostate. Notice however, that, contrary to the case where the system is classical, there exists different bases in which the measurement defining the coarse-graining could be performed in principle. Each choice of basis defines a particular coarse-graining, associated to a macrostate space and an algebra of accessible observables.

In order to lay the ground to the generalization to more sophisticated coarse-graining schemes collecting information from several non-commuting measurements, we say formally that the accessible subalgebra characterizing the coarse-graining considered in this section is the commutative 
von Neumann algebra\footnote{See App.~\ref{app: algerba review} for a brief reminder of the mathematical properties of the von Neumann algebras.} $\mathcal{A}^c \subset \mathcal{B}(\mathcal{H})$ generated by the set of self-adjoint and orthogonal projectors $\{ \Pi_{J} \}$. Any operator $a \in \mathcal{A}^c$ can be written as 
\be
    a = \sum_{J} a_{J} \Pi_{J}, \qquad a_{J} \in \mathbb{C}, 
\ee
and its expectation value $\mean{a}_\rho =\text{Tr}\{a\rho\}$ can therefore be computed directly from the probability distribution $p_J$, or equivalently, from the macrostate. In this sense, the algebra describes the set of observables accessible to an observer through a given measurement scheme.

\vspace{0.3 cm}

One important point is that if there is in general no inverse to the micro-macro channel \eqref{eq: macromap} as there exists many microstates compatible with one given macrostate.
Nevertheless, one can associate with each macrostate $\bar{\rho} \in \mathcal{B}(\bar{\mathcal{H}})$ a \emph{coarse-grained state} in $\mathcal{B}(\mathcal{H})$, which provides a representative microstate compatible with the available macroscopic information \cite{BuscemiSchindlerSafranek2023}.
One possible approach consists in selecting, among all states compatible with the constraints, the one of maximal entropy, following the standard maximum-entropy prescription of Jaynes \cite{Jaynes1957a}. In this approach, the coarse-grained entropy is defined as
\be\label{d:cgentropy}
S^{\text{cg}}(\rho) = \max_{\rho' \in \mathbb{D}(\mathcal{H})} \{ S_{\text{v.N}}(\rho') \quad \lvert \quad \forall J, \quad \tr{(\rho' \Pi_J)} = p_J \}
\ee
where the maximization is taken over the set $\mathbb{D}(\mathcal{H})$ of density matrices on ${\cal H}$\footnote{Since $\mathbb{D}(\mathcal{H})$ is compact, this maximum exists.}. 
The coarse-grained state is defined as the state reaching this maximal von Neumann entropy, which is \cite{BuscemiSchindlerSafranek2023, Wehrl1978}
\be \label{eq: coarse-grained state}
    \rho^{\text{cg}} := \sum_{J} p_{J} \frac{\Pi_{J}}{V_{J}}
\ee
where 
\be \label{eq: defvolume}
    V_{J} := \tr{\Pi_J}
\ee
are the ranks of the projectors generating the commutative algebra $\mathcal{A}^c$, and, as we see later, play the role of the macrostate volumes. As discussed later, the system is expected to be in the coarse-grained state \eqref{eq: coarse-grained state} if the observables that commute with the algebra $\mathcal{A}^c$, i.e. the internal observables inaccessible with the considered measurement scheme, have reached some equilibrium. In contrast, the external variables encoded in $\mathcal{A}^c$ can be arbitrarily far from equilibrium. 

An alternative way to obtain the coarse-grained state is to use the \textit{Petz recovery map} \cite{Petz1986, Petz1988, Santoso2023Petz} defined for any CPTP map $C$ as
\be
    P_{C, \t}(\cdot) = \t^\f12 C^\dag (C(\t)^{- \f12} \cdot C(\t)^{- \f12}) \t^\f12,
\ee
where $C^\dagger$ denotes the adjoint map of $C$. $P_{C, \t}$ is also CPTP \footnote{If the operator $C(\t)$ is invertible} and can be seen as the quantum version of Bayes inversion theorem, where $\t$ is the reference state satisfying $P_{C, \t} (C(\t)) = \t$
which plays the role of the prior of Bayesian inference. 
The Petz recovery map transforms any macrostate $\bar{\rho} \in \mathcal{B}(\bar{\mathcal{H}})$ into a coarse-grained state $\rho^{\mathrm{cg}} \in \mathcal{A}^c$, belonging to the algebra accessible through the measurement scheme. More precisely, the composition of the micro-macro channel $C$ and the Petz recovery map prepares the same coarse-grained state as obtained via the entropy maximization principle attached to $C$, i.e. Eq.~\eqref{eq: coarse-grained state}, provided we take a reference state proportional to the identity. We then define the coarse-graining map

\begin{align} \label{eq: coarse-grained state map}
    \mathcal{B}(\mathcal{H}) &\longrightarrow \mathcal{A}^c \nn \\
    \rho &\longrightarrow \rho^{\text{cg}} = (P_{C, u} \circ C)(\rho) = \sum_{J} \tr{(\rho \Pi_{J})}\frac{\Pi_{J}}{V_{J}}, \qquad u = \frac{\Id_{\mathcal{H}}}{D},
\end{align}
Finally, notice that for any operator $a$ in the accessible algebra $\mathcal{A}^c$, one has 
\be \label{eq: mean value a cg states}
    \langle a \rangle_\rho = \tr{(a \rho)} = \tr{(a \rho^{\text{cg}})} = \langle a \rangle_{\rho^{\text{cg}}}.
\ee
Conversely, for any operator $o \in \mathcal{B}(\mathcal{H})$, there exists an operator $o_{\mathcal{A}^c}$ of the accessible $\mathcal{A}^c$ having the same average value as $o$ in all the coarse-grained states, i.e.: 
\be\label{d:representative_comm}
    \langle o \rangle_{\rho^{\text{cg}}} = \tr{(o \rho^{\text{cg}})} = \tr{(o_{\mathcal{A}^c} \rho^{\text{cg}})} = \langle o_{\mathcal{A}^c} \rangle_{\rho^{\text{cg}}}, \quad \forall \rho^{cg}\in {\cal A}^c.
\ee
where $o_{\mathcal{A}^c}  = \sum_J \tr{\left(o \frac{\Pi_J}{V_J}\right)} \Pi_J$. We call this operator $o_{\mathcal{A}^c}$ the \emph{representative} of the operator $o$ in the algebra $\mathcal{A}^c$.

\begin{remark}
    One could attach another prior $\tau \neq u$ to the Petz map defining the expected state of the degrees of freedom that one cannot control (associated to the operators commuting with the algebra $\mathcal{A}^c$). For instance, one could expect a thermal (canonical) distribution of the microstate population instead of the microcanonical distribution, based on the knowledge of the temperature of the system. This would correspond to imposing additional constraints on the entropy maximization \cite{Schindler:2025zfu}.  However, note that only the uniform distribution ensures that the coarse-grained state has a von Neumann entropy that is larger than the one of the initial state for any initial state. Nonetheless, all the results of this paper can be 
    extended to an arbitrary choice of prior $\tau$. 
\end{remark}

\begin{remark}
    To go beyond the idealized case and describe realistic measurements, one typically employs a POVM, i.e., a complete set of positive operators $M_J \geq 0$ satisfying $\sum_J M_J = \Id_{\mathcal H}$, rather than a set of orthogonal projectors. Unlike projective measurements, however, the elements of a generic POVM are not, in general, closed under multiplication and therefore do not themselves constitute a subalgebra $\mathcal{A} \subset \mathcal{B}(\mathcal{H})$. Moreover, the expectation values $p_J := \tr(\rho M_J)$ are generally not preserved under repeated measurements. We leave the general investigation of this question for future work,
    and simply mention in App.~\ref{app: POVM} the aspects of our formalism which directly extend to arbitrary POVMS.
\end{remark}

\subsubsection{Observational entropy}

In the case of a projective measurement that we have considered so far, the coarse-grained entropy defined in Eq.\eqref{d:cgentropy} corresponds to the so-called observational entropy \cite{vsafranek2021brief} whose significance for thermodynamics was already discussed in \cite{vsafranek2021brief, Schindler:2025zfu, RubinoBruknerManzano2026}. 
Since the observational entropy is the maximal von Neumann entropy compatible with the outcomes of the measurement, one has 
\be \label{eq: obentropy1}
    S^{\text{ob}}(\rho) = S_{\text{v.N}}(\rho^{\text{cg}}) \geq S_{\text{v.N}}(\rho)
\ee
It is enlightening to recast the observational entropy \eqref{eq: obentropy1} as 
\be \label{eq: obentropy2}
    S^{\text{ob}}(\rho) := - \sum_{J} p_{J} \ln{\frac{p_{J}}{V_{J}}} = \underbrace{-\sum_{J} p_{J} \ln{p_{J}}}_\text{Shannon/Gibbs entropy} + \underbrace{\sum_{J} p_{J} \ln{V_{J}}}_\text{Boltzmann entropy}
\ee
letting us identify an informational (Shannon) and a coarse-graining (Boltzmann) contributions. The first term is the entropic term usually present in stochastic and quantum (stochastic) thermodynamics 
\cite{Elouard17, ElouardMohammady2018, Manzano18}. If  
one assumes no coarse-graining (i.e. $\forall J, V_{J} = 1$), it reduces to the von Neumann entropy \footnote{More precisely, the von Neumann entropy of the average state after the measurement. It corresponds to the von Neumann entropy before the measurement if one chooses to measure the system in the eigenbasis of the density operator.}. In the presence of coarse-graining, it quantifies the uncertainty about which \emph{macrostate} the system occupies. The second term is induced by the coarse-graining, and represents the number of microstates compatible with the observed macrostate. Therefore, it carries the statistical-mechanical definition of entropy that we are used to in (equilibrium) macroscopic thermodynamics. 

\vspace{0.3 cm}

The observational entropy \eqref{eq: obentropy2} captures two different notions of entropy productions: i) 
due to stochastic (i.e. non-unitary) contributions to the dynamics of the system {\it at the macrostate level}, leading to changes of populations in the Shannon/Gibbs part of the observational entropy, 
ii) the entropy production related to a coarse-graining, more specifically to the information lost about the internal variables (which microstate the system occupies) when restricting our description to a subalgebra $\mathcal{A}^c \subset \mathcal{B}(\mathcal{H})$.
One can make an analogy to classical physics, for which the entropy production is related to a variation of the density of states in the phase space. This volume variation has two possible origins: either the dynamics is not Hamiltonian or either a \textit{coarse-graining} is performed in phase space because of our inability to perform fine-grained measurements, leading to some entropy production even if the dynamics is Hamiltonian. If none of these processes occurs, Liouville theorem ensures that the state density behaves like an incompressible fluid in phase space and there is no entropy production. Being sensitive to both forms of irreversibility, observational entropy is suitable to construct a coarse-grained quantum thermodynamic framework \cite{strasberg2021first,RubinoBruknerManzano2026}.

\subsection{Non commutative case}\label{sec: noncomm cg entropy}

The coarse-graining introduced in the previous section is based on the measurement of a single observable (or a set of mutually commuting observables), leading to an essentially classical macrostate. The freedom in quantum mechanics to perform a measurement in different bases implies that we may 
formulate several such possible coarse-graining schemes. Based on the coarse-graining procedure of section~\ref{sec: coarse graining commutative} however, those different measurements would lead to different parallel thermodynamic description.

 Can we do better and combine the information from several incompatible measurements to obtain a single, fully quantum, thermodynamic story? In the absence of coarse-graining, we are used to combine results from a tomographically-complete set of  measurements to re-build a density operator. Can we do the same with coarse measurements? One case where it is clearly possible is when the coarse-graining simply corresponds to tracing out some degrees of freedom (i.e. when the space of accessible and inaccessible states generates the full Hilbert space via a tensor product). In that case, a tomographic set of local measurements enables one to reconstruct the reduced density operator of the accessible degrees of freedom. But an arbitrary coarse-graining may not be formulated as a mere partial trace, since the accessible observable do not necessarily appear in tensor product with respect to the internal ones. This happens in the case of a single coarse measurement discussed in the previous section: the macrostates cannot be obtained from a partial trace over the microstates, unless the ranks $V_J$ of the projectors $\Pi_J$ are equal for all $J$.
More generally, a sufficient condition for the accessible observables to admit a well-defined reduced quantum theory is that they form a von Neumann subalgebra \cite{BratteliRobinson1987}($\mathcal{A} \subset B(\mathcal{H})$) of the full algebra of observables (see App.~\ref{app: algerba review} for a brief intuitive introduction to the algebraic framework of quantum theory and the classification of von Neumann algebras).

 In this section, we  
 define a generalization of the macrostates and of the observational entropy reflecting the knowledge an observer can gather from a set of non-commuting measurements. In this approach, the macrostates are now fully quantum states on a reduced Hilbert space, and their entropy interpolates between the von Neumann entropy and the observational entropy to reflect additional knowledge from combining several measurement bases.

Let us consider that the observer can measure the observables in the type I von Neumann subalgebra $\mathcal{A} \subset \mathcal{B}(\cal H)$. It can be proven \cite{TakesakiI, KadisonRingroseII} that any type I von Neumann algebra $\mathcal{A} \subset \mathcal{B}(\mathcal{H})$ takes the form \footnote{
In Eq.~\eqref{eq: algIdecompo}, the direct sum may also be continuous for infinite dimensional Hilbert space, but we consider only the discrete case for simplicity.}
\be \label{eq: algIdecompo}
    \mathcal{A} \simeq \bigoplus_{J} \mathcal{B}(\mathcal{H}_J) \otimes \Id_{\mathcal{H}_J^{'}} := \bigoplus_{J} {\cal A}_J,
\ee
where the symbol $\simeq$ means equal up to a global unitary transformation. We have introduced ${\cal H}_J$ and ${\cal H}'_J$ which are Hilbert subspaces of ${\cal H} = \bigoplus_J({\cal H}_J\otimes{\cal H}'_J)$. We can see that, up to a change of basis (i.e. a unitary transformation $U$), the operators in the algebra $\mathcal{A}$ are block-diagonal, with the individual blocks belonging to the subspaces ${\cal H}_J\otimes{\cal H}_J'$. The latter generalize the subspaces associated with the projector $\Pi_J$ in the single measurement case
(green shaded rectangles in Fig.~\ref{fig1}b)
-- we refer to them as \emph{sectors}. 
Crucially, in each sector, the observer can now access all the observables in ${\cal B}(H_J)$ rather than only the projectors $\Pi_J$, while observables of ${\cal H}'_J$ remain inaccessible. Denoting the sector dimensions $V_J = D_J V'_J$ where $D_J = \text{dim}{\cal H}_J$ and $V'_J = \text{dim} {\cal H}'_J$, the commutative case is retrieved when we take $D_J=1$ such that ${\cal A}_J= \mathds{C} \otimes \Id_{\mathcal{H}_J^{'}}$. The only accessible observables are then proportional to $\Pi_J = \Id_{\mathcal{H}_J^{'}}$, and $V'_j=V_J$. Conversely, if $D_J \geq 2$ for at least one sector $J$, the subalgebra ${\cal A}$ becomes non-commutative. Consequently, measuring the observables in ${\cal A}$ provides more information about the system's state than measuring the projectors $\Pi_J$ alone, thereby allowing for a partial tomography of $\rho$. In other words, the reduced description of the sector $J$ is a $D_J-$dimensional quantum system rather than a single quantum level.
As a consequence, the macrostates gathering all the information must now be described by a full density operator $\rho_J\in {\cal B}(H_J)$ rather than simply a number $p_J$.  In the extreme case $V'_J=1$ and $D_J=V_J$, all observables in sector $J$ are accessible.

\vspace{0.3 cm}

It is useful to introduce the \textit{commutant} $\mathcal{A}'$ of the algebra $\mathcal{A}$ which corresponds to the set of operators in $\mathcal{B}(\mathcal{H})$ commuting with all operators of $\mathcal{A}$, i.e.
\be
    \mathcal{A}':= \{ a' \in \mathcal{B}(\mathcal{H}) \lvert \quad \forall a \in \mathcal{A}, \quad [a,a'] = 0 \}. 
\ee
It is straightforward to check that $\mathcal{A}' \subset \mathcal{B}(\mathcal{H})$ is also a type I von Neumann algebra. In addition, 
\be
    \mathcal{A}' \simeq \bigoplus_{J} \Id_{\mathcal{H}_J} \otimes \mathcal{B}(\mathcal{H}_J^{'}).
\ee
For instance, the commutant of the commutative algebra $\mathcal{A}^c$ generated by the set of projectors $\{ \Pi_J \}$ is the non-commutative algebra 
\be
    \mathcal{A}^{c '} = \{ a' = \sum_J \Pi_J o \Pi_J \quad \lvert \quad \forall o \in \mathcal{B}(\mathcal{H}) \}
\ee
and consists of all block-diagonal operators that commute with the $\{ \Pi_J \}$. These operators can resolve different microstates associated with the same macrostate and therefore correspond to observables that become inaccessible under coarse-graining. Similarly, in the general noncommutative setting, the commutant $\mathcal{A}'$ encodes the observables inaccessible to the chosen measurement scheme.

\vspace{0.3 cm}

In the general setting we therefore define a coarse-graining as being associated to a specific choice of subalgebra of observables $\mathcal{A} \subset \mathcal{B}(\mathcal{H})$ verifying
\be\label{eq: vN alg U J}
   \mathcal{A} = U\left( \bigoplus_{J} \mathcal{B}(\mathcal{H}_J) \otimes \Id_{\mathcal{H}_J^{'}}\right) U^\dag,
\ee
with a given unitary $U$ set by the measurement setup. This implies that the observer has access to all the mean values of the observables belonging to ${\cal A}$, that is all the numbers
\be\label{eq:constraintava}
    \qquad  \tr{(a \rho)} = \langle a \rangle_\rho, \qquad \forall a \in \mathcal{A}.
\ee
We can now extend the considerations of the previous section by looking for the state that maximizes the von Neumann entropy given the constraints Eq.~\eqref{eq:constraintava}, and define the coarse-grained entropy associated with the accessible algebra ${\cal A}$ as
\be \label{eq: algebraicentropyI}
    S_{\mathcal{A}}(\rho) = \max_{\rho' \in \mathbb{D}(\mathcal{H})} \{ S_{\text{v.N}}(\rho') \quad \lvert \quad \forall a \in \mathcal{A}, \quad \tr{(a \rho')} = \langle a \rangle_\rho \}.
\ee
This maximization problem is simply solved for the state $\rho^{\text{cg}}_{\mathcal{A}} \in \mathcal{A}$ given by (see the proof in App.~\ref{app: proofs})
\begin{align} \label{eq: coarse-grained state ga}
    \rho^{\text{cg}}_{\mathcal{A}} &= U \left( \sum_J   \tr_{\mathcal{H}_J^{'}}{(\Pi_J U^{-1} \rho U \Pi_J)} \otimes\frac{\Id_{\mathcal{H}_J^{'}}}{V_J^{'}} \right) U^{-1} \nn \\
    &= U \left( \sum_J   \rho_J \otimes \frac{\Id_{\mathcal{H}_J^{'}}}{V_J^{'}} \right) U^{-1}.
\end{align}
where
\bea
\rho_J := \tr_{\mathcal{H}_J^{'}}{(\Pi_J U^{-1} \rho U \Pi_J)}
\eea
is an un-normalized density operator in ${\cal B}(H_J)$ whose trace verifies $\tr{\rho_J}=p_J \geq 0$, such that $\sum_J \tr{\rho_J} = 1$. With this choice of normalization, the operator $\bar\rho =(\sum_J\rho_J) \in \bar{\cal H} = \bigoplus_J {\cal H}_J$ generalizes the notion of a macrostate introduced in the previous sections through
Equation \eqref{eq: macromap}, see Fig.~\ref{fig1}b. 
 The coarse-grained states given by \eqref{eq: coarse-grained state ga} are arbitrarily out of equilibrium with respect to the observables contained in the algebra $\mathcal{A}$,  while being micro-canonically distributed with respect to the complementary degrees of freedom encoded in its commutant algebra $\mathcal{A}'$. They are therefore appropriate tools to interpolate between quantum and macroscopic thermodynamics whenever a suitable choice of accessible algebra ${\cal A}$ can discriminate between fast internal degrees of freedom and slow macroscopic observables.
 
From \eqref{eq: algebraicentropyI}, we obtain the following expression for the algebra-dependent coarse-grained entropy 
\be \label{eq: algebraicentropyII}
    S_{\mathcal{A}}(\rho) = \underbrace{- \sum_J p_J \ln{p_J}}_{\text{Shannon/Gibbs entropy}} + \underbrace{\sum_J p_J S_{\text{v.N}}(\tilde{\rho}_J)}_{\text{von Neumann entropy}} + \underbrace{\sum_J p_J \ln{V_J^{'}}}_{\text{Boltzmann entropy}}
\ee
where $\tilde{\rho}_J = \frac{\rho_J}{p_J}$ is normalized state. 
$S_{\mathcal{A}}(\rho)$ decomposes into three contributions. The first term is the Shannon entropy associated with the probability distribution $p_J$ over the sector $J$. This contribution is classical in nature, reflecting the fact that the algebra $\mathcal{A}$ contains no operators capable of probing coherences between distinct sectors. 
The second term of \eqref{eq: algebraicentropyII} is the von Neumann entropy associated with the accessible subpart of sector $J$.
The third term accounts for the inaccessible degrees of freedom and is given by the maximal von Neumann entropy compatible with their Hilbert-space dimension, that is the Boltzmann entropy constrained to the space ${\cal H}'_J$ not resolved by algebra ${\cal A}$. Notice that $S_{\text{v.N}}(\tilde{\rho}_J) + \ln V'_J \leq \ln V_J$ reflects the additional knowledge gathered by measuring several noncommuting observables in the sector rather than just $\Pi_J$. 
Note that a similar definition of algebra-dependent entropy was proposed in \cite{Facchi2021}. There, the last term of Eq.~\eqref{eq: algebraicentropyII} was dropped as it is dependent on the particular representation of the algebra \footnote{Namely, the multiplicity-free representation of the finite dimensional $\mathbb{C}^\ast$-algebra considered.}. In contrast, here, the representation of the algebra is imposed by the physical situation under study, and this last term plays a significant role for thermodynamics as already discussed.

\begin{remark} 
    We have assumed a uniform distribution over the inaccessible degrees of freedom associated with the spaces $\mathcal{H}'_J$, corresponding to a principle of maximum ignorance. However, as in the commutative case, one may consider distributions other than microcanonical, corresponding to additional constraints on the inaccessible degrees of freedom. 
    
    One such possibility is to assume that the degrees of freedom associated with the commutant algebra $\mathcal{A}'$ satisfy a KMS condition \cite{Haag1967} with respect to a prescribed one-parameter group of automorphisms $\alpha_t$ preserving the commutant, $\alpha_t(\mathcal{A}')=\mathcal{A}'$. Formally, for suitable analytic elements $a',b'\in\mathcal{A}'$, this condition reads
    \be
    \langle \alpha_t(a') b' \rangle_\rho = \langle b' \alpha_{t+i\beta}(a') \rangle_\rho,
    \qquad t\in\mathbb{R}.
    \ee
    More generally, any alternative notion of equilibrium with respect to the inaccessible degrees of freedom can be incorporated through a nonuniform prior $\tau$. Equation \eqref{eq: coarse-grained state ga} is then replaced by
    \be
    \rho^{\mathrm{cg}}_\tau =
    U\left(
    \sum_J \rho_J\otimes\tau_J
    \right)U^{-1},
    \ee
    where $\tr{\tau_J}=1$. The corresponding algebraic entropy becomes
    \be \label{eq: other prior alg entropy formula}
    S_{\mathcal{A}}^{\tau}(\rho) =
    -\sum_J p_J\ln{p_J}
    +
    \sum_J p_J S_{\mathrm{v.N}}(\tilde{\rho}_J)
    +
    \sum_J p_J S_{\mathrm{v.N}}(\tau_J).
    \ee
    The uniform choice $\tau_J=\frac{\Id_{\mathcal{H}'_J}}{V_J^{'}}$ is recovered as the special case corresponding to maximum ignorance within each inaccessible sector.
\end{remark}

\vspace{0.3cm}

As for the commutative case, we introduce the coarse-graining map
\begin{align} 
    \mathcal{B}(\mathcal{H}) &\longrightarrow \mathcal{A} \subset \mathcal{B}(\mathcal{H}) \nn \\
     \rho &\longrightarrow \rho^{\text{cg}}_\mathcal{A} = U \left( \sum_J \rho_J \otimes \frac{\Id_{\mathcal{H}_J^{'}}}{V_J^{'}} \right) U^{-1}
\end{align}
which is CPTP, as it is a direct composition of elementary CPTP maps.  
In particular, one can then write (see Appendix \ref{app: proofs} for a proof) 
\be \label{eq: petzmapcgena}
    \rho^{\text{cg}}_{\mathcal{A}} = (P_{C_\mathcal{A}, u} \circ C_\mathcal{A})(\rho),\quad \bar{\rho} = C_\mathcal{A}(\rho) = \sum_J \rho_J
\ee
 where $C_{\cal A}$ is the CPTP map projecting a microstate onto a macrostate, extending the notions of Section \ref{sec:cgdef}. Importantly, as in the commutative case, the coarse-graining map $P_{C_\mathcal{A}, u} \circ C_\mathcal{A}$ is a projector. Therefore, we also have that 
 for any operator $a \in \mathcal{A}$, $\langle a \rangle_\rho = \langle a \rangle_{\rho_{\mathcal{A}}^{\text{cg}}}$. Hence, we can straightforwardly extend the notion of representatives. For any $o \in \mathcal{B}(\mathcal{H})$, there exists an operator $o_{\mathcal{A}} = (P_{C_\mathcal{A}, u} \circ C_\mathcal{A})(o)  \in \mathcal{A}$ so that 
\be\label{d:representative_noncomm}
    \langle o \rangle_{\rho_\mathcal{A}^{\text{cg}}} = \langle o_{\mathcal{A}} \rangle_{\rho_\mathcal{A}^{\text{cg}}}
\ee
for all $\rho^{\text{cg}}_{\mathcal{A}} \in \mathcal{A}$.

\vspace{0.3 cm}

Finally, we can prove that (see Appendix \ref{app: proofs})
\be \label{eq: entropy hierarchy}
    S_{{\cal A}^c}(\rho)=S^{\text{ob}}(\rho)\geq  S_{\mathcal{A}}(\rho) := S_{\text{v.N}}(\rho_{\mathcal{A}}^\text{cg}) \geq S_{\text{v.N}}(\rho)
\ee
where $\mathcal{A}^c$ is the commutative subalgebra generated by the projectors $\Pi_J$ on the sectors of $\mathcal{A}$. Moreover,
the entropy associated with any von Neumann subalgebra $\mathcal{A}_0 \subset \cal A$  verifies 
\be \label{eq: entropy hierarchy 2}
     S_{\mathcal{A}_0}(\rho) := S_{\text{v.N}}(\rho_{\mathcal{A}_0}^\text{cg}) \geq S_{\mathcal{A}}(\rho) := S_{\text{v.N}}(\rho_{\mathcal{A}}^\text{cg})
\ee
In \eqref{eq: entropy hierarchy} and \eqref{eq: entropy hierarchy 2}, the entropy increases upon restriction to smaller algebras (i.e. coarse-graining). This reflects the interpretation of entropy as a measure of missing information: restricting the set of accessible observables reduces the information available to the observer and therefore increases the entropy.

\subsection{Internal equilibrium}
\label{sec: internal equilibrium}

Equilibrium statistical mechanics is rooted in the principle of equiprobability, according to which a system at equilibrium is assumed to explore all microstates compatible with the imposed macroscopic constraints with equal probability. This assumption provides the microscopic foundation of thermodynamics by allowing an enormous number of microscopic degrees of freedom to be described in terms of only a few macroscopic variables, such as energy, particle number, or volume. In conventional thermodynamics, external constraints may be modified or relaxed, leading to changes in the coarse-grained energy and entropy of the system as it evolves toward a new equilibrium state \cite{Callen91}. By contrast, common approaches to quantum thermodynamics typically assume complete access to the system and describe its state by the full density matrix, without explicitly distinguishing between accessible and inaccessible degrees of freedom of the system \cite{Alicki79,Esposito10}. The huge advantage is that deriving thermodynamic laws do not require the system to be at equilibrium; but this comes at the price that every detail of the system must be taken into account in the description.

Our framework combines the advantages of both approaches by allowing the distinction between what are macroscopic/external variables and microscopic/internal variables to be tuned, via the choice of the accessible subalgebra $\mathcal{A}\subset\mathcal{B}(\mathcal{H})$. In this setting, one expects the standard laws of thermodynamics to emerge in their usual form only when the discarded information has not too much impact on the macroscopic scale. 

This requires introducing a notion of internal equilibrium for the degrees of freedom that are inaccessible to observation and independent of $\mathcal{A}$, namely those encoded in the commutant algebra $\mathcal{A}'$. 
This requirement mirrors the situation encountered in macroscopic thermodynamics, where microscopic details are assumed to have equilibrated and therefore become irrelevant to the macroscopic description. Operationally, we say that a system is in \textit{internal equilibrium relative to the algebra $\mathcal{A}$}
whenever the state of the system is equal to the coarse-grained state given by Eqs.\eqref{eq: coarse-grained state} and \eqref{eq: coarse-grained state ga}, i.e. $\rho=\rho^\text{cg}_\mathcal{A}$.
This assumption may be viewed as the algebraic analogue of the Boltzmann hypothesis that the microscopic degrees of freedom explore all states compatible with the macroscopic constraints. Under such assumption of ergodicity, the unresolved degrees of freedom encoded in $\mathcal{A}'$ can therefore be described by the maximum-entropy distribution compatible with the expectation values of observables in $\mathcal{A}$. 

If one thinks of the equilibration of a closed system, this condition may appear rather stringent, as no equilibration theorem ensures that all those observables will be close to their microcanonical average at any time. Only local and coarse-grained observables are typically shown to equilibrate \cite{Deutsch91,Linden09,Short11,Goldstein15,Gogolin16}. However, several additional properties help enforcing this requirement in many practical situations for many-body systems.

First, the evolution of large system is rarely completely unitary. In contrast, even on time-scale over which dissipation in the environment may be neglected, noise from boundaries of the system typically leads to an internal variable dynamics which is non-unitary and unital (i.e. a pure-dephasing channel), which can enforce this equilibration.

Second, the timescale relevant to describe the macroscale is typically much slower than that of the microscale; then $\rho(t)$ shouldn't be seen as the instantaneous microstate but rather to be averaged over time steps characterizing the macroscopic dynamics. Once again, this averaging can lead to a non-unitary dynamics and justify that this object may match its coarse-grained counterpart \cite{Linden09,Short11}.

Finally, internal variables may be out of equilibrium but dynamically decoupled from the macroscopic ones. In that case, while the system is not strictly speaking at internal equilibrium, one can still safely ignore the internal variables without expecting any apparent violations of the laws of thermodynamics, at least for a large class of initial states. 

The combination of all three effects delineates a large class of situations where the concepts of thermodynamics can be directly applied to quantum many-body systems. We stress however that our framework is not strictly restricted to internal equilibrium. In contrast, it allows us to express quantitative corrections to the usual laws of thermodynamics when departing from this condition.

\section{Commutative algebra: Second law and fluctuation theorem for the observational entropy}
\label{sec: two point thermodynamics}

\begin{figure}[th!]
\begin{framed}
\centering
\includegraphics[width=0.70\textwidth]{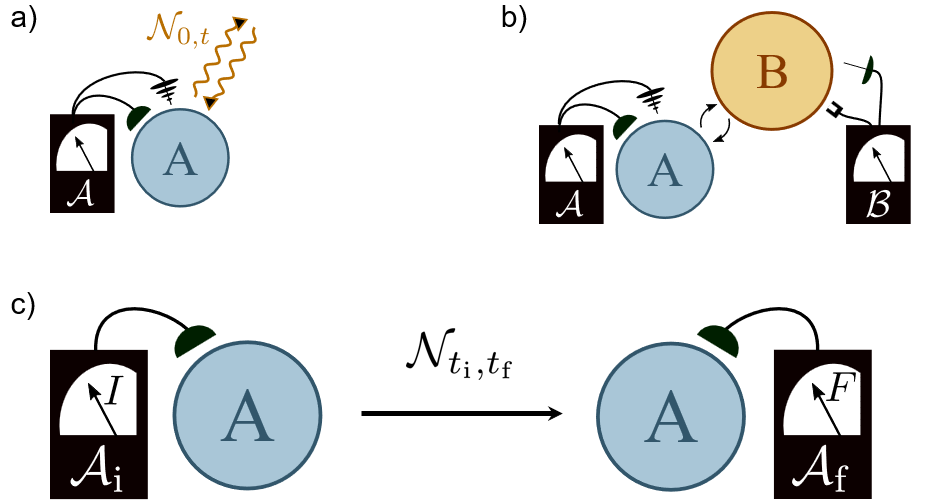}

\caption{
{\bf Probing a thermodynamic process}. a),b) In this article, we take two different standpoints to formulate our thermodynamic description: a) {\bf Open-system viewpoint}: $A$ is the (possibly many-body) quantum system of interest. Its evolution is assumed to be given by an unknown map ${\cal N}_{0,t}$ which may be unitary (in particular if the system is isolated) or not (if the system is in contact with an inaccessible environment). The observer has access to a measurement setup allowing to monitor the observables generating a subalgebra ${\cal A}$ of the full set ${\cal B}({\cal H}_A)$ of operators for $A$.  b) {\bf Autonomous viewpoint}: $A$ and $B$ are two quantum systems isolated from the rest of the world. Their joint evolution is therefore unitary. The observer's measurement setup allow to measure observables from subalgebras ${\cal A}_A$ and ${\cal A}_B$ of the total sets of operators for the system A $\cup$ B. In both cases, our framework defines entropy functions and formulates the laws of thermodynamics based on the obtained measurement outcomes. c) {\bf
Coarse-grained two-point measurement protocol}. The building block of a thermodynamic protocol consists in probing the state of the system at the beginning and at the end of a process. The initial and final accessible algebras need not to be the same, allowing the observer to probe the system across different incompatible bases and therefore reach a genuinely quantum thermodynamic description of the process. Average and stochastic thermodynamic quantities can be related to the distribution of initial and final macrostates, labeled by the stochastic variables $I$ and $F$, respectively.}
\label{fig12}
\end{framed}
\end{figure}

For the sake of pedagogy, throughout this section and Section \ref{sec: heat and work}, we focus on the case of commutative accessible algebras ($\mathcal{A} = \mathcal{A}^c \subset \mathcal{B}(\mathcal{H})$) generated by sets of projectors, i.e. the coarse-graining results from the projective measurement of a set of commuting observables. The thermodynamics of the non-commutative setting is treated in Section \ref{sec: non commutative subalgebras}.

\subsection{Two point measurement and average second law}

The main ingredients of our thermodynamic formulation consist of a system A, described by a Hilbert space $\mathcal{H}_{\text{A}}$ of dimension $D_{\text{A}}$, and an environment B, described by a Hilbert space $\mathcal{H}_{\text{B}}$ of dimension $D_{\text{B}}$. We will consider two standpoints (see Fig.~\ref{fig1}a-b): an open-system viewpoint where only the system A is assumed to be accessible, and an autonomous viewpoint where A and B are assumed to form a closed system and both can be measured separately, with possibly different coarse-graining schemes.

The system and the environment interact over the time interval $(t_\mathrm{i},t_\mathrm{f})$. We assume that this process is probed via a two-point measurement scheme: A quantum measurement $C_{\mathrm{i}}$ is performed at the initial time $t_\mathrm{i}$ on A, and possibly on B, while a second quantum measurement $C_{\mathrm{f}}$ is performed on A, and possibly on B, at the final time $t_\mathrm{f}$ (see Fig.~\ref{fig1}c).
The initial and final measurements are arbitrary, possibly different and non-commuting, projective measurements. 

Performing initial and final projective measurements is equivalent to specifying the joint values of sets of commuting observables at the initial and final times. These observables generate the respective commutative subalgebras $\mathcal{A}^c_{\text{A},\mathrm{i}}$ and $\mathcal{A}^c_{\mathrm{A},\mathrm{f}} \subset \mathcal{B}(\mathcal{H}_{\text{A}})$ for system A. In the open-system approach, where no confusion can arise, the labels A will be omitted for ${\cal A}^c_{\mathrm{A}}$, ${\cal H}_\mathrm{A}$ and $D_\mathrm{A}$. In the autonomous approach, one must additionally consider the initial and final accessible subalgebras $(\mathcal{A}^c_{\text{B},\mathrm{i}}, \mathcal{A}^c_{\text{B},\mathrm{f}}) \subset \mathcal{B}(\mathcal{H}_{\text{B}})$ for environment B.

\subsubsection{Open system viewpoint}\label{sec: 2nd law commut open system}

In this subsection, we take the open system viewpoint and therefore focus on the dynamics of system A alone, assuming that we have no access to its environment B (see Fig.~\ref{fig1}b). The latter only appears in the fact that between the two measurements  $C_\ii$ and $C_\ff$ performed by the observer, we assume a possibly non-unitary evolution for system A.
Importantly, we \emph{do not} require this dynamics (resulting from the combined action of B, the internal dynamics of A and the measurements) to be exactly known.  
In contrast, we only assume that the observer has access to the macrostate fixed point of the total dynamics of A \footnote{One can characterize this fixed point via tomography, after repeating the protocol (measurements and evolution) many times until the system has relaxed to this macroscopically stationary state.}

From the open system viewpoint, the fixed point of the dynamics can be seen as the state in which the system has exhausted all available resources. Indeed, if the system has reached this macrostate, repeating the same protocol does not change any accessible observable of the system; consequently, system A and its effective environment composed of B and of the measurement setup can be considered in relative equilibrium, from what the observer can judge. Consequently, the entropy production associated with the protocol, as estimated by this observer, must vanish when the system is in the fixed point $\s$ at $t_\mathrm{i}$\footnote{In this open viewpoint, we consider that no distinction can be made between equilibrium of A with its environment and a nonequilibrium steady state of A, since distinguishing between the two requires access to information about the environment, such as the currents flowing into and out of the system. This limitation will be addressed within the autonomous viewpoint introduced in Section \ref{subsubsection: bipartite approach}.}.

\vspace{0.3 cm}

Similar observations have led to the definition of entropy production, in the context of quantum open systems, as the variation of the relative entropy between the actual state of the system and the stationary state \cite{lindblad1975completely,Spohn78,Alicki79,Petz86,Landi21}
\be \label{eq: entropy prod 1}
    \Sigma_{t_\mathrm{i}, t_\mathrm{f}} := S(\rho(t_\mathrm{i}) \lvert \lvert \s) - S(\rho(t_\mathrm{f}) \lvert \lvert \s),
\ee 
where $\rho(t)$ is the density operator of the system at time $t$ and $\rho(t_\ff)={\cal N}_{t_\ii,t_\ff}[\rho(t_\ii)]$, with ${\cal N}_{t_\ii,t_\ff}$ the map describing the open system dynamics, and $\sigma={\cal N}_{t_\ii,t_\ff}(\sigma)$ is a fixed point of  ${\cal N}_{t_\ii,t_\ff}$ assumed to be full rank\footnote{The set of density operators that have are not full rank (i.e. the populations of some subspaces are exactly zero) has measure zero and can then be excluded without loss of generality.}.
If the map 
${\cal N}_{t_\ii,t_\ff}$ is (C)PTP, the fixed point $\sigma$ is ensured to exist \cite{Watrous2018Theory, Tumulka2024FixedPoints} and $\Sigma_{t_\ii,t_\ff} \geq 0$. This is, in particular, the case if the system and environment are initially uncorrelated.
If in addition it is (C)P-divisible, then the dynamics is said to be Markovian. For such Markovian dynamics, one can often introduce the generator of the dynamics, i.e. a GKLS master equation, and derive a positive entropy production rate $\dot\Sigma = -\frac{d}{dt}S(\rho(t)\|\sigma)$ \cite{Spohn78,Alicki79} . For general, possibly non-Markovian dynamics, however, information backflow from the environment can cause this quantity to increase \cite{Strasberg19,Landi21,Colla2025Thermodynamic,Picatoste26,Theret2026Entropy}.

When taking into account the possibility of initial correlations with the environment B, one can instead prove (see Appendix \ref{app: proofs}):
\be \label{eq: entropy prod rate}
    \Sigma_{t_\mathrm{i}, t_\mathrm{f}} = \Delta S_{\text{v.N}} - \Delta \langle K \rangle_\rho \geq - I_{AB}(t_\mathrm{i})
\ee
where
\be
    K := - \ln{\sigma}
\ee
is the \textit{modular Hamiltonian} attached to 
a fixed point $\s \in \mathcal{D}(\mathcal{H})$ of the CPTP map induced by initially factorized states \footnote{The CPTP character of the map induced by initially factorized states guarantees the existence of at least one such fixed point $\sigma$. (see the related proofs in Appendix \ref{app: proofs} for details)}, and $I_{AB}(t_\mathrm{i})$ is the initial quantum mutual information between A and B. $\Delta \langle K \rangle_\rho = \text{Tr}\{K[\rho(t_{\text{f}})-\rho(t_{\text{i}})]\}$ therefore plays the role of the entropy flow (i.e. $\frac{Q}{T_{R}}$ for a monothermal process) for a general map.

\vspace{0.3 cm}

\begin{figure}[t]
\begin{framed}
\centering
\includegraphics[width=0.96\textwidth]{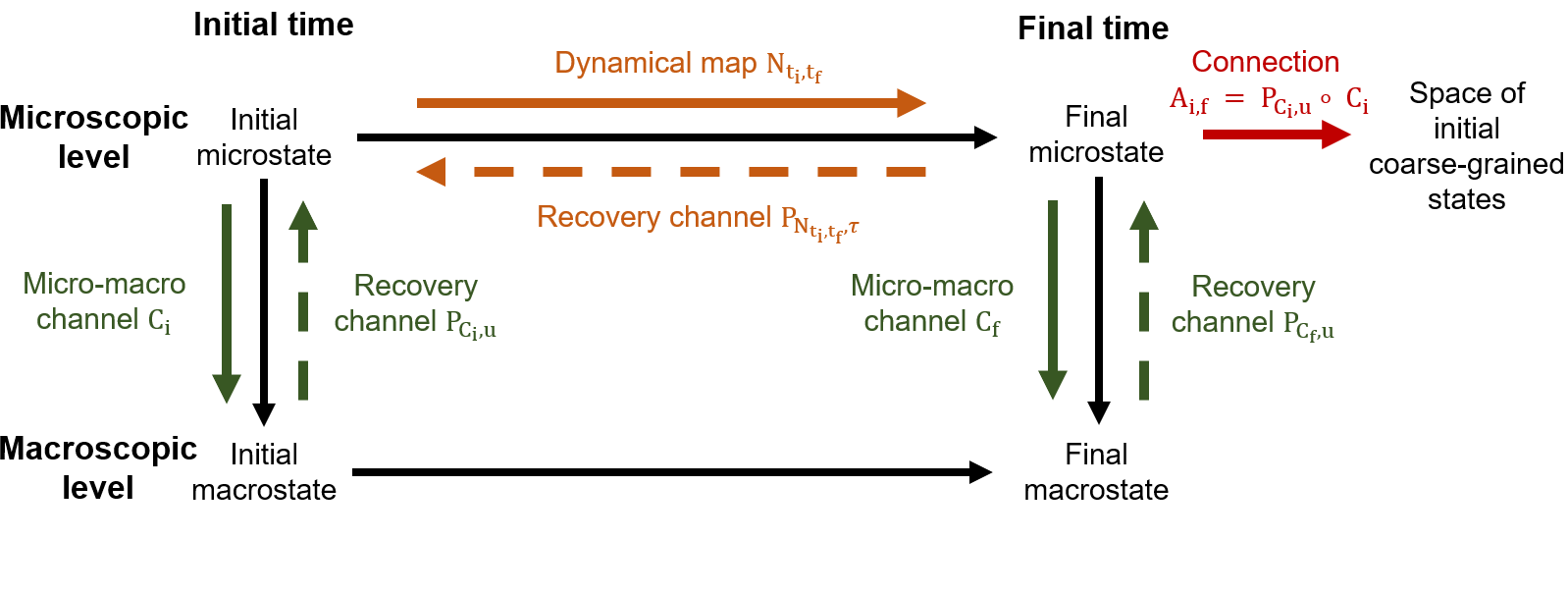}
\caption{This diagram provides a visual representation of the framework. The vertical direction is kinematical, corresponding to successive levels of coarse-graining, whereas the horizontal direction is dynamical, corresponding to time evolution. The natural orientation of the diagram is therefore towards the right (time direction) and downwards (coarse-graining or "spatial" direction). Any vertex can be connected to a neighboring vertex by an arrow representing a quantum channel. The elementary arrows are given by the microscopic evolution map $\mathcal{N}_{t_{\ii},t_{\ff}}$ and the coarse-graining maps $C_{\ii}$ and $C_{\ff}$. More general arrows are obtained by composing these elementary maps and, when necessary, reversing their direction. The composition of arrows corresponds to the composition of the associated quantum channels, while a reversed arrow is represented by the Petz recovery map associated with the original channel. As an illustration, the map connecting the initial macrostate to the final microstate is obtained by composing three elementary arrows. The corresponding quantum channel is $ C_{\ff} \circ \mathcal{N}_{t_{\ii},t_{\ff}} \circ P_{C_{\ii}, u} $.
}
\label{fig: diagcovfram}
\end{framed}
\end{figure}

However, Eq.\eqref{eq: entropy prod rate} assumes that the observer has access to the full algebra $\mathcal{B}(\mathcal{H})$, and therefore only quantifies entropy production arising from the stochasticity of the dynamics.
 Therefore, we are looking for an extension of Eq.\eqref{eq: entropy prod rate} to the case where one 
 has only access to the initial and final measurement outcome statistics. In this setting, one \textit{has to} treat on the same footing the part of the entropy production coming from the dynamics and the part of the entropy production coming from the measurement and coarse-graining. We reunite those two types of irreversibility in a unifying covariant picture where the total evolution of the system \emph{macrostate} is decomposed into a kinematical part (associated to the initial and final coarse-grainings), and a dynamical part (see Fig.~\ref{fig: diagcovfram}).
  
Namely, denoting by $\{\Pi_I\}$ ($\{\Pi_F\}$) the set of projectors associated with the initial measurement $C_\ii$ (final measurement $C_\ff$), and $\ket{I}$ ($\ket{F}$) the corresponding bases of marcrostate, we can build the map connecting the initial microstate $\rho(t_\mathrm{i})$ to the final macrostate $\bar{\rho}(t_\ff) = C_{\text{f}}\circ \mathcal{N}_{t_\mathrm{i}, t_\mathrm{f}}(\rho(t_\ii))$.
However, it is not possible to invert the initial map $C_{\text{i}}$ to connect the initial macrostate $\bar{\rho}(t_\ii)$ to the initial microstate $\rho(t_\mathrm{i})$. One can only infer the coarse-grained state associated with the initial macrostate (see Section \ref{sec:cgdef}). We therefore expect that our second law will contain a correction term associated with the distance between the initial state $\rho(t_\mathrm{i})$ and its associated coarse-grained state $\rho^{\text{cg}}(t_\mathrm{i})$ given by \eqref{eq: coarse-grained state}, which should vanish if the appropriate internal equilibrium conditions are satisfied at the initial time, for the chosen measurement scheme.
 
 Another obstacle is to extend the notion of fixed points in the case in which the commutative algebras $\mathcal{A}_{\ii}^c$ and $\mathcal{A}_{\ff}^c$, generated by the set of projectors $\{ \Pi_I \}$ and $\{ \Pi_F \}$ respectively, are not identical (i.e., the initial and final measurement bases do not commute). Basically, it corresponds to asking what the notion of equilibrium means in such a context. 
 Let us consider that the protocol composed of the initial measurement, the evolution by map ${\cal N}_{t_\ii,t_\ff}$ and the final measurement) is repeated many times. One can compare the macrostate obtained in one iteration of the protocol and in the next one. Then we consider that no resource can be extracted anymore from the ``environment'' of the system when the same macrostate is found in consecutive iterations. As the dynamics at the macrostate level arise from both the interaction with the actual environment B and the measurements, this consideration extends the notion of equilibrium to become a property relative to the overall protocol. 
This idea is formalized by introducing the notion of \emph{connection}
 \begin{align} \label{eq: connection}
    \text{A}_{\ii,\ff}: \mathcal{A}^c_\ff &\longrightarrow \mathcal{A}^c_\ii \nn \\
    \rho^{\text{cg}}_\ff &\longrightarrow \rho^{\text{cg}}_\ii = P_{C_\ii, u} \circ C_\ii (\rho^{\text{cg}}_\ff)
 \end{align}
 mapping the final state of an iteration of the protocol to the initial coarse-grained state of the next iteration of the protocol. We can now use as an equilibrium state a fixed point $\sigma_{\text{i}} = \tilde{\Lambda}_{t_\mathrm{i}, t_\mathrm{f}}(\sigma_{\text{i}}) \in \mathcal{A}^c_\ii$ of the map
  \begin{align} \label{eq: tildelambda}
        \tilde{\Lambda}_{t_\mathrm{i}, t_\mathrm{f}}: \mathcal{A}^c_\ii &\longrightarrow \mathcal{A}^c_\ii \nn \\
    \rho^{\text{cg}}_\ii(t_\ii) &\longrightarrow \rho^{\text{cg}}_\ii(t_\ff) = \text{A}_{\ii,\ff} \circ \Lambda_{t_\mathrm{i}, t_\mathrm{f}} (\rho^{\text{cg}}_\ii(t_\ii))
 \end{align}
 where 
 \begin{align} \label{eq: Lambdatitf}
    \Lambda_{t_\mathrm{i}, t_\mathrm{f}}: \mathcal{B}(H) &\longrightarrow \mathcal{A}^c_{\text{f}} \nn \\
    \rho(t_\mathrm{i}) &\longrightarrow  \rho^{\text{cg}}(t_\ff) =  P_{C_\text{f}, u} \circ C_\text{f} \circ \mathcal{N}_{t_\mathrm{i}, t_\mathrm{f}}(\rho(t_\mathrm{i}))
 \end{align}
maps an arbitrary initial state $\rho(t_\ii)$ at time $t_\mathrm{i}$ within $\mathcal{B}(\mathcal{H})$ into the final coarse-grained state at time $t_\mathrm{f}$. If $\mathcal{N}_{t_\ii,t_\ff}$ is CPTP, as is the case, for instance, when the system and environment are initially uncorrelated, then \eqref{eq: Lambdatitf} is also CPTP and therefore admits a fixed point. 
 We also introduce the image of the fixed point of \eqref{eq: tildelambda} under the map \eqref{eq: Lambdatitf}, $\s_\ff = \Lambda_{t_\mathrm{i}, t_\mathrm{f}}(\s_\ii)$.  
The quantum state $\s_\ii$ (resp. $\s_\ff$) belongs to the algebra $\mathcal{A}^c_{\ii}$ (resp. $\mathcal{A}^c_\ff$) so that the corresponding modular Hamiltonian 
\be
K_{\ii}:=-\ln\sigma_\ii \quad(\text{resp.}\;K_{\ff}:=-\ln\sigma_\ff) 
\ee
is fully accessible from the chosen initial (final) measurement schemes.

 Within this context, and now allowing the potential presence of initial correlations between the system and the environment, we prove (see Appendix \ref{app: proofs}) a coarse-grained version of Eq.\eqref{eq: entropy prod rate} involving the variation of observational entropy 
 \begin{align} \label{eq: generic second law commalg opens}
     \Delta S^{\text{ob}}(\rho) - \Delta \langle K_{\ii,\ff} \rangle_\rho \geq - I_{\text{ext}}(t_\mathrm{i}) - I_{\text{int}}(t_\mathrm{i})
 \end{align}
 where 
 \begin{align}
     \Delta S^{\text{ob}}(\rho) &:= S^{\text{ob}}_{\ff}(\rho(t_{\ff})) - S^{\text{ob}}_{\ii}(\rho(t_{\ii})) \nn \\
     \Delta \langle K_{\ii,\ff} \rangle_\rho &:= \langle K_\ff \rangle_{\rho(t_{\ff})} - \langle K_\ii \rangle_{\rho(t_{\ii})}.
 \end{align} 
 Eq.~\eqref{eq: generic second law commalg opens} is the main result of this section. The left-hand side only involves quantities accessible in the measurement scheme. This inequality mimics the Clausius relation, up to the lower bounds
 \begin{align} \label{eq: Ilowerbounds}
     I_{\text{ext}}(t_\mathrm{i}) &:= I_{AB}(t_\mathrm{i}) = S(\rho_{AB}(t_\mathrm{i}) \lvert \lvert \rho(t_{\ii}) \otimes \rho_B(t_\mathrm{i})) \geq 0 \nn \\
     I_{\text{int}}(t_\mathrm{i}) &:= S(\rho(t_{\ii}) \lvert \lvert \rho^{\text{cg}}(t_\ii)) \geq 0
 \end{align}
 so that 
 \be
    I_{\text{ext}}(t_\mathrm{i}) + I_{\text{int}}(t_\mathrm{i}) = S(\rho_{AB}(t_\mathrm{i}) \lvert \lvert \rho^{\text{cg}}(t_\ii) \otimes \rho_B(t_\mathrm{i})) \geq 0
 \ee
 The lower bounds in \eqref{eq: Ilowerbounds} can be interpreted as additional resources that may be expended to reduce entropy. The mutual information term $-I_{\mathrm{ext}}(t_{\mathrm{i}})$ (already present in the form $-I_{AB}(t_i)$ in \eqref{eq: entropy prod rate}) provides a first example. Its presence on the right-hand side of Eq.~\eqref{eq: generic second law commalg opens} signals the well-established possibility of apparent violations of the second law associated with the consumption of initial system-environment correlations \cite{Sagawa2012, Sagawa2013, Parrondo2015}. Such correlations constitute a thermodynamic resource, allowing the relative entropy to increase during an evolution consuming them. This process is a hallmark of information thermodynamics, and the term $-I_\text{ext}(t_\ii)$ is simply its signature in our coarse-grained framework. 
 
 However, since the observer has only access to the initial macrostate rather than the underlying microstate, the mismatch between the microscopic state $\rho(t_{\ii}) \in \mathcal{B}(\mathcal{H})$ and its coarse-grained (equilibrium) counterpart $\rho^{\text{cg}}(t_\ii) \in \mathcal{A}_{\ii}^c \subset \mathcal{B}(\mathcal{H})$ also potentially constitutes an additional hidden resource. More precisely, the relative entropy $I_{\text{int}}(t_\mathrm{i})$ quantifies the correlations with the inaccessible variable, which may affect the macroscopic evolution and lead to a decrease of the macroscopically accessible entropy, and hence an apparent violation of the coarse-grained second law.   
 The term $-I_\text{int}(t_\ii)$ vanishes if the system is initially at internal equilibrium (see section.~\ref{sec: internal equilibrium}), but also if the dynamics of the inaccessible degrees of freedom do not impact the accessible dynamics. The latter condition can be expressed via
 \be \label{eq: faithful map}
 C_{\mathcal{A}_\text{f}} \circ \mathcal{N}_{t_{\text{i}}, t_{\text{f}}} = C_{\mathcal{A}_\text{f}} \circ \mathcal{N}_{t_{\text{i}}, t_{\text{f}}} \circ (P_{C_{\mathcal{A}_\text{i}}, u} \circ C_{\mathcal{A}_\text{i}}),
\ee
and may hold under a variety of circumstances.
If \eqref{eq: faithful map} holds, then we will say in the following that the dynamics induced by the map $\mathcal{N}_{t_\ii, t_\ff}$ is \emph{scale divisible} with respect to the initial and final algebras $\mathcal{A}_\ii$ and $\mathcal{A}_\ff$ \footnote{Note that the formulation Eq.~\eqref{eq: faithful map} holds equally for the case where the subalgebras are non-commutative, hence the formulation. In the commutative setting, take $\mathcal{A}_\ii=\mathcal{A}_\ii^c$ and $C_{\mathcal{A}_\ii}=C_\ii$,  $\mathcal{A}_\ff=\mathcal{A}_\ff^c$ and $C_{\mathcal{A}_\ff}=C_\ff$.}. 

\begin{remark}
    More precisely, we do not have to require the dynamics to be scale divisible for \emph{all} states in $\mathcal{B}(\mathcal{H})$, but only for an \text{overwhelming majority} of them. Indeed, thermodynamic laws such as the second law may be violated for highly fine-tuned initial states, while remaining valid for the vast majority of physically relevant states.
\end{remark}

\begin{remark}
The notion of a scale divisible map can be understood in close analogy with that of a CPTP map. If the system evolves as if it were initially uncorrelated with its environment at $t_\ii$, then its reduced dynamics is described by a CPTP map. Likewise, a dynamics is scale divisible if it proceeds as if the system were initially in its coarse-grained state with respect to the algebra $\mathcal{A}_{\ii}$, i.e. as if the inaccessible internal degrees of freedom were at equilibrium. In this sense, while CPTP reduced dynamics is associated with the absence of information backflow from the environment to the system, scale divisible dynamics is associated with the absence of information backflow from the ignored microscopic degrees of freedom to the observed macroscopic ones.
\end{remark}

Notice that when performing the initial projective coarse-grained measurement on the initial state $\rho(t_{\ii})$, the measurement postulate of quantum mechanics yields an updated system state
 \be \label{eq: perturbmleasurement}
    \rho(t_{\ii}) \longrightarrow \rho(t_{\ii}^+) = \sum_I \Pi_I \rho(t_{\ii}) \Pi_I.
 \ee
Taking into account this back-action (which assumes that both the final and initial measurements are performed in each run of the protocol) requires replacing $\rho(t_{\ii})$ with $\rho(t_{\ii}^+)$, and $I_{\text{in}}(t_\mathrm{i})$ in \eqref{eq: Ilowerbounds} by $I_{\text{in}}(t_\mathrm{i}^+) = S(\rho(t_{\ii}^+) \lvert \lvert \rho^{\text{cg}}(t_{\ii}))$. Typically, one expects $I_{\text{in}}(t_\mathrm{i}^+)\leq I_{\text{in}}(t_\mathrm{i})$ as the measurement back-action suppresses coherence between the different blocks $I$ which is absent in the coarse-grained state. In this case, one can prove a bound on the possible violations of the second law due to coarse-graining, namely (see App.~\ref{app: proofs}) 
 \be \label{eq: lower boundpivi}
    I_{\text{int}}(t_\mathrm{i}^+) \leq \sum_I p_I \ln{V_I}.
 \ee
 The bound depends only on the initial probability distribution $\{ p_I \}$ characterizing the initial macrostate and the initial projector ranks $V_I$. A toy example illustrating the necessity of the term $-I_{\text{int}}(t_\mathrm{i})$ in Eq.~\eqref{sec: 2nd law commut open system} and the tightness of bound Eq.~\eqref{eq: lower boundpivi} is provided in App.~\ref{app: Maxwell demons}. 
 
As expected, Eq.~\eqref{eq: generic second law commalg opens} captures entropy production even if the evolution of the system is unitary, owing to coarse-graining. This is illustrated in App.~\ref{app: spontaneous emission} on the example of the spontaneous emission, where the dependency on the accessible algebra is analyzed.

\begin{remark}
In Ref.~\cite{RubinoBruknerManzano2026}, a coarse-grained second law similar to Eq.~\eqref{eq: generic second law commalg opens} was obtained. There, the lower bounds are set to zero by assuming (i) the absence of initial correlation with the environment, and (ii) that the system is \emph{reprepared} in coarse-grained state by the observer after the initial measurement, i.e. whenever outcome $I$ is obtained, the system is prepared in state $\Pi_I/V_I$. This protocol therefore enforces internal equilibrium.
\end{remark}

We finish by mentioning that so far, the formulation of the second law via Eq.~\eqref{eq: generic second law commalg opens} relates purely informational quantities. The connection to energy transfers and the first law will be discussed in section~\ref{sec: heat and work}.

\subsubsection{Autonomous viewpoint}
\label{subsubsection: bipartite approach}

The open system formulation of the second law \eqref{eq: generic second law commalg opens} introduced in Sec.~\ref{sec: 2nd law commut open system} has two major weaknesses. First, regarding the role of the system-environment mutual information. The lower bound involves the initial amount of mutual information rather than the variation of mutual information as in other existing formulations of the second law in the absence of coarse-graining \cite{Jennings10,Sagawa2012,Horowitz14,Landi21,Elouard23}. This is because one cannot quantify the final mutual information from a measurement on the system only, and there is therefore no way to check whether mutual information was consumed during the process. 
Second, the thermodynamic description based on the system only is somehow incomplete as several environments may lead to the same reduced system dynamics, while being associated to radically different thermodynamic balances. For instance, one could not distinguish an environment that is composed of several reservoirs at different temperatures from an environment able to perform work via a chemical potential. The actual resources consumed can be read unambiguously in the dynamics of environment B \cite{Elouard23}. Moreover, quantum stochastic thermodynamics for open systems is also typically based on measuring the environment \cite{Campisi11,Elouard17,Manzano18}. While perfectly tracking the environment state is often intractable, it is often possible to perform some limited measurements (e.g. measure currents from the system). Our coarse-grained thermodynamic description is well suited to address those scenarios and then solve the aforementioned issues.

To do so, we consider an autonomous situation where the system and its environment (or simply, two systems of interest) are forming a closed system. One can then apply the approach of the previous subsection by considering that A $\cup$ B plays the role of the former system, and taking the map ${\cal N}_{t_\ii,t_\ff}$ to be the unitary evolution due to the Hamiltonian of A$\cup$ B.
Moreover, we consider that at any time, the accessible algebra ${\cal A}^c = \mathcal{A}^c_{\text{A}} \otimes \mathcal{A}^c_{\text{B}}$ attached to the join system A $\cup$ B is spanned by a set of projectors $\{ \Pi_{J_A} \otimes \Pi_{J_B} \}$ so that 
\be \label{eq: macro volume joint def}
    V_{J_A J_B} = \tr{(\Pi_{J_A} \otimes \Pi_{J_B})} = V_{J_A} V_{J_B}
\ee
while in general
\be \label{eq: macro proba joint def}
    p_{J_A J_B} = \tr{(\rho_{AB}(\Pi_{J_A} \otimes \Pi_{J_B}))} \neq \tr{((\rho_A \otimes \rho_B)(\Pi_{J_A} \otimes \Pi_{J_B}))} = p_{J_A} p_{J_B}.
\ee
Then, one can define the observational mutual information as 
\begin{align}
    I^{\text{ob}}_{AB} 
    &:= S(\rho^{\text{cg}}_{AB} \lvert \lvert \rho^{\text{cg}}_{A} \otimes \rho^{\text{cg}}_{B} ) =  \sum_{J_A, J_B} p_{J_A J_B} \ln{\frac{p_{J_A J_B}}{p_{J_A} p_{J_B}}} \geq 0
\end{align}
measuring the classical correlations between the macrostates.  Of course, the commutative subalgebra $\mathcal{A}^c_{\text{A}, \ff} \otimes \mathcal{A}^c_{\text{B}, \ff}$ (resp. $\mathcal{A}^c_{\text{A}, \ii} \otimes \mathcal{A}^c_{\text{B}, \ii}$) considered at the final time $t_\ff$ (resp. initial time $t_{\ii}$) is generated by the set of projectors $\{ \Pi_{F_A} \otimes \Pi_{F_B} \}$ (resp. $\{ \Pi_{I_A} \otimes \Pi_{I_B} \}$). Since the joint system $A \cup B$ is isolated, the joint dynamic is unitary and the map $\Lambda_{t_\ii,t_\ff}$ defined in Eq.~\eqref{eq: Lambdatitf}
is unital; we can therefore take $\sigma = u=\frac{\mathds{1}_{\mathcal{H}_\text{A} \otimes \mathcal{H}_\text{B}}}{D_A D_B}$ as a fixed point, yielding
\be \label{eq: total entropy}
    \Delta S^{\text{ob}}(\rho_{AB}) \geq - I_{\text{int}}(t_\mathrm{i}^+)
\ee
as a particular application of \eqref{eq: generic second law commalg opens}, where explicitly 
\be
    \Delta S^{\text{ob}}(\rho_{AB}) := S^{\text{ob}}_{\mathcal{A}_{A,\ff}^c \otimes \mathcal{A}_{B,\ff}^c}(\rho_{AB}(t_{\ff})) - S^{\text{ob}}_{\mathcal{A}_{A,\ii}^c \otimes \mathcal{A}_{B,\ii}^c}(\rho_{AB}(t_{\ii}))
\ee
and $I_{\text{int}}(t_\mathrm{i}^+) = S(\rho_{AB}(t_\mathrm{i}^+) \lvert \lvert \rho^{\text{cg}}_{AB}(t_\mathrm{i}) ) \leq \sum_{I_A I_B} p_{I_A I_B} \ln{V_{I_A I_B}}$ from \eqref{eq: lower boundpivi}. If the joint system $A \cup B$ is in internal equilibrium at the initial time $t_\mathrm{i}$, or alternatively if the map ${\cal N}_{t_\ii,t_\ff}$ is scale divisible with respect to the algebras considered at $t_{\ii}$ and $t_{\ff}$, the right hand side of \eqref{eq: total entropy} can be set to zero. We assume this to be the case from now on. Then, in this setting, \eqref{eq: total entropy} is equivalent to
\be \label{eq: bipartite ob scnd law}
    \Delta S^{\text{ob}}(\rho_A) + \Delta S^{\text{ob}}(\rho_B) = \Delta S^{\text{ob}}(\rho_{AB}) + \Delta I^{\text{ob}}_{AB} \geq \Delta I^{\text{ob}}_{AB} 
\ee
where
\begin{align}
    \Delta S^{\text{ob}}(\rho_A) &:= S^{\text{ob}}_{\mathcal{A}_{A,\ff}^c}(\rho_A(t_{\ff})) - S^{\text{ob}}_{\mathcal{A}_{A,\ii}^c}(\rho_A(t_{\ii})) \nn \\
    \Delta S^{\text{ob}}(\rho_B) &:= S^{\text{ob}}_{\mathcal{A}_{B,\ff}^c}(\rho_B(t_{\ff})) - S^{\text{ob}}_{\mathcal{A}_{B,\ii}^c}(\rho_{B}(t_{\ii}))
\end{align}
and 
\begin{align}
    \Delta I^{\text{ob}}_{AB} &=  S(\rho^{\text{cg}}_{AB}(t_\ff) \lvert \lvert \rho^{\text{cg}}_{A}(t_\ff) \otimes \rho^{\text{cg}}_{B}(t_\ff) ) -  S(\rho^{\text{cg}}_{AB}(t_\ii) \lvert \lvert \rho^{\text{cg}}_{A}(t_\ii) \otimes \rho^{\text{cg}}_{B}(t_\ii) ) .
\end{align}
Equation \eqref{eq: bipartite ob scnd law} implies that the time variation of the observational entropies attached to system A and environment B is positive up to the variation of the observational mutual information. This is in line with the known fact that negative apparent entropy production may be observed if some initial correlations with the environment are consumed 
\cite{Sagawa2012, Sagawa2013, Horowitz2014, Parrondo2015}.

As a first advantage over Eq.~\eqref{eq: generic second law commalg opens} is that the variation in mutual information between A and B appears, rather than the initial mutual information with the environment. The bound is therefore tighter, e.g. predicting a positive entropy production when there is a nonzero initial mutual information but the dynamics cannot consume it.

The second law from the autonomous viewpoint Eq.~\eqref{eq: bipartite ob scnd law} (just as Eq.~\eqref{sec: 2nd law commut open system}) still involves only informational quantities. A first connection to the energy balance and therefore the first law is however now possible owing to the information available about the state of environment B \cite{Elouard23}. To do so, we express the initial coarse-grained state of environment B (assumed to be of full-rank) $\rho^{\text{cg}}_{B}(t_i)$ in terms of its modular Hamiltonian
\be \label{eq: mod ham Bti 2p measurement}
    K_{\text{B}, \ii} = - \ln{\rho^{\text{cg}}_{B}(t_i)} \in \mathcal{A}^c_{\text{B}, \ii}. 
\ee
Since the modular Hamiltonian \eqref{eq: mod ham Bti 2p measurement} is a self-adjoint (coarse-grained) operator, we can write it as a linear combination of a set of operators $\{ \Id_{\text{B}}, H_{B, \text{i}}, O_{B, \text{i}}^1, \cdots, O_{B, \text{i}}^N \}$ that form an orthogonal basis of the initial algebra $\mathcal{A}_{\text{B},\ii}^c$, assumed to be finite dimensional. These operators represent system observables that can be manipulated at the macroscopic level to perform work on the system. In particular, $\Id_{\text{B}}$ is the identity operator on $\mathcal{H}_{\text{B}}$ that is always part of $\mathcal{A}^c_{\text{B}, \ii}$ and 
\be
    H_{B, \text{i}} := \sum_{I_B} H_{I_B} \Pi_{I_B}, \qquad H_{I_B} := \tr{\left(H_{\text{B}}(t_{\ii}) \frac{\Pi_{I_B}}{V_{I_B}}\right)}
\ee
is the representative of the Hamiltonian $H_{\text{B}}(t_{\ii})$ of bath B of the initial algebra $\mathcal{A}^{c}_{\text{B}, \ii}$ attached to bath B. Likewise, we have 
\be
   \forall k \in [1,N], \qquad O_{B, \text{i}}^k := \sum_{I_B} O_{I_B}^k \Pi_{I_B}, \qquad O_{I_B}^k := \tr{ \left(O_{B, \text{i}}^k \frac{\Pi_{I_B}}{V_{I_B}}\right)}. 
\ee
Thus, one has
\be \label{eq: KBt in op basis}
    K_{\text{B}, t_{\ii}} := \beta(t_{\ii}) \left(H_{B, \text{i}} + \sum_{k=1}^N \l_{k}(t_{\ii}) O_{B, \text{i}}^k\right) + \ln{\left[\tr{\left(e^{- \beta(t_{\ii}) (H_{B, \text{i}} + \sum_k \l_{k}(t_{\ii}) O_{B, \text{i}}^k)}\right)}\right]} \Id_{\text{B}}
\ee
where the coefficient in front of the identity operator $\Id_{\text{B}}$ is always chosen for the state $\rho_{\text{B}, \ii}^{\text{cg}}$ to be normalized. The coefficients $\beta(t_\ii)$ and $\{\lambda_k(t_\ii) \}$ of the expansion are then set by the initial coarse-grained state of B.
Of course, $\beta(t_\ii)$ must be interpreted as the effective inverse temperature of B, while the other parameters $\{ \l_k(t_\ii) \}$ conjugated to the other accessible observables are other effective intensive parameters needed to describe the macroscopically out-of-equilibrium state of B. In addition, we require the temperature $T(t_\ii) = \frac{1}{\beta(t_\ii)}$ to be positive. Of course, when parameterizing an arbitrary 
quantum state, 
nothing prevents the coefficient multiplying the Hamiltonian from being negative. In that case however, as discussed in App.~\ref{app: negativetemperature}, one can always add an operator $O_{\text{B},\ii}^0$ proportional to the Hamiltonian $H_{B, \text{i}}$ in the linear combination \eqref{eq: Kt in op basis} with a coefficient $\lambda_0$ acting as a positive chemical potential, and then take a positive temperature coefficient.  In App.~\ref{app: negativetemperature}, we propose a systematic construction to identify those two coefficients inspired by the autonomous formulation of thermodynamics of Ref.~\cite{Elouard23}. 
Hence, 
\be \label{eq: bipartite ob scnd law 2}
    \rho^{\text{cg}}_{B, t_{\ii}} := e^{- K_{\text{B}, t_{\ii}}} = \frac{e^{- \beta(t_{\ii}) (H_{B, \text{i}} + \sum_k \l_{k}(t_{\ii}) O_{B, \text{i}}^k)}}{\tr{(e^{- \beta(t_{\ii}) (H_{B, \text{i}} + \sum_k \l_{k}(t_{\ii}) O_{B, \text{i}}^k)}})} = \sum_{I_B} \frac{e^{- \beta(t_{\ii}) (H_{I_B} + \sum_k \l_{k}(t_{\ii}) O_{I_B}^k)}}{\sum_{I_B} V_{I_B} e^{- \beta(t_{\ii}) (H_{I_B} + \sum_k \l_{k}(t_{\ii}) O_{I_B}^k)}} \Pi_{I_B}
\ee 
Injecting this ansatz in \eqref{eq: bipartite ob scnd law} leads to 
\begin{align} \label{eq: variation and law bipartite}
    \Delta I^{\text{ob}}_{AB} \leq \Delta S^{\text{ob}}(\rho_{AB}) + \Delta I^{\text{ob}}_{AB} + S(\rho^{\text{cg}}_{B, t_{\ff}} \lvert \lvert \rho^{\text{cg}}_{B, t_{\ii}}) &= \Delta S^{\text{ob}}(\rho_A) + \beta(t_{\ii}) \left(\Delta \langle H_{B, \text{i}} \rangle_{\rho^{\text{cg}}_B} + \sum_k \l_{k}(t_{\ii}) \Delta \langle  O_{B, \text{i}}^k \rangle_{\rho^{\text{cg}}_B}\right). 
\end{align}
with
\begin{align}
     \Delta \langle H_{B, \text{i}} \rangle_{\rho^{\text{cg}}_B} &:= \tr{((\rho_{B, t_{\ff}}^{\text{cg}} - \rho_{B, t_{\ii}}^{\text{cg}})  H_{B, \text{i}})}, \qquad 
     \Delta \langle O_{B, \text{i}}^k \rangle_{\rho^{\text{cg}}_B} := \tr{((\rho_{B, t_{\ff}}^{\text{cg}} - \rho_{B, t_{\ii}}^{\text{cg}})  O_{B, \text{i}}^k)}.
\end{align}
Now, interpreting Eq.~\eqref{eq: variation and law bipartite} as Clausius' relation, 

\be\label{eq: Clausius sec 3}
 \Delta S^{\text{ob}}(\rho_A) + \beta(t_{\ii}) \mathcal{Q}_{B,t_\ii, t_\ff} \geq 0
\ee
we see that the term 
\be \label{eq: heat def sec 3}
\mathcal{Q}_{B,t_\ii, t_\ff} := \Delta \langle H_{B, \text{i}} \rangle_{\rho^{\text{cg}}_B} + \sum_k \l_{k}(t_{\ii}) \Delta \langle  O_{B, \text{i}}^k \rangle_{\rho^{\text{cg}}_B},
\ee
plays the role of the heat flow into environment B, and
\bea \label{eq: work def sec 3}
    \mathcal{W}_{\mathcal{B}, t_\ii, t_\ff} &= \langle H_{\text{B}, \ff} \rangle_{\rho_\text{B}(t_\ff)} - \langle H_{\text{B}, \ii} \rangle_{\rho_\text{B}(t_\ii)} - \mathcal{Q}_{B,t_\ii, t_\ff}\nn\\
    &= \mean{(H_{B,\ff}-H_{B,\ii})}_{\rho_B^\text{cg}(t_\ii)} - \sum_k \lambda_k \Delta \langle  O_{B, \text{i}}^k \rangle_{\rho^{\text{cg}}_B},
\eea
that of the work received by environment B. The latter contains two expected contributions, related to the variation of the Hamiltonian of B and to the variation of the accessible observables $O_{B,\ii}^k$. While this identification may appear somewhat arbitrary at this stage, the notions of heat and work emerging from our formalism, together with their physical interpretation, are discussed in detail in section.~\ref{sec: heat and work}.

\begin{remark}
\label{remarkoninnerproduct}
We have not explicitly specified the inner product with respect to which the operator basis is orthogonal. When the underlying Hilbert space is finite-dimensional, a natural choice is the Hilbert-Schmidt inner product $\langle a,b\rangle=\operatorname{Tr}(a^\dagger b)$. In particular, the Hamiltonian can always be made orthogonal to the identity by subtracting an appropriate constant. However, when the underlying Hilbert space is infinite-dimensional, this inner product may diverge. One possibility is to restrict attention to sectors with finite volume $V_J=\Tr{(\Pi_J)}<\infty$, assuming that sectors with infinite volume have vanishing occupation probabilities in the state, as they would otherwise lead to divergent Boltzmann entropy. Alternatively, since the accessible algebra $\mathcal{A}_{B,\mathrm{i}}$ is finite-dimensional, it always admits a faithful representation on a finite-dimensional Hilbert space on which the ordinary Hilbert-Schmidt inner product can be defined. The Hilbert space of macrostates $\bar{\mathcal{H}}_B$, which is finite-dimensional when the accessible algebra is finite-dimensional, provides an example of such a faithful representation. This allows us to define a well-defined inner product on the full accessible algebra, independently of the dimension of the underlying Hilbert space.
\end{remark}

\subsection{Stochastic version of the second law}
\label{sec: stochastic two pt open system }

We are now equipped with two versions of the second law for the observational entropy, with the open and autonomous viewpoints. Both are formulated as average laws. However, the much more powerful approach of stochastic thermodynamics aims to characterize thermodynamic laws beyond the average by attaching an entropy production (and eventually a heat and work transfer) to each single realization of the protocol of interest. In our setting, the latter can be seen as stochastic trajectories in the space of macrostates, from an initial macrostate $\ket{I}$ to a final macrostate $\ket{F}$ obtained in a single realization of the initial and final measurements.  
In this section, we investigate the stochastic version of the algebraic approach to quantum thermodynamics, still focusing on commuting initial and final algebras. A non-commutative version is discussed in Section \ref{sec: non commutative subalgebras}. As was the case for the average law, one distinguishes here the open system viewpoint from the autonomous approach. 

\subsubsection{Open system viewpoint}\label{sec: stoch open}

We focus first on the open system scenario, so that the measurements are performed on system A only. To get the stochastic version, we need to look at the ratio of the forward and backward process probabilities. To simplify the discussion, one can take a vanishing initial mutual information between the system and the environment, but nothing prevents us from treating this case with the simplest appropriate modifications \cite{Manzano18}. First, we have to look at the probability of the forward and backward processes. Let $\ket{i}$ ($\ket{f}$) be one of the eigenvectors of $\Pi_I$ ($\Pi_F$) for some measurement result $I$ ($F$) of the initial (final) measurement. Then, the  probability attached to the trajectory $\g_{if}$, starting in the state $\ket{i}$ and ending in the state $\ket{f}$, is given by 
\be \label{eq: forward prob}
    p(\g_{if}) := p_i \bra{f} \Lambda_{t_\ii,t_\ff} [\ket{i} \bra{i}]\ket{f}
= p_i \bra{f} P_{C_\ff, u} \circ C_\ff \circ \mathcal{N}_{t_\mathrm{i}, t_\mathrm{f}} [\ket{i} \bra{i}] \ket{f}, \qquad p_i = \tr{(\rho(t_\ii) \ket{i} \bra{i})}
\ee 
We define the time-reversed trajectory ${\g}^R_{if}$ as starting from the final state $\ket{f}$ and ending in state $\ket{i}$ through the time-reversed map $\Lambda^R_{t_\mathrm{i}, t_\mathrm{f}}$. From Fig.~\ref{fig: diagcovfram}, it is clear that one can construct such a map, using only the information coming from the accessible observables, owing to the Petz recovery map associated with the forward map $\Lambda_{t_\ii,t_\ff}$.
This is exactly equivalent to the prescription for time-reversal introduced in Ref.~\cite{Crooks99}. The reference state of the Petz map should be the state in which there is no average entropy production, that is, macroscopic equilibrium, because it will enforce equal forward and backward trajectories in that state. From the discussion of Section \ref{sec: 2nd law commut open system}, we take as a reference state $\sigma_\ii$, i.e. $\Lambda^R_{t_\mathrm{i}, t_\mathrm{f}}=P_{\Lambda_{t_\mathrm{i}, t_\mathrm{f}}, \s_\ii}$, yielding:
\be \label{eq: backward prob}
    p({\g}^R_{if}) := p_f \bra{i} P_{\Lambda_{t_\mathrm{i}, t_\mathrm{f}}, \s_\ii} [\ket{f}\bra{f}]\ket{i} = p_f  \bra{i} \Lambda_{t_\ii,t_\ff}^\dag [\ket{f}\bra{f}]\ket{i}  e^{K_{\ff,F} - K_{\ii,I}}, \qquad p_f = \tr{\big( \Lambda_{t_\ii, t_\ff}(\rho(t_\ii)) \ket{f} \bra{f} \big)}
\ee
where $\Lambda_{t_\ii,t_\ff}^\dag$ is the adjoint map of $\Lambda_{t_\ii,t_\ff}$ and $K_{\ii,I}=\text{Tr}\left(K_\ii\frac{\Pi_I}{V_I}\right)$ [$K_{\ff,F}=\text{Tr}\left(K_\ff\frac{\Pi_F}{V_F}\right)$] is the stochastic value of the modular Hamiltonian associated with outcome $I$ at $t_\ii$ ($F$ at $t_\ff$). Following the usual theory of (quantum) stochastic thermodynamics 
\cite{Seifert12,Crooks99} we define the stochastic coarse-grained entropy production term attached to the trajectory $\g_{if}$ as
\be \label{eq: entropy creation term stoch}
    s_c(\g_{if}) := \ln{\frac{p(\g_{if})}{p({\g}_{if}^R)}} = s_f - s_i - K_{\ff,F} + K_{\ii,I} 
\ee
where $s_i := - \ln{p_i}$ and $s_f := - \ln{p_f}$. Since the image of $\Lambda_{t_\ii,t_\ff}$ being a coarse-grained state, we set $p_f = \frac{p_F}{V_F}$. The stochastic entropy production term \eqref{eq: entropy creation term stoch} can be positive or negative. However, its average  
\be \label{eq: average entropy stoch}
    \langle s_c(\g_{if}) \rangle = \sum_{if} p(\g_{if}) \ln{\frac{p(\g_{if})}{p({\g}^R_{if})}} \geq 0 
\ee
is non-negative, as it is given by a relative entropy. In addition, one has 
\be
    \langle s_c(\g_{if}) \rangle = S^{\text{ob}}_\ff (\rho(t_\ff)) - S_{\text{v.N}}(\rho(t_\ii^+)) - \Delta \langle K_{\ii, \ff} \rangle_{\rho} = S(\rho(t_\mathrm{i}^+) \lvert \lvert \s_\ii) - S(\rho^{\text{cg}}(t_\ff) \lvert \lvert \s_\ff) \geq 0,
\ee
which is equivalent to the monotonicity of relative entropy that we used to obtain \eqref{eq: generic second law commalg opens}. \\

The advantage of the stochastic approach is that it leads to equalities known as fluctuation theorems. We derive the so-called integral fluctuation theorem \cite{Seifert05} from
\be\label{eq:FT1}
\sum_{if} e^{-s_c(\gamma_{if})} p(\gamma_{if}) = \sum_{if | p(\g_{if})\neq 0}  p({\gamma}_{if}^R) := \xi \leq 1, 
\ee
where the case $\xi < 1$ occurs when there exist possible backward trajectories attached to a direct trajectory with strictly zero probability of occurrence (e.g. starting in an initially forbidden region of phase space). Such a phenomenon has been described as \emph{absolute irreversibility} \cite{Murashita2014, Funo2015}.

To interpret Eq.~\eqref{eq:FT1} as an operational fluctuation theorem, we need to express it in terms of the outcomes of the initial measurement actually performed $C_\ii$. Using that each microstate $i$ belongs to a single coarse-grained outcome $I$ and introducing  
\be \label{eq: pgammaIF}
    p(\gamma_{IF}) := \sum_{i\in I, f \in F}p(\gamma_{if})
\ee
and $s_f = s_F = -\ln{\frac{p_F}{V_F}}$, one has:
\be\label{eq:FT2}
\sum_{IF} p(\gamma_{IF}) \sum_{i\in I, f \in F} e^{-s_c(\gamma_{if})}\frac{p(\gamma_{if})}{p(\gamma_{IF})}  = \sum_{IF}  p(\gamma_{IF})e^{-s_F+s_I+K_{\ff,F}-K_{\ii,I}} \sum_{i\in I}e^{-s_I-\ln p_i}\frac{p(\gamma_{iF})}{p(\gamma_{IF})}   = \xi. 
\ee
where $s_I = - \ln{\frac{p_I}{V_I}}$. We finally introduce the stochastic quantity
\be\label{eq:cgstochEP}
s_{IF}^\text{cg} := -\ln \sum_{i\in I}e^{-s_I-\ln p_i}\frac{p(\gamma_{iF})}{p(\gamma_{IF})}.
\ee
which reflects the distance between the actual initial microstate distribution and the initial coarse-grained state and can be seen as a fluctuating counterpart to $I_\text{int}(t_\ii^+)$. Indeed, we prove (see App.~\ref{app: proofs}) 
\be \label{eq: average coarse grained stat}
    0\leq \langle s_{IF}^{\text{cg}}
    \rangle \leq I_{\text{int}}(t_\mathrm{i}^+).
\ee
Finally, we get
\be\label{eq:FT3}
\sum_{IF} p(\gamma_{IF}) e^{-s_F+s_I+K_{\ff,F}
-K_{\ii,I}-s_{IF}^\text{cg}}  = \mean{e^{-s_F+s_I+K_{\ff,F}
-K_{\ii,I}-s_{IF}^\text{cg}}} = \xi
\ee
which is the final form of the fluctuation theorem. Notice that convexity of the exponential implies that 
\be
    \langle -s_F+s_I+K_{\ff,F}
-K_{\ii,I}-s_{IF}^{\text{cg}} - \ln{\xi} \rangle \leq 0
\ee
which is equivalent to 
\be
    \Delta S^{\text{ob}}_{\ii,\ff}(\rho) - \Delta \langle K_{\ii, \ff} \rangle_\rho \geq - \langle s_{IF}^{\text{cg}} \rangle - \ln{\xi}
\ee
where $\ln{\xi} \leq 0$ so that the lower bound obtained from the stochastic approach is 
always tighter than that obtained at the average level.

\subsubsection{Autonomous viewpoint}

The autonomous viewpoint also admits a stochastic formulation leading to fluctuation theorems. 
In this setting, one takes $\mathcal{N}_{t_\mathrm{i}, t_\mathrm{f}} = {\cal U}_{t_\mathrm{i}, t_\mathrm{f}}$ to be a unitary map acting on the isolated composite system A $\cup$ B. Then, $i_{AB}$ and $f_{AB}$ now label a pair of initial and final microstates of the joint system $A \cup B$. One gets from \eqref{eq: forward prob}
\be \label{eq: autonompus forward prob}
    p(\g_{i_{AB}f_{AB}}) := p_{i_{AB}}  \bra{f_{AB}} P_{C_\ff, u} \circ C_\ff \circ {\cal U}_{t_\mathrm{i}, t_\mathrm{f}} [\ket{i_{AB}} \bra{i_{AB}}] \ket{f_{AB}}
\ee
while the reverse process is given by 
\be \label{eq: autonompus backward prob}
    p(\g_{i_{AB}f_{AB}}^R) := p_{f_{AB}} \bra{i_{AB}} P_{P_{C_\ff, u} \circ C_\ff \circ {\cal U}_{t_\mathrm{i}, t_\mathrm{f}}, u} [\ket{f_{AB}}\bra{f_{AB}}]\ket{i_{AB}}
\ee
since the maximally mixed state $u$ of the joint system A $\cup$ B is a fixed point of the unital CPTP map $P_{C_\ff, u} \circ C_\ff \circ {\cal U}_{t_\mathrm{i}, t_\mathrm{f}}$. To simplify the discussion, in this paragraph, one assumes that $\rho_{AB}(t_\mathrm{i}^+) = \rho_{AB}^{\text{cg}}(t_\ii)$, so that one will not take into account the coarse-grained entropy term $s_{IF}^{\text{cg}}$ in the discussion, since it has already been treated in the former subsection \footnote{One can retrieve this contribution from the discussion there by taking $A \cup B$ as the system }. 
Then, the entropy production term is simply given by 
\be \label{eq: entropduction auto bipartite}
    s_c(\g_{i_{AB}f_{AB}}) := \ln{\frac{p(\g_{i_{AB}f_{AB}})}{p(\g_{i_{AB}f_{AB}}^R)}} = \ln{\frac{p_{i_{AB}}}{p_{f_{AB}}}} = s_{f_{AB}}- s_{i_{AB}}  
\ee
where $s_{f_{AB}} := - \ln{p_{f_{AB}}} = - \ln{\frac{p_{F_{AB}}}{V_{F_{AB}}}}$ and $s_{i_{AB}} := - \ln{p_{i_{AB}}} = - \ln{\frac{p_{I_{AB}}}{V_{I_{AB}}}}$ since one starts from the initial coarse-grained state. Of course, the initial and final macro probabilities and volumes have been defined in Equations \eqref{eq: macro proba joint def} and \eqref{eq: macro volume joint def} and are applied here for the initial ($J = I$) and final ($J = F$) joint algebras. We also introduce the initial and final stochastic mutual information
\begin{align} \label{eq: mutual information element autonomous}
    i_{f_{AB}} &:= \ln{ \frac{p_{f_{AB}}}{p_{f_A} p_{f_B}}} = \ln{ \frac{p_{F_{AB}}}{p_{F_A} p_{F_B}}} := i_{F_{AB}}\nn \\
    i_{i_{AB}} &:= \ln{ \frac{p_{i_{AB}}}{p_{i_A} p_{i_B}}} = \ln{ \frac{p_{I_{AB}}}{p_{I_A} p_{I_B}}} := i_{I_{AB}}
\end{align}
where the second equality results from the assumption $\rho_{AB}(t_\mathrm{i}^+) = \rho_{AB}^{\text{cg}}(t_\ii)$. We then obtain from \eqref{eq: entropduction auto bipartite} and \eqref{eq: mutual information element autonomous}
\be \label{eq: ent prod element auto}
    s_c(\g_{i_{AB}f_{AB}}) = i_{I_{AB}} - i_{F_{AB}} + s_{F_A} - s_{I_A} + s_{F_B} - s_{I_B} = -\Delta i_{I_{AB},F_{AB}} + \Delta s_{I_A,F_A} + \Delta s_{I_B,F_B}
\ee
where $s_{J_A} := - \ln{\frac{p_{J_A}}{V_{J_A}}}$ and $s_{J_B} := - \ln{\frac{p_{J_B}}{V_{J_B}}}$ for $J = I,F$. Of course, as in the open-system viewpoint, the entropy production element \eqref{eq: ent prod element auto} is not necessarily positive, but positivity of relative entropy ensures that its average is
\be
    \langle s_c(\g_{i_{AB}f_{AB}})  \rangle = \sum_{i_{AB} f_{AB}} p(\g_{i_{AB}f_{AB}}) \ln{\frac{p(\g_{i_{AB}f_{AB}})}{p(\g_{i_{AB}f_{AB}}^R)}} \geq 0
\ee
implying that 
\be
    \Delta \langle s_{I_A,F_A} \rangle + \Delta \langle s_{I_B,F_B} \rangle \geq \Delta \langle i_{I_{AB},F_{AB}} \rangle
\ee 
which is of course the same inequality as \eqref{eq: bipartite ob scnd law}. 

In full generality, following the analysis led in the previous subsection for the open system but applied to the joint system A $\cup$ B, one readily obtains a fluctuation theorem
\be
    \mean{e^{-s_{F_A} +s_{I_A} - s_{F_B} + s_{I_B} + i_{F_{AB}} - i_{I_{AB}} -s_{IF}^\text{cg}}} =1
\ee
where we have reinstated the coarse-grained entropy contribution
$s_{IF}^{\text{cg}}$ of the joint system A $\cup$ B, defined from Eq.~\eqref{eq:cgstochEP}, which is present if A $\cup$ B is initially out of internal equilibrium. We have also assumed $\xi = 1$, corresponding to the absence of absolute irreversibility.
Now, to obtain a fluctuation theorem involving the heat transfer between system A and B, we need to connect $\Delta s_{I_B,F_B} = s_{F_B} - s_{I_B}$ to a stochastic heat exchange $q_{I_B, F_B}$. However, even if one could introduce a parametrization like Eq.~\eqref{eq: bipartite ob scnd law 2} to express $s_{I_B}$ and $s_{F_B}$, there is no canonical way to combine those two terms in a stochastic heat transfer. Such a construction is, however, possible when considering a quasi-static evolution, as explained in section~\ref{sec: heat and work}.

\section{Commutative case: Quasi-static transformation, heat and work.}
\label{sec: heat and work}

In the previous section, the second law was established for initial and final commutative algebras (not necessarily identical).
However, the connection to the first law and to the notions of work and heat was only possible in the autonomous viewpoint and at the average level. This follows from the fact that the evolution was modeled by a single map taking the initial state directly to the final state over a finite time interval $t_\ff-t_\ii$. As is well known in thermodynamics, heat and work are path-dependent quantities: different dynamical maps can share the same fixed points while corresponding to different amounts of heat and work. Consequently, although the variation of the modular Hamiltonian, $\Delta \langle K_{\ii,\ff} \rangle_\rho$ in \eqref{eq: generic second law commalg opens}, can be heuristically interpreted as an information-theoretic analogue of heat (up to an unspecified temperature scale, since $K_{\ii}$ and $K_{\ff}$ are dimensionless), it does not yet admit a genuine thermodynamic interpretation in the sense that it does not explicitly relate information/entropy and energy. 

In this section, we show that a consistent notion of heat and work emerges once a path is specified. More precisely, we introduce the analogue of the quasi-static transformations of macroscopic thermodynamics by resolving the evolution into a sequence of infinitesimal intermediate steps. During each step, the algebra to which the system macrostate belongs can be varied infinitesimally by changing the measurement basis according to an infinitesimal rotation. 
Along such paths, we can define heat and work infinitesimal increments. In the open-system viewpoint, heat is associated with relaxation toward the steady state in the current observation basis, whereas, in the autonomous viewpoint, it is identified with the energy variation of environment B that is proportional to its observational entropy variation.
Work can come from three mechanisms: (i) in the open system viewpoint, work can be exerted by the environment generating the map ${\cal N}$. This includes time-dependent contributions in the system's Hamiltonian modeling couplings to external classical fields \cite{Alicki79, Pusz1978}, and environment-induced energy transitions not fulfilling the detailed balance condition \cite{Cuetara15,Szczygielski13,Forastiere22}. (ii) In the autonomous approach, the environment B may behave as a source of work when it is not in a canonical thermal state \cite{Elouard23} (iii) In both approaches, the time-dependency of the algebra (i.e. of the observation basis) can be associated with a work performed by the observer. The latter mechanism is specifically identified by our approach. This can be understood physically by noting that a change in the algebra reflects a change in the constraints imposed on the system by the observer, which is a common source of work exchange in macroscopic thermodynamics \cite{Callen91}. Remarkably, for quantum mechanical systems, it includes an energy flow coming directly from the measuring apparatus when the measurement basis is rotated with respect to the energy eigenbasis \cite{Elouard17}.

\subsection{Continuous variation of the accessible algebra}

More formally, we consider that the observer has access at time $t$ to a coarse-grained algebra of observables ${\cal A}^c(t)$ attached to system A, via a measurement scheme (micro-macro channel) $C_t$. The latter is associated with a set of projectors $\{\Pi_J(t)\}$. The measurement scheme, and its attached micro-macro channel, can change from $t$ to $t+dt$, determining a coherent trajectory in the Hilbert space. Such a prescription allows one to follow some predictable macroscopic coherent evolution of the system, say a macroscopic oscillation, without artificially turning it into a classical stochastic process (and thus overestimating irreversibility). It is crucial to ensure that at any time $t$, the relevant ``slow variables'' (which may be out of equilibrium) are being tracked, and only quickly equilibrating variables are ignored.

To easily connect any observable to the accessible algebra $\mathcal{A}^c(t)$, it is useful to use the notion of \emph{representative} introduced in the paragraph \ref{sec:cgdef}. 
Since the observables belonging to ${\cal A}^c$ are the only observables allowed, we simplify the notation $O_{\mathcal{A}^c(t)}$ and simply write the representatives
as 
\be\label{d: representative simple}
    O_{\mathcal{A}^c(t)} := O_t.
\ee
In particular, we denote the coarse-grained state at time $t$ as $\rho_t= \rho^{\text{cg}}_{\mathcal{A}^c(t)}=\sum_J p_J(t)\Pi_J(t)/V_J(t)$. It is important to keep in mind that at any time $t$, one has 
\be
    \langle O_t \rangle_{\rho(t)} = \langle O_t \rangle_{\rho_t}
\ee
from \eqref{eq: mean value a cg states}, even if the system is not in internal equilibrium (i.e $\rho(t) \neq \rho_t$).\\

For the sake of simplicity, we consider that the evolution in time of the measurement scheme is simply given by a unitary:
\be 
\forall J, \quad\Pi_J(t+dt) = U_{\gamma(t)}\Pi_J(t)U_{\gamma(t)}^\dag,
\ee
with $U_{\gamma(t)}=e^{iG_{\gamma(t)} dt}$ is a family of elementary unitaries. This implies that the volume of the macrostates are conserved:
\be
\forall J, \quad V_J(t+dt)=\tr{\Pi_J(t+dt)}=\tr{\Pi_J(t)}=V_J(t).
\ee
The case in which the rank of the projectors can vary, and the definition of heat and work in this context, are left for future work.
It is useful to note that, at first order
\begin{align} \label{eq: projectorproductorder2}
    \tr{\left[\Pi_K(t+dt) \Pi_J(t)\right]} &= \tr{\left[(\Pi_K(t) + i (G_{\g(t)} \Pi_K(t) - \Pi_K(t) G_{\g(t)})dt + O(dt^2)) \Pi_J(t) \right]} \nn \\
    &= \tr{\left[ \d_{IJ} \Pi_K(t) + i (G_{\g(t)} \d_{KJ} \Pi_K(t) - \Pi_K(t) G_{\g(t)} \Pi_J(t)) dt + O(dt^2) \right]} \nn \\
    &= \tr{\left[\Pi_K(t) \Pi_J(t)\right]} + O(dt^2) = V_K \d_{K,J} + O(dt^2).
\end{align}

\subsection{Average thermodynamic laws}

\subsubsection{Work from varying the algebra}\label{sec:alg_work}

First, let us consider a system at internal equilibrium at time $t$ ($\rho(t)= \rho_t$) and let us examine the case where the system undergoes no dynamics other than those induced by the measurements themselves, (i.e., in the open approach, we take the map ${\cal N}_{t_\ii,t_\ff}$ to be the identity channel). Of course, if the system has a non-vanishing Hamiltonian, it also undergoes an internal dynamics. However, since this dynamics does not contribute to the energetics considered here, we neglect it in the following, or equivalently work in the interaction picture.

\begin{figure}[htp]
\begin{framed}
\centering
\includegraphics[scale=0.42]{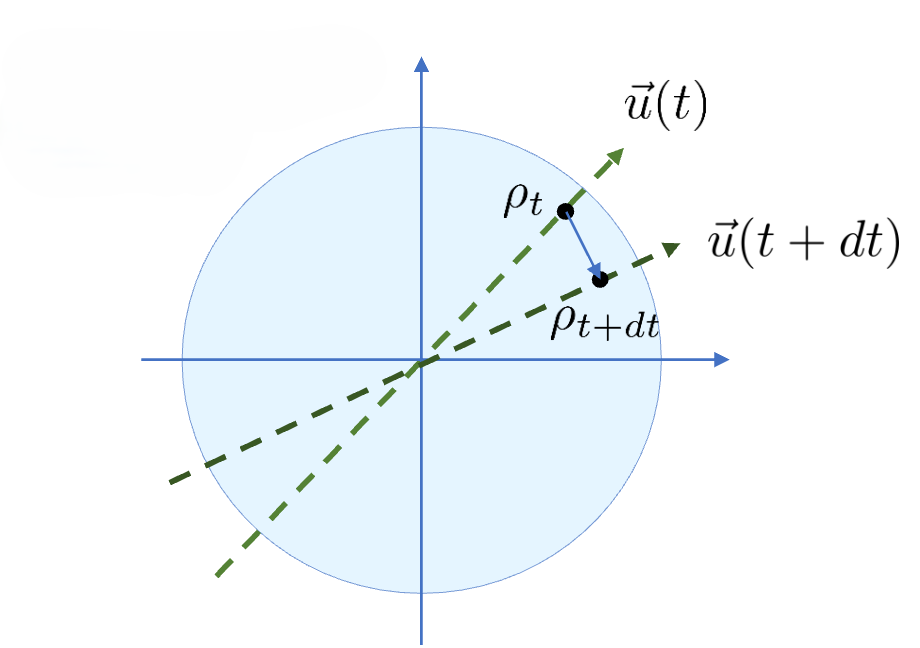}
\caption{\textbf{Measurement induced work transfer}: A qubit is measured
along the axis $\vec{u}(t)$ at time $t$ and along a neighboring axis
$\vec{u}(t+dt)$ at time $t+dt$. An arbitrary qubit state is conveniently
represented by a vector $\vec{r}$ in the Bloch sphere. If $\vec{u}(t)$ and
$\vec{u}(t+dt)$ have different $\theta$-components on the Bloch sphere, the
mean energy of the qubit changes as a result of the measurement. Since the
process is reversible and the associated energy transfer is isentropic, this
transfer must be interpreted as work, in agreement with our formalism.}
\label{blochspherefree}
\end{framed}
\end{figure}

Property Eq.~\eqref{eq: projectorproductorder2} implies that after the measurement $C_{t+dt}$ at time $t+dt$, the system is at internal equilibrium in the algebra used at $t+dt$. Namely, the post-measurement state is $\sum_J \Pi_J(t+dt) \rho_t \Pi_J(t+dt) = \sum_J \frac{p_J(t)}{V_J(t)}\Pi_J(t+dt) + O(dt^2) = \rho_{t+dt}$, with the same set of populations and volumes (and therefore, no variation of observational entropy). Actually, the evolution of the system is unitary, given by $U_{\gamma(t)}$, and reversible \footnote{Had we retained the free internal dynamics, the result would be unchanged to first order, since unitary evolution does not modify the populations of an initially diagonal state at this order.}. It can be reversed by simply measuring the system again according to $C_t$. In addition, the energy variation of the system takes the form:
\bea\label{eq:algebraic_work}
\text{d}E &=& \tr{\left[H(\rho_{t+dt}-\rho_t)\right]}\nonumber\\
&=& \tr{\left[H_{t+dt}\rho_{t+dt}-H_t\rho_t)\right]}\nn\\
&=& \tr{\left[(H_{t+dt}-H_t)\rho_t\right]}.
\eea
In the second line, we have introduced the representatives $H_t$ and $H_{t+dt}$ of $H$. In the third line, we have used that $p_J(t+dt)=p_J(t)$ and property \eqref{eq: projectorproductorder2}.
Eq.~\eqref{eq:algebraic_work} allows us to identify a reversible work increment associated to an infinitesimal variation of the algebra in which the system is described, leading to an apparent change of its Hamiltonian. This work exchange is directly equal to the energy variation during the infinitesimal unitary induced by the rotated measurement (see App. \ref{app: qubit} for a detailed application on the case of a qubit system). A simple visual toy model is provided by the projective measurement of the qubit illustrated in Fig.~\ref{blochspherefree}. This contribution to the work is a consequence of our framework, and appears on an equal footing with work provided by the environment of the system, as we will see in the remainder of this section.

\subsubsection{Open system viewpoint}\label{sec:WH_bipartite_open_av}

We now consider that between the measurements at $t$ and $t+dt$, the system evolves according to a map ${\cal N}_{t,t+dt}$. We can therefore apply result \eqref{eq: generic second law commalg opens} by taking $t_\ii=t$ and $t_\ff=t+dt$.
The roles of $\sigma_\ii$ and $\sigma_\ff$ are played by an instantaneous fixed point $\sigma_t$ of $\tilde{\Lambda}_{t,t+dt}=\text{A}_{t,t+dt}\circ\Lambda_{t,t+dt}$\footnote{A fixed point $\sigma_t \in \mathcal{D}(\mathcal{H})$ is guaranteed to exist whenever the map ${\cal N}_{t,t+dt}$ is (C)PTP. Otherwise, its existence must be assumed. In the absence of such a fixed point, it becomes difficult to formulate a meaningful thermodynamic description with the open-system viewpoint, while the autonomous viewpoint still holds.}, where $\text{A}_{t,t+dt}$ is the connection \eqref{eq: connection},  and its image $\Lambda_{t,t+dt}(\sigma_t)$, respectively. Consequently:
\bea
\text{d} \mean{K_{t,t+dt}} = -\text{Tr}\left(\rho_{t+dt}\ln\Lambda_{t,t+dt}(\sigma_t)\right)+\text{Tr}\left(\rho_t\ln\sigma_t\right).
\eea
However, property \eqref{eq: projectorproductorder2} implies that (see a proof in App.~\ref{app: proofs})
\be \label{eq: first order fixed point projector}
    \tr{\left(\rho_{t+dt} \ln{\Lambda_{t,t+dt}(\s_{t})}\right)} = \tr{(\rho_{t+dt} \ln{\s_{t})}} + O(dt^2),
\ee
leading to the entropy production rate: 

\be \label{eq: second law loc t t+dt 2}
    \Sigma_{t,t+dt}:= \text{d} S^{\text{ob}}_{t, t+dt} - \tr{((\rho_{t+dt} - \rho_t) K_t)} 
\ee
Eq.~\eqref{eq: second law loc t t+dt 2} is the time-local version of Eq.~\eqref{eq: entropy prod 1} and is therefore interpreted as a definition of the entropy production rate. One sees from \eqref{eq: generic second law commalg opens} that a sufficient condition for the entropy production rate to be positive is to consider $I_{\text{ext}}(t) = 0$ and $I_{\text{int}}(t) = 0$ at any time $t \in (t_\ii, t_\ff)$. However, $I_{\text{ext}}(t)$ can be set to zero in \eqref{eq: generic second law commalg opens} if there is no backflow of information between $t$ and $t+dt$, which means that the map ${\cal N}_{t,t+dt}$ is CPTP and therefore ${\cal N}_{t_\ii,t_\ff}$ is CP-divisible (see for instance \cite{Colla2025Thermodynamic, Picatoste26, Theret2026Entropy}). In analogy, $I_{\text{int}}(t)$ can be set to zero in \eqref{eq: generic second law commalg opens} at any time along the path if there is no backflow of information \emph{from the microscopic degrees of freedom towards the macroscopic degrees of freedom} such that we can (safely) take the state at any time $t$ to be the coarse-grained state $\rho_t$. When this property holds, we say that the map ${\cal N}_{t_\ii,t_\ff}$ is \emph{locally scale divisible} (${\cal N}_{t,t+dt}$ is scale divisible for any $t$). Therefore, if $\mathcal{N}_{t,t+dt}$ is CP-divisible and locally scale divisible, the entropy production term \eqref{eq: second law loc t t+dt 2} is positive.

To connect the informational equality Eq.~\eqref{eq: second law loc t t+dt 2} to energetic quantities, we then follow the procedure developed in subsection \ref{subsubsection: bipartite approach} and decompose the modular Hamiltonian of the instantaneous fixed point $K_t = - \ln{\s_t} \in \mathcal{A}^c(t)$ of the local dynamics using an orthogonal basis $\{ \Id, H_t, O_t^1, \cdots O_t^N \}$ of the algebra $\mathcal{A}^c(t)$.
Here, $H_t$ is the representative of the Hamiltonian of the system $H$ in the algebra $\mathcal{A}^c(t)$. Therefore, one has
\be \label{eq: Kt in op basis}
    K_t := \beta(t) \left(H_t + \sum_{k} \l_k(t) O_t^k\right) + \ln{\left\{\tr{\left[e^{-\beta(t) (H_t + \sum_{k} \l_k(t) O_t^k)}\right]}\right\}} \Id.
\ee
Exactly as in \eqref{eq: KBt in op basis}, $\beta(t)$ must be interpreted as the effective inverse temperature, while the other parameters $\{ \l_k(t) \}$, conjugated to the other accessible observables $\{ O_t^k \} \in \mathcal{A}^c(t)$, are other effective intensive parameters. 
From \eqref{eq: second law loc t t+dt 2}, the entropy production now reads 
\be\label{eq: second law loc t t+dt 3}
    \Sigma_{t,t+dt}:= \text{d} S^{\text{ob}}_{t, t+dt} - \beta(t) (\text{d} \langle H_t \rangle_\rho + \sum_k \l_k(t) \text{d} \langle O_t^k \rangle_\rho)
\ee
from which we identify a heat-like increment
\be\label{d:heatincrementopen}
    \d \mathcal{Q}_{t, t+dt} := \text{d} \langle H_t \rangle_\rho + \sum_k \l_k(t) \text{d} \langle O_t^k \rangle_\rho.
\ee
From $\text{d} E_{t,t+dt} = \tr{(\rho_{t+dt} H_{t+dt})} - \tr{(\rho_{t} H_{t})}$, we deduce the work-like increment
\bea\label{d:workincrementopen}
    \d \mathcal{W}_{t, t+dt} &=& \text{d} E_{t,t+dt} - \d \mathcal{Q}_{t, t+dt}\nn\\ &=& \tr{\rho_t (H_{t+dt} - H_t)} - \sum_k \l_k(t) \text{d} \langle O_t^k \rangle_\rho \label{d:dW}.
\eea
The work increment $\d \mathcal{W}_{t, t+dt}$ contains several contributions. First, the time-variation of the local Hamiltonian $H_t$ (appearing in the first term of the last line) can be due to a coupling to a classical system varying the potential energy landscape experienced by the system such that its Hamiltonian is time-dependent. This is a common form of work exchange considered in quantum thermodynamics \cite{Alicki79,Esposito09}. However, in our framework, the time-dependency of $H_t$ may also stem from the change in time of the observation algebra (as considered in section \ref{sec:alg_work}). Both contributions cannot be distinguished without further information, such as the knowledge of the microscopic Hamiltonian $H(t)$. Finally, the last term in Eq.\eqref{d:dW} arises when the environment is out of thermal equilibrium, or exchanges conserved charges with the system \cite{Rigol2007Relaxation, Horowitz16,Vidmar2016Generalized, Ilievski2015Complete, Essler2015Generalized, Manzano18b,Fukai2020Noncommutative}. Notably, this form of non-unitary work appears in the context of quantum open systems for nanoscale thermoelectric setups \cite{Benenti17}. It is also the usual expression for the work increment appearing in macroscopic thermodynamics \cite{Callen91}. Equation \eqref{d:dW} interpolates between the different notions of work typically used in quantum and macroscopic thermodynamics and treats them on an equal footing. On the other hand, the heat increment is related to the elementary variation of observational entropy as expected in macroscopic thermodynamics, i.e., both coincide whenever the transformation is reversible, that is $\Sigma_{t,t+dt}=0$ at any $t$, which requires that the macrostate of the system is equal to the instantaneous fixed point $\sigma_t$ all along the evolution.

\begin{figure}[htp]
\begin{framed}
\centering
\includegraphics[scale=0.5]{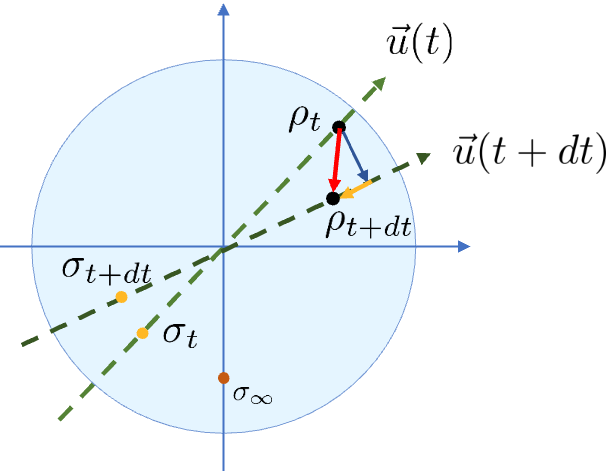}
\caption{\textbf{Thermodynamic analysis of a dissipative qubit}: 
We assume that the dynamics of the qubit is ruled by a GKLS equation (described explicitly in App.~\eqref{app: qubit}, Eq.~\eqref{lindbladref}) inducing thermalization. The latter generates a flow in $\mathbb{D}(\mathcal{H})$ driving any initial state towards the thermal fixed point $\sigma_\infty$ at inverse temperature $\beta$, located on the $z$-axis. 
At time $t$, a measurement is performed in the basis defined by the direction $\vec{u}(t)$, while at time $t+dt$ the measurement basis is rotated to $\vec{u}(t + dt)$. 
The qubit energy variation during $dt$ is split into two parts. The radial (entropy-varying) component of the energy variation corresponds to the heat increment defined in Eq.~\eqref{d:heatincrementopen} (depicted by the yellow arrow), while the unitary (radius preserving) part corresponds to the work increment defined in Eq.~\eqref{d:workincrementopen}, (depicted by a blue arrow). For every state $\rho_t\in\mathcal{A}^c(t)$, repeated application of $\tilde{\Lambda}_{t,t+dt}$ (corresponding to the protocol composed of a measurement at time $t$ along $\vec{u}(t)$, the GKLS evolution over interval $[t,t+dt]$, the second measurement at time $t+dt$ along $\vec{u}(t+dt)$, followed by a last measurement along the initial axis $\vec{u}(t)$) irreversibly drives the system towards the local equilibrium state, i.e. $\underset{N \rightarrow + \infty}{\lim}\tilde{\Lambda}^N_{t,t+dt}(\rho_t) = \s_t$, 
(see details in App.~\ref{app: qubit}.).}
\label{fig:BlochSphere}
\end{framed}
\end{figure}

To illustrate those notions, we analyze in Appendix \ref{app: qubit} the case of a qubit system whose dynamics obeys a GKLS equation modeling a weak coupling to a bath at temperature $T$, measured in the eigenbasis of its density operator every $dt$ (see also Fig.~\ref{fig:BlochSphere}). Our approach allows us to decompose its dynamics into a deterministic part, associated with the controlled rotation of its eigenbasis and an uncontrolled evolution of the populations. The former is associated with a work increment, as the representative of the qubit Hamiltonian evolves in time together with the measurement basis. This energy transfer is controllable and could be reversed by measuring the qubit back in its initial eigenbasis.
The latter is irreversible and associated with heat dissipation and entropy production while the state progresses towards the fixed point attached 
to the instantaneous orientation of the measurement basis.

 \begin{remark}
     Notice that, although a basis of operators $\{ \Id, H_t, O^1_t, \cdots, O_t^N\}$ is used to decompose the self-adjoint modular Hamiltonian, the division of energy transfer between heat and work does not depend on the particular choice of orthogonal basis elements $\{ O_t^k \}$ (and their associated conjugate intensive parameters $\l_k(t)$) in the subspace orthogonal to the Hamiltonian.
 \end{remark}

\begin{remark}
    The present description emphasizes that heat and work, as well as entropy production, are inherently observer-dependent quantities as already argued in \cite{RubinoBruknerManzano2026, Faist2018Fundamental}. In our framework, this dependency is made explicit via the choice of the algebra of available observables to describe and control the system.
\end{remark}

\begin{remark}
    Confining the system to a particular algebra despite its dynamics is associated with a work cost which can be identified in our framework. Let us consider a two-step cycle starting in the state $\sigma_t$. The first step is the process described above (measurement $C_t$, evolution by ${\cal N}_{t,t+dt}$, measurement $C_{t+dt}$, which together define the map $\Lambda_{t,t+dt}$), and the second step is another measurement $C_t$. Together, these two steps define a map $\tilde\Lambda_{t,t+dt}$ that leaves the state $\sigma_t$ unchanged. The second step corresponds to the reversible process considered in section \ref{sec:alg_work}, which is associated with the work increment $\text{Tr}\{\rho_{t+dt}(H_{t}-H_{t+dt})\} = -\text{Tr}\{\sigma_{t}(H_{t+dt}-H_{t})\}$. 
    This work is the amount of work needed to maintain the state in the initial algebra ${\cal A}^c(t)$, while the dynamical map $\Lambda_{t+dt}$
    maps an arbitrary state
    to a different algebra ${\cal A}^c(t+dt)$. When further restricting to the case where the environment provides no work, one can identify this contribution with the stationary work needed to maintain the system in the algebra $C_t$, while the environment tries to prepare a stationary state with respect to 
    a different algebra (see the example of the qubit in App.~\ref{app: qubit})
\end{remark}

\subsubsection{Autonomous viewpoint}
\label{sec: paragraph autonomous inf wh}

In the open-system standpoint, the distinction between heat and work can only be based on the system dynamics, which could be in principle induced by different environments which may realize different thermodynamic scenarios. The thermodynamic description of Section.~\eqref{sec:WH_bipartite_open_av} constitutes in some sense the ideal case where no resource is wasted 
from B. However, verifying which resources were actually paid to induce a given system dynamics requires to do a similar analysis in the environment, which we perform here by extending the autonomous viewpoint of section~\ref{subsubsection: bipartite approach} to continuous-time dynamics. Formally, for each instant $t \in (t_\ii, t_\ff)$, we consider a two-point measurement performed on the composite system A $\cup$ B between $t_\mathrm{i} = t$ and $t_\mathrm{f} = t+dt$. The measurements are defined with respect to the local algebras $\mathcal{A}_{\text{A}}^c(t) \otimes \mathcal{A}_{\text{B}}^c(t)$ and $\mathcal{A}_{\text{A}}^c(t+dt) \otimes \mathcal{A}_{\text{B}}^c(t+dt)$, each generated by their own family of projectors. As in the open-system case, we assume that these two algebras are related by an infinitesimal unitary transformation $U_{\g(t)}$ acting on the Hilbert space $\mathcal{H}_{\text{A}} \otimes \mathcal{H}_{\text{B}}$. 

Under the assumption of internal equilibrium at time $t$ (or, more generally, that the dynamics is locally scale divisible with respect to the initial and final algebras, see Sec.~\ref{sec: 2nd law commut open system}), 
we may apply Eq.~\eqref{eq: variation and law bipartite} between $t$ and $t+dt$ to get
\bea \label{eq: variation and law bipartite 2}
    \text{d} S_{t,t+dt}^{\text{ob}}(\rho_A) + \beta(t) (\text{d} \langle H_{B,t} \rangle_{\rho_B} + \sum_k \l_k(t) \text{d} \langle O_{B,t}^k \rangle_{\rho_B})\nn\\
    = \text{d} S_{t,t+dt}^{\text{ob}}(\rho_{AB}) + \text{d} I^{\text{ob}}_{AB, t,t+dt} + S(\rho_{B,t+dt} \lvert \lvert \rho_{B,t}) \geq \text{d} I^{\text{ob}}_{AB,t,t+dt}
\eea
where, of course, all the variations $\text{d}$ are taken between times $t$ and $t+dt$ and are the infinitesimal analogs of those appearing in Eq.~\eqref{eq: variation and law bipartite}. From the identity (proven in App. \ref{app: proofs})
\be \label{eq: relative entropy odt2}
    S(\rho_{B,t+dt} \lvert \lvert \rho_{B,t}) = O(dt^2)
\ee
we deduce that at first order in $dt$, Eq.~\eqref{eq: variation and law bipartite 2} becomes 
\bea \label{eq: variation and law bipartite 3}
    \text{d} S_{t,t+dt}^{\text{ob}}(\rho_A) + \beta(t) (\text{d} \langle H_{B,t} \rangle_{\rho_B} + \sum_k \l_k(t) \text{d} \langle O_{B,t}^k \rangle_{\rho_B}) \nonumber\\
    = \text{d} S_{t,t+dt}^{\text{ob}}(\rho_{AB}) + \text{d} I^{\text{ob}}_{AB,t,t+dt} + O(dt^2) \geq \text{d} I^{\text{ob}}_{AB,t,t+dt}.
\eea
Therefore, the heat-like increment in B can be read directly from the second law \eqref{eq: variation and law bipartite 3}
\be
    \d \mathcal{Q}_{t,t+dt}^B = \text{d} \langle H_{B,t} \rangle_{\rho_B} + \sum_k \l_k(t) \text{d} \langle O_{B,t}^k \rangle_{\rho_B}
\ee
so that we deduce work-like increment from the first law
\be \label{eq: work from B}
    \d \mathcal{W}_{B,t,t+dt} = \text{d} E_{B,t,t+dt} - \d \mathcal{Q}_{B,t,t+dt} = \tr{(\rho_{B,t}(H_{B,t+dt} - H_{B,t}))} - \sum_k \l_k(t) \text{d} \langle O_{B,t}^k \rangle_{\rho_B}
\ee
attached to environment B.

As a consequence of Eq.~\eqref{eq: relative entropy odt2}, the heat increment is directly proportional to the variation of the observational entropy of environment B at first order, i.e.
\be
    \d \mathcal{Q}_{t,t+dt}^B = T(t) \text{d} S^{\text{ob}}_{t,t+dt}(\rho_B) 
\ee
where $T(t) = \beta(t)^{-1}$, and vanishes for isentropic transformations of the reservoir B. Moreover, the work increment corresponds to the part of the energy variation that is not associated with an entropy variation. Instead, as follows from \eqref{eq: work from B}, it is related to two distinct contributions: (i) the time dependence of the Hamiltonian of B, arising either from external driving or from a change in the observation algebra of B, and (ii) the dynamical consumption of the local resources stored in the state $\rho_{B,t}$ that quantify its departure from a thermal state with positive temperature $T(t)$. Compared with the naive definitions of heat and work introduced in \eqref{eq: heat def sec 3} and \eqref{eq: work def sec 3} for the two-point measurement scheme, the continuous scheme offers two main advantages: (i) it clearly identifies heat provided to A from B to an entropy variation in system B, and (ii) it relates the resources consumed in B to the work needed to induce the evolution of A.  As we show in the next section, a third advantage is the possibility to provide a stochastic-thermodynamic interpretation of those quantities, and associated fluctuation theorems.

\subsection{Stochastic thermodynamic laws}\label{sec: stoch thermo comm}

In this section, we rely on the individual quantum trajectories defined by the continuous measurements to formulate a stochastic thermodynamic description, in which quantities such as heat and work flows can be defined directly at the trajectory level and obey fluctuation theorems.

\subsubsection{Open system viewpoint}\label{sec: stoch open quasistatic}

For the evolution of the open system during an infinitesimal interval, we define the forward and backward probability transitions as in \eqref{eq: forward prob} and \eqref{eq: backward prob} by taking $t_\mathrm{i} = t$ and $t_\mathrm{f} = t+dt$, and write the stochastic entropy production increment as in \eqref{eq: entropy creation term stoch} to get
\be \label{eq: entropy prod local stoch 2}
    \delta s_c{(\g_{if})} = s_f - s_i - K_{t,F} + K_{t,I} + O(dt^2),
\ee
with $K_{t,J}=\text{Tr}\{K_t\frac{\Pi_J(t)}{V_J}\}$ is the $J$-th component of modular Hamiltonian associated with the fixed point $\s_t$ of the map $\tilde{\Lambda}_{t,t+dt}$. To prove \eqref{eq: entropy prod local stoch 2} from \eqref{eq: entropy creation term stoch}, we used the identities (proven in App.~\ref{app: proofs})
\be \label{eq: 3 identities mod proj ttdt}
    \text{Tr}\{ \Pi_J(t)\ln{\Lambda_{t,t+dt}(\sigma_t)\}} =\text{Tr}\{\Pi_J(t)\ln{\sigma_t} \}+O(dt^2) = \text{Tr}\{ \Pi_J(t+dt)\ln{\Lambda_{t,t+dt}(\sigma_t)\}}+O(dt^2)
\ee 
allowing us to neglect the variation of the modular Hamiltonian operator $K_t$ between $t$ and $t+dt$, being of order $O(dt^2)$. This implies that $K_{t,I}$ and $K_{t,F}$ will be expressed in terms of a single basis of observables $O_{t,J}^k$ via decomposition \eqref{eq: Kt in op basis}, i.e.
\begin{align}
    K_{t, J} &= \beta(t) (H_{t, J} + \sum_k \l_k(t) O_{t, J}^k) + \ln{\tr{e^{-\beta(t) (H_{t} + \sum_k \l_k(t) O_{t}^k)}}} 
\end{align}
and we can recast Eq.~\eqref{eq: entropy prod local stoch 2} as
\be \label{eq: entropy prod local stoch 3}
   \delta s_c{(\g_{if})} = s_f - s_i - \beta(t) \left[(H_{F,t} - H_{I,t}) + \sum_k \l_k(t)(O^k_{t, F} - O^k_{t, I})\right].
\ee
Then, the positivity of relative entropy \eqref{eq: average entropy stoch} reproduces the positivity of the average entropy production $\langle \delta s_c{(\g_{if})} \rangle = \Sigma_{t,t+dt} \geq 0$ during the time interval $(t, t+dt)$.
More importantly, following exactly the analysis of Sec.~\ref{sec: stochastic two pt open system } now applied between the times $t$ and $t+dt$, one obtains a fluctuation theorem involving the stochastic heat increment $\delta Q_{t,(I,F)}=(H_{t,F} - H_{t,I}) + \sum_k \l_k(t) (O_{t,F}^k - O_{t,I}^k)$ associated with time interval $(t,t+dt)$
\begin{align}\label{eq:FT4}
1 &= \sum_{IF} p(\gamma_{IF}) e^{-s_F+s_I+ \beta(t) \delta Q_{t,(I,F)}
-s_{IF}^\text{cg}}  \nn \\
&= \mean{e^{-s_F+s_I+\beta(t)[(H_{t,F} - H_{t,I}) + \sum_k \l_k(t) (O_{t,F}^k - O_{t,I}^k)]-s_{IF}^\text{cg}}}.
\end{align}
We have once again neglected absolute irreversibility ($\xi =1$).

We can now construct a finite trajectory between $t_\mathrm{i}$ and $t_\mathrm{f}$ by dividing the time interval $\Delta t= t_\mathrm{f} - t_\mathrm{i}$ into $N$ infinitesimal steps, each of duration $dt = \frac{\Delta t}{N} \ll \Delta t$. Moreover, we define the stochastic trajectory $\g_{1, \cdots, N}$ specified by the sequence of $N$ local states visited by the system during the time interval $(t_\mathrm{i}, t_\mathrm{f})$. We assume that the system is in internal equilibrium at each step of the process, or, more generally, that the dynamics is locally scale divisible with respect to the algebras under consideration, so that the coarse-grained contribution $-s_{IF}^{\text{cg}}$ either vanishes or plays no role in the analysis. Under this assumption, transitions between microstates over an infinitesimal interval $t$ and $t+dt$ become equivalent to transitions between the corresponding macrostates. Consequently, the fluctuation theorem derived for microstate transitions can be applied directly to the macrostate dynamics. Then, of course,
\be
    p(\g_{1, \cdots, N}) = p_1 \prod_{i=1}^{N - 1} p(i+1 \lvert i)
\ee
where $i \in [1, \cdots, N]$ labels macrostates, while the probability of the backward trajectory is given by
\be
    p({\g}^R_{1, \cdots, N}) = p_N \prod_{i=1}^{N - 1} p(N-i \lvert N-i+1)
\ee
which gives us 
\begin{align} \label{eq: ent prod stoch 1N}
    s_{\g_{1, \cdots, N}} = \ln{\frac{p(\g_{1, \cdots, N})}{p({\g}^R_{1, \cdots, N})}} &= s_{N} - s_{1} + \sum_{k=1}^{N-1} (K_{t_k, J_{k+1}} - K_{t_k, J_k}) \nn \\
    &= s_{N} - s_{1} + \sum_{k=1}^{N-1} \beta(t_k)\delta Q_{t_k,(J_k,J_{k+1})}
\end{align}
That is, the total entropy production attached to the path $\g_{1, \cdots, N}$, which can be clearly decomposed into the difference between the initial and final entropy of the macrostate and the heat flux attached to the trajectory. In \eqref{eq: ent prod stoch 1N}, the paths $\g_{1 \cdots N}$ is characterized by the the set of macrostates $\{ J_k \lvert k = 1, \cdots, N\}$.  In particular, one has $s_1 = s_{J_1} = - \ln{\frac{p_{J_1}}{V_{J_1}}}$ while $s_N = s_{J_N} = - \ln{\frac{p_{J_N}}{V_{J_N}}}$. We therefore obtain the integral fluctuation theorem
\begin{align} \label{eq: fluctuationtheoremint1}
   1 &= \sum_{1 \cdots N} p(\gamma_{1 \cdots N}) e^{-s_N+s_1+ \sum_{k \in [1, \cdots N]}\beta(t_k) \delta Q_{t_k,(J_k,J_{k+1})}}  \nn \\
    &= \langle e^{-s_\text{f} + s_\text{i} + \int_\gamma \beta(t) \dot{Q}(t) dt}\rangle
\end{align}
once the continuous limit ($N \rightarrow + \infty$, $\Delta t$ fixed) has been taken, still ignoring absolute irreversibility. We retrieve the integral fluctuation theorem expected for a system coupled to an environment with time dependent temperature $\beta(t)$, here derived for an arbitrary system dynamics provided it fulfills the assumption of local scale divisibility (or ensures internal equilibrium at all times).

\subsubsection{Autonomous viewpoint}
\label{sec: autonomou sapproach continuous stoch}

To complete the picture, we also develop a stochastic thermodynamic description from the autonomous viewpoint, under continuous measurement of the joint system $A \cup B$ via the set of projectors $\{ \Pi_{J_A}(t) \otimes \Pi_{J_B} (t)\}$ at any time $t \in (t_\ii, t_\ff)$. To construct a continuous path, we assume that at any time $t$, the joint system $A \cup B$ is in internal equilibrium (or, as the reader is used to now, that the dynamics of the joint system is locally scale divisible), and we therefore ignore the coarse-grained entropy term $s_{IF}^{\text{cg}}$, as we did in the previous subsection for the open system viewpoint. 

If the joint system A $\cup$ B starts in the microstate $i_{AB}$ attached to the macrostate $I_{AB}$ at $t_\ii = t$ and ends in the microstate $f_{AB}$ attached to the macrostate $F_{AB}$ at $t_\ff = t+dt$, the entropy production along this particular trajectory is given by Eq.~\eqref{eq: entropduction auto bipartite}. This leads to the same relations Eqs.~\eqref{eq: entropduction auto bipartite}, \eqref{eq: mutual information element autonomous} and \eqref{eq: ent prod element auto} as in section.~\ref{sec: stoch open quasistatic} but now evaluated for the infinitesimal time interval $(t,t+dt)$ rather than $(t_\ii,t_\ff)$.
We then write
\begin{align} \label{eq: entropy B element}
    s_{I_B} &= K_{t,I_B} = \b(t)(H_{t,I_B} - \sum_k \l_k(t) O^{k}_{t,I_B}) + \ln{(\sum_{I_B} V_{I_B} e^{- \b(t)(H_{t,I_B} - \sum_k \l_k(t) O^{k}_{t,I_B})}) } \nn \\
    s_{F_B} &= K_{t+dt,F_B} = \b(t+dt)(H_{t+dt,F_B} - \sum_k \l_k(t+dt) O^{k}_{t+dt,F_B}) \nn \\
    &+ \ln{(\sum_{F_B} V_{F_B} e^{- \b(t+dt)(H_{t+dt,F_B} - \sum_k \l_k(t+dt) O^{k}_{t+dt,F_B})}) }
\end{align}
from which we deduce the natural decomposition
\begin{align} \label{eq: entropyprod}
    \d s_c(\g_{i_{AB}f_{AB}}) &= -s_{I_A} + s_{F_A} + K_{t+dt,F_B} - K_{t,I_B} + i_{I_{AB}} - i_{F_{AB}} \nn  \\ 
    &= -s_{I_A} + s_{F_A} + \beta(t) \d Q_{t,t+dt,(I_B,F_B)} + D_{t,F_B} dt + i_{I_{AB}} - i_{F_{AB}}
\end{align}
where the stochastic mutual information elements $i_{I_{AB}}$, $i_{F_{AB}}$ have been defined in Eq.~\eqref{eq: mutual information element autonomous} and with
\begin{align} 
    \beta(t) \d Q_{t,t+dt,(I_B,F_B)} &:= K_{t,F_B} - K_{t,I_B} = \beta(t)[(H_{t,F_B} - H_{t,I_B}) + \sum_k \l_k(t) (O_{t,F_B}^{k} - O_{t,I_B}^{k})] \nn \\
    D_{t,F_B} &:= \p_t K_{t,F_B}.
\end{align}
The term $ D_{t,F_B}$ is an additional drift term arising from the fact that environment $\text{B}$ is itself out of equilibrium. Notice that this term vanishes at first order if the state of the environment $\text{B}$ evolves solely due to a unitary and/or by projective measurements in a slowly rotating basis. This is because in such cases
the probability distributions $\{ p_{i_B} = \frac{p_{I_B}}{V_{I_B}} \}$ and $\{ p_{f_B} = \frac{p_{F_B}}{V_{F_B}} \}$, evaluated at $t$ and $t + dt$, respectively, are identical at order $O(dt)$, and so are the modular Hamiltonians attached to the distribution. In other words, this drift term does not arise from the continuous variation of the set of projectors $\{ \Pi_{J_B}(t) \}$ into $\{ \Pi_{J_B}(t+dt) \}$, but instead from the non-unitary part of the internal dynamics of reservoir B. Notice also that at any time $t$
\be \label{eq: mean value of D}
    \langle D_{t,F_B} \rangle = 0,
\ee
such that the average law Eq.~\eqref{eq: variation and law bipartite 3} is not modified, as expected (see App.~\ref{app: proofs} for a proof).

Taking the case $\xi$ = 1 (no absolute irreversibility), we get the fluctuation theorem for a single elementary step
\be \label{eq: FTinfbob}
    1 = \sum_{i_{AB}f_{AB}} p(\g_{i_{AB}f_{AB}}^R) = \sum_{i_{AB}f_{AB}} p(\g_{i_{AB}f_{AB}}) e^{- \d s_c(\g_{i_{AB}f_{AB}})} = \langle e^{s_{I_A} - s_{F_A} - \beta(t) \d Q_{t,t+dt,(I_B,F_B)} - D_{t, F_B} dt + i_{I_{AB}} - i_{F_{AB}}}\rangle.
\ee
Considering a continuous path $\g_{1, \cdots, N}$ where $N$ is very large (and $\Delta t = t_\ff - t_\ii$ fixed), we finally obtain the integral fluctuation theorem relevant for the  autonomous situation under continuous measurement
\be
    1 = \langle e^{s_{I_A} - s_{F_A} + i_{I_{AB}} - i_{F_{AB}} + \int_{\g} (\beta(t) \dot{Q}_B(t) + D_B(t) )dt} \rangle.
\ee
Of course, we have set $J_1 = I$ and $J_N = F$ there. In addition, notice that convexity of the exponential implies that 
\begin{align}
    0 &\geq \langle s_{I_A} - s_{F_A} + i_{I_{AB}} - i_{F_{AB}} + \int_{\g} (\beta(t) \dot{Q}_B(t) + D_B(t))dt  \rangle \nn \\
    &= \langle s_{I_A} \rangle - \langle s_{F_A} \rangle + \langle i_{I_{AB}} \rangle - \langle i_{F_{AB}} \rangle + \int_{\g} \beta(t) \langle \dot{Q}_B(t) \rangle dt 
\end{align}
where we used \eqref{eq: mean value of D} to get the last line.

In summary, in Section.~\ref{sec: heat and work}, we have used a sequence of measurement bases to define a continuous path, along which the system(s) can considered to be at any point at internal equilibrium, that is the equivalent of a quasi-static transformation of macroscopic thermodynamics. From there, we were able to identify general notions of heat and work increments unifying the quantum and macroscopic definitions and connecting the informational second law and fluctuation theorems derived in Section.~\ref{sec: two point thermodynamics} to the first law and heat fluctuations. The same fluctuation theorems can be expressed in terms of work fluctuations owing to the first law.

\section{Thermodynamics with arbitrary accessible subalgebras}
\label{sec: non commutative subalgebras}

In Sections.~\ref{sec: two point thermodynamics}-\ref{sec: heat and work}, we have developed our framework by considering coarse-graining based on commutative subalgebras, i.e. potentially coarse-grained projective measurements in a single given basis at each time. Consequently, the resulting thermodynamic description depended on the choice of initial and final measurement bases. Repeating the experiment with different bases would in general lead to distinct, competing thermodynamic interpretations. For instance, measuring in the eigenbasis of the instantaneous system state would generally make the macrostate dynamics appear less irreversible than measuring in another, incompatible basis. Moreover, the macrostates associated with the measurement outcomes were, by construction, classical diagonal distributions. Although the possibility to develop an arbitrary two-point or continuous measurement protocols can capture certain quantum thermodynamic properties of a system, they do not provide a unified view of a process, as the density-matrix based formulation of open system thermodynamics does \cite{Alicki79}. Leveraging our algebra-dependent coarse-grained entropy defined in Sec.~\eqref{sec: noncomm cg entropy}, we provide here the most general version of our thermodynamic formalism where the coarse-graining schemes can be arbitrary type I von Neumann algebras. In particular, we show that the tools presented for the commuting case can be extended to derive average thermodynamic laws and fluctuation theorems, in both the open and autonomous standpoints.

\subsection{General setting: open system viewpoint}\label{sec: noncomm open}

We now consider two (not necessarily commutative) type I von Neumann subalgebras, denoted by $\mathcal{A}_{\ii} \subset \mathcal{B}(\mathcal{H})$ and $\mathcal{A}_{\ff} \subset \mathcal{B}(\mathcal{H})$, which describe the observables accessible of a physical system A at the initial and final stages of the process, respectively. 
As we did in the previous sections, the induced microscopic map acting on the system between $t_\ii$ and $t_\ff$ is denoted $\mathcal{N}_{t_\ii, t_\ff}$ and one has $\rho(t_\ff) = \mathcal{N}_{t_\ii, t_\ff}(\rho(t_\ii))$.

However, one does not have access to the initial and final states, but only to the outcomes of the accessible observables generating the algebras $\mathcal{A}_\ii$ and $\mathcal{A}_\ff$. Therefore, one must instead consider the coarse-grained states
\be
    \rho^\text{cg}_{\mathcal{A}_{\ii}} = (P_{C_{\mathcal{A}_{\ii}}, u} \circ C_{\mathcal{A}_{\ii}})(\rho), \qquad \rho^\text{cg}_{\mathcal{A}_{\ff}} = (P_{C_{\mathcal{A}_{\ff}}, u} \circ C_{\mathcal{A}_{\ff}})(\rho)
\ee
attached to the initial and final algebras $\mathcal{A}_{\ii}$ and $\mathcal{A}_{\ff}$. The map \eqref{eq: Lambdatitf} describing the macroscopic dynamics is thus straightforwardly generalized to
\be \label{eq: evolutionmap gena}
    \Lambda_{t_\mathrm{i},t_\mathrm{f}} := P_{C_{\mathcal{A}_{\ff}}, u} \circ C_{\mathcal{A}_{\ff}} \circ \mathcal{N}_{t_\mathrm{i},t_\mathrm{f}}. 
\ee
Following our open-system approach, we define a fixed point of our two-time protocol, which we interpret as the macroscopic equilibrium state of the system. 
In close analogy with the construction used for projector algebras, we introduce the \emph{connection} transporting the states from the final algebra to the initial algebra
\begin{align} 
    \text{A}_{\ii, \ff}: \mathcal{A}_\ff &\longrightarrow \mathcal{A}_\ii \nn \\
    \rho^{\text{cg}}_{\mathcal{A}_\ff} &\longrightarrow \rho^{\text{cg}}_{\mathcal{A}_\ii} =  (P_{C_{\mathcal{A}_{\ii}}, u} \circ C_{\mathcal{A}_{\ii}})(\rho^{\text{cg}}_{\mathcal{A}_\ff}) .
\end{align}
Then it is meaningful to consider a fixed point $\sigma_\ii := e^{-K_\ii} \in \mathcal{A}_\ii$ (assumed to be of full rank so that one can define the modular Hamiltonian $K_\ii$) of the map 
\begin{align} \label{eq: evolutionmap gena tilde}
    \tilde{\Lambda}_{t_\mathrm{i},t_\mathrm{f}}: \mathcal{B}(\mathcal{H}) &\longrightarrow \mathcal{A}_\ii \nn \\
    \rho &\longrightarrow \text{A}_{\ii, \ff} \circ \Lambda_{t_\mathrm{i},t_\mathrm{f}}(\rho),
\end{align}
associated with repeated two-point measurement protocols, with the two measurements separated by the evolution ${\cal N}_{t_\ii,t_\ff}$\footnote{Of course, as in the commutative case, if $\mathcal{N}_{t_\ii,t_\ff}$ is not (C)PTP, the fixed point $\s_i \in \mathcal{A}_\ii$ is not ensured to exist.}. In exact analogy with \eqref{eq: generic second law commalg opens}, we prove in App.~\ref{app: proofs} the non-commutative version of the second law
\be \label{eq: secondlaw mean value genalg}
    \Delta S_{\mathcal{A}_{\ii,\ff}}(\rho) - \Delta \langle K_{\ii,\ff} \rangle_{\rho} \geq - S(\rho_{AB} (t_\mathrm{i})\lvert \lvert \rho^{\text{cg}}_{\mathcal{A}_\ii}(t_\mathrm{i}) \otimes \rho_B(t_\mathrm{i})) = - I_{AB}(t_\mathrm{i}) - S(\rho(t_\mathrm{i}) \lvert \lvert \rho^\text{cg}_{\mathcal{A}_\ii}(t_\mathrm{i}))
\ee
where
\begin{align}
     \Delta S_{\mathcal{A}_{\ii,\ff}}(\rho) &:= S_{\mathcal{A}_{\ff}}(\rho(t_\mathrm{f})) - S_{\mathcal{A}_{\ii}}(\rho(t_\mathrm{i})) \nn \\
     \Delta \langle K_{\ii,\ff} \rangle_{\rho} &:= \langle K_\ff \rangle_{\rho(t_\mathrm{f})} - \langle K_\ii \rangle_{\rho(t_\mathrm{i})},
\end{align}
and $K_\ii$ is the modular Hamiltonian associated with a fixed point of the map \eqref{eq: evolutionmap gena tilde} induced by microscopic initially factorized states\footnote{As in the commutative case, when system A and environment B are initially uncorrelated, the induced microscopic map $\mathcal{N}_{t_\ii,t_\ff}$ is CPTP and therefore the map \eqref{eq: evolutionmap gena tilde} is CPTP. Hence, in this case, a fixed point of \eqref{eq: evolutionmap gena tilde} always exists.}, and $K_\ff = - \ln{\s_\ff} = - \ln{\Lambda_{t_\ii, t_\ff}(\s_\ii)}$. Note that the role of entropy is now played by the algebraic entropy rather than the observational entropy \eqref{eq: algebraicentropyII}.
Of course, the quantity $- S(\rho (t_\mathrm{i})\lvert \lvert \rho^{\text{cg}}_{\mathcal{A}_\ii}(t_\mathrm{i})) \leq 0$ measures, as $I_\text{int}(t_\ii)$ did before, the degree to which the initial state $\rho(t_\mathrm{i})$ departs from internal equilibrium with respect to the accessible initial subalgebra. Now, however, the coarse-grained initial state is characterized by a larger set of observables, giving the freedom to better characterize the relevant out of equilibrium observables, even if they do not commute, thereby minimizing the contribution $S(\rho(t_\mathrm{i}) \lvert \lvert \rho^\text{cg}_{\mathcal{A}_\ii}(t_\mathrm{i}))$ and leading to a tighter formulation of the second law. This is expected to be particularly useful in typical situations where the eigenbasis of $\rho(t_i)$ is not easily predictable, and hence the relevant information is expected to be scattered over different observables.

\subsection{The autonomous viewpoint}
\label{sec: two point m bipartite approach gena}

To generalize the autonomous formulation to the non-commuting algebra case, we consider the enlarged system A $\cup$ B composed of the original system A and environment B and the accessible von Neumann algebra
\be
    \mathcal{A}_{\text{A} \cup \text{B}} := \mathcal{A}_{\text{A}} \otimes \mathcal{A}_{\text{B}}
\ee
where $\mathcal{A}_{\text{A} \cup \text{B}} \subset \mathcal{B}(\mathcal{H}_{\text{A}}) \otimes \mathcal{B}(\mathcal{H}_{\text{B}})$, $\mathcal{A}_{\text{A}} \subset \mathcal{B}(\mathcal{H}_{\text{A}})$ and $\mathcal{A}_{\text{B}} \subset \mathcal{B}(\mathcal{H}_{\text{B}})$. These algebras correspond to the observables that one is able to probe on system A and on environment B. The representation theorem of von Neumann algebras \eqref{eq: algIdecompo} implies 
\be\label{eq: auto AB subalgebra decomp}
    \mathcal{A}_{\text{A}} \otimes \mathcal{A}_{\text{B}} \simeq \bigoplus_{J_A J_B} \left[\mathcal{B}(\mathcal{H}_{J_A}) \otimes \mathcal{B}(\mathcal{H}_{J_B}) \otimes \left(\Id_{\mathcal{H}_{J_A}^{'}} \otimes \Id_{\mathcal{H}_{J_B}^{'}}\right) \right] 
\ee
 Moreover, we define the mutual information between A and B as seen from the accessible algebras $\mathcal{A}_{\text{A}}$ and $\mathcal{A}_{\text{B}}$ as
\be
    I_{\mathcal{A}_{\text{A}} \otimes \mathcal{A}_{\text{B}}}(\rho) := S\left(\rho_{\mathcal{A}_{\text{A}} \otimes \mathcal{A}_{\text{B}}}^{\text{cg}} \lvert \lvert \rho^{\text{cg}}_{\mathcal{A}_{\text{A}}} \otimes \rho^{\text{cg}}_{ \mathcal{A}_{\text{B}}}\right).
\ee
This quantity interpolates between the quantum mutual information between A and B, $I_{AB} = S(\rho_{AB} \lvert \lvert \rho_A \otimes \rho_B)$, if $\mathcal{A}_{\text{A}} = \mathcal{B}(\mathcal{H}_{\text{A}})$ and $\mathcal{A}_{\text{B}} = \mathcal{B}(\mathcal{H}_{\text{B}})$, and the classical coarse-grained mutual information $I_{AB}^{\text{ob}} = \sum_{J_A J_B} p_{J_A J_B} \ln{\frac{p_{J_A J_B}}{p_{J_A} p_{J_B}}}$ if the two algebras $\mathcal{A}_{\text{A}}$ and $\mathcal{A}_{\text{B}}$ are commutative.

\vspace{0.3 cm}

From the autonomous viewpoint, the evolution map for system $\text{A} \cup \text{B}$ is a unitary map $\mathcal{N}_{t_\ii, t_\ff} = \mathcal{U}_{t_\ii, t_\ff}$ so that the map \eqref{eq: evolutionmap gena} is CPTP and unital.  If the initial state of the joint system A $\cup$ B is in internal equilibrium, or if the dynamics of the joint system A $\cup$ B is assumed to be scale divisible, we thus obtain
\be \label{eq : gen bipartite second law}
    \Delta S_{\mathcal{A}_{\text{A}}}(\rho) + \Delta S_{\mathcal{A}_{\text{B}}}(\rho) = \Delta S_{\mathcal{A}_{\text{A}} \otimes \mathcal{A}_{\text{B}}} (\rho) + \Delta I_{\mathcal{A}_{\text{A}} \otimes \mathcal{A}_{\text{B}}}(\rho) \geq \Delta I_{\mathcal{A}_{\text{A}} \otimes \mathcal{A}_{\text{B}}}(\rho) .
\ee

Exploiting the procedure of Eq.~\eqref{eq: KBt in op basis}, we can work further on \eqref{eq : gen bipartite second law} to introduce an average effective heat and work transfers directly extending Eqs.~\eqref{eq: heat def sec 3}  and \eqref{eq: work def sec 3}. Rather than expanding on this here, we proceed with the quasi-static description providing more robust notions of work and heat increments. 

\subsection{Quasi-static process, heat and work}\label{sec:algebraic_heat_and_work}

To obtain a systematic definition of heat and work, we introduce a continuous path $\gamma(t)$ in the space of von Neumann subalgebras $\mathcal{A}(t)$, defined by
\be
    \mathcal{A}(t+dt) := \mathcal{U}_{\g(t)} \mathcal{A}(t) \mathcal{U}^\dag_{\g(t)}, \qquad \mathcal{A}_\ii = \mathcal{A}(t_\mathrm{i}), \qquad \mathcal{A}_\ff = \mathcal{A}(t_\mathrm{f}). 
\ee
Therefore, all the algebras considered in the time interval $(t_\mathrm{i}, t_\mathrm{f})$ have the same block decomposition \eqref{eq: algIdecompo}, and only the global unitary ($U$ in Eq.~\eqref{eq: vN alg U J}) is assumed to change.
Then, as in the case of commutative algebras, the prescription differs depending on whether one has access to system A only or also to (a subalgebra of) environment B. In both cases, the average thermodynamics of the non-commutative case follows straightforwardly from the ideas discussed in the previous sections.

\subsubsection{Open system viewpoint}

When only the system is accessible, we consider the map Eq.~\eqref{eq: evolutionmap gena} for an infinitesimal interval and denote by $\s_t$ a fixed point of the map \eqref{eq: evolutionmap gena tilde} with $t_\ii = t$ and $t_\ff = t+dt$, to obtain the increment of entropy production.
\be \label{eq: ent prod alg inf}
    \Sigma_{t,t+dt}:= S(\rho_{t} \lvert \lvert \s_t) - S(\rho_{t+dt} \lvert \lvert \Lambda_{t, t+dt}(\s_t)) 
\ee
which is positive if the dynamics is CP-divisible and locally scale divisible (see Section.~\ref{sec:WH_bipartite_open_av}). 
Above, $(\rho_{t}, \s_t) \in \mathcal{A}(t)$ and $(\rho_{t+dt}, \Lambda_{t,t+dt}(\s_{t})) \in \mathcal{A}(t+dt)$ are coarse-grained states (we keep using the simplified notation Eq.~\eqref{d: representative simple} for representatives of operators in the accessible subalgebras). Then, the entropy production \eqref{eq: ent prod alg inf} can be decomposed into  
the infinitesimal variation of algebraic entropy
\be
    \text{d} S_{\mathcal{A}(t)}(\rho) := S_{\mathcal{A}(t+dt)}(\rho(t+dt)) - S_{\mathcal{A}(t)}(\rho(t))
\ee
and the variation of modular Hamiltonian
\be
    \text{d} \langle K_{t, t+dt} \rangle_\rho := \tr{\rho_t \ln{\s_t}} - \tr{\rho_{t+dt} \ln{\Lambda_{t,t+dt}(\s_{t})}}.
\ee
However, we can prove (see Appendix \ref{app: proofs}) that 
\be \label{eq: trivial connection geenral setting}
    \tr{\rho_{t+dt} \ln{\Lambda_{t,t+dt}(\s_{t})}} = \tr{\rho_{t+dt} \ln{\s_{t}}} + O(dt^2)
\ee
so that at first order we get
\be
    \Sigma_{t,t+dt} = \text{d} S_{\mathcal{A}(t)}(\rho) - \text{d} \langle K_{t} \rangle_\rho 
\ee
where $K_t = - \ln{\s_t}$ is the instantaneous modular Hamiltonian and
\begin{align}
     d \langle K_{t} \rangle_\rho &:= \tr{((\rho_{t+dt} - \rho_{t}) K_t)}
\end{align}
As in the commutative case, we can decompose the modular Hamiltonian $K_t$ in an orthogonal basis $\{ \Id, H_{\mathcal{A}_t}, O^1_{\mathcal{A}_t}, \cdots O^N_{\mathcal{A}_t} \}$ of the algebra $\mathcal{A}_t$, i.e.
\be
    K_t = \beta(t) \left(H_{\mathcal{A}_t} + \sum_k \l_k(t) O^k_{\mathcal{A}_t}\right) + \ln{\left\{\tr{\left[e^{-\beta(t) (H_{\mathcal{A}_t} + \sum_k \l_k(t) O^k_{\mathcal{A}_t})}\right]}\right\}} \Id \in \mathcal{A}_t
\ee
for some intensive parameters $\{ \ln{\tr{(e^{-\beta(t) (H_{\mathcal{A}_t} + \sum_k \l_k(t) O^k_{\mathcal{A}_t})})}},  \beta(t), \l_1(t), \cdots, \l_N(t) \}$ conjugated to the observables $\{ H_{\mathcal{A}_t}, O^1_{\mathcal{A}_t}, \cdots, O^N_{\mathcal{A}_t} \}$. Of course, $H_{\mathcal{A}_t}$ is the representative of the microscopic Hamiltonian $H(t)$ in the algebra $\mathcal{A}_t$. The heat increment between $t$ and $t+dt$ is then identified as 
\be
    \d \mathcal{Q}_{t, t+dt} := T(t) \text{d} \langle K_{t} \rangle_\rho
\ee
with $T(t) = \beta^{-1}(t)$ is the effective temperature, as usual. Introducing
\be
     \text{d} E_{\mathcal{A}_{t,t+dt}} := \langle H_{\mathcal{A}_{t+dt}} \rangle_{\rho(t+dt)} - \langle H_{\mathcal{A}_{t}} \rangle_{\rho(t)}
\ee
the work increment is defined as 
\bea
    \d \mathcal{W}_{t, t+dt} &=& \text{d} E_{\mathcal{A}_{t,t+dt}} - \d \mathcal{Q}_{t, t+dt}\nn\\
   &=&  \tr{\left[(H_{t+dt}-H_t)\rho_t\right]}-\sum_k \lambda_k(t)\text{d}\mean{O_{{\cal A}_t}^k}.
\eea
All the interpretations provided in Section.~\ref{sec:WH_bipartite_open_av} extend to those quantities, including the notion of work associated with the time-dependency of the algebra.
\begin{remark}
The considerations concerning the choice of inner product on the algebra, discussed in Remark~\ref{remarkoninnerproduct} apply equally to the non-commutative setting.
\end{remark}

\subsubsection{Autonomous viewpoint}
\label{subsec: autonomous viewpoint gen}

\begin{figure}[htp]
\begin{framed}
\centering
  \includegraphics[scale=0.29]{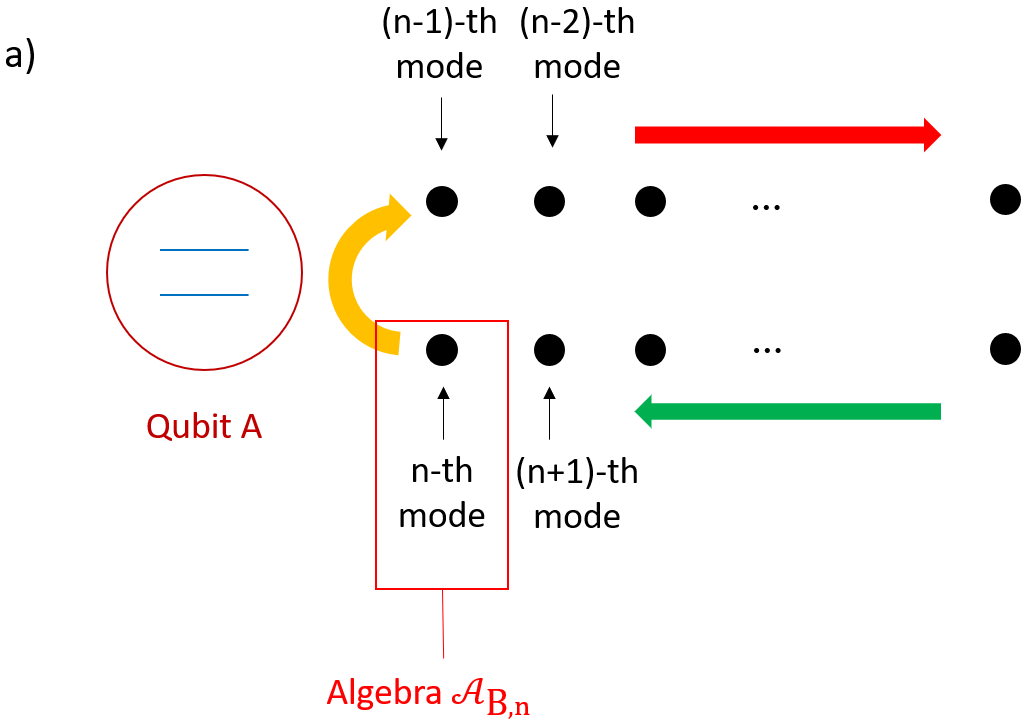}
  \hspace{0.5em}
  \includegraphics[scale=0.29]{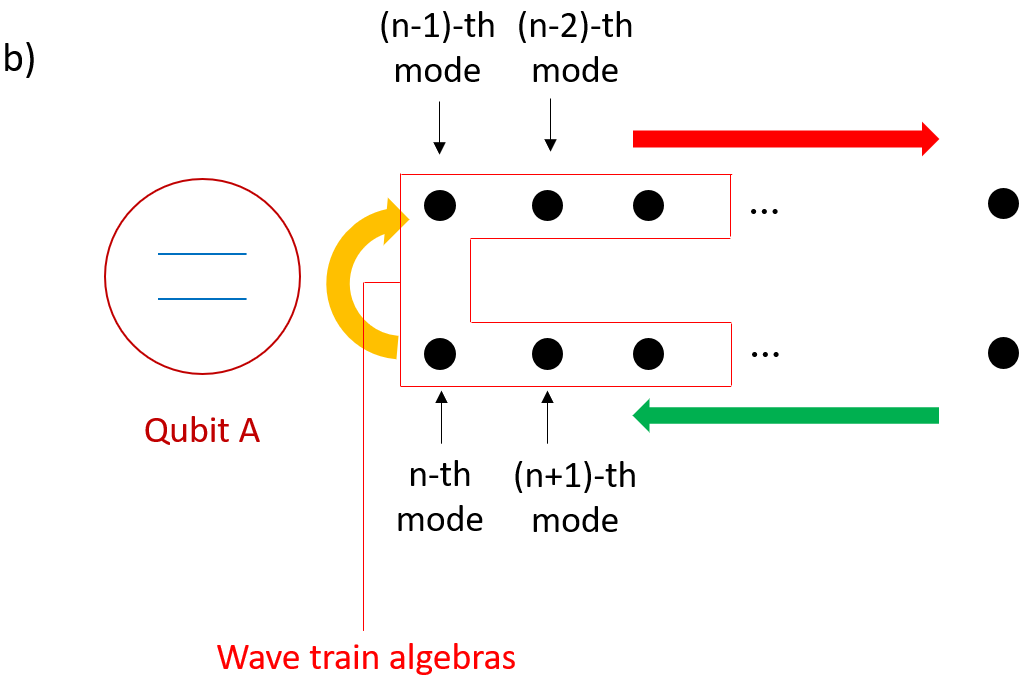}
\caption{\textbf{Application to an emitter coupled to a 1D propagating field.} In the weak emitter-field coupling regime, the situation can be mapped onto a collision model \cite{Prasad2026}, where the total algebra of the electromagnetic field is decomposed into temporal modes (depicted as black spheres), which are initially uncorrelated. The modes freely propagate toward the emitter A, with which they sequentially interact weakly, before being transported away. One may consider as the accessible subalgebra of the field either a) the algebra $\mathcal{A}_{\mathrm{B},n}$ associated with a single temporal mode, or b) an algebra associated with a collection of several temporal modes, which we refer to as a \emph{wave train}. The notions of work, heat and entropy production introduced in our framework are sensitive to this choice, reflecting the different resources available when using different control/measurement setups. In particular, accessing the coherences of a wave-train allows to identify the work contribution associated to the coherent part of the emission (See App.~\ref{app: collisional model}). 
}
\label{fig:collmodel}
\end{framed}
\end{figure}

If the observer has access to a set of continuously varied subalgebras $\mathcal{A}_{\text{B}}(t)$ of environment B on top of $\mathcal{A}_{\text{A}}(t)$, one can build on Section.~\ref{sec: two point m bipartite approach gena} to define work and heat increments. We are interested, at $t_\ii = t$, in a subalgebra $\mathcal{A}_{\text{A} \cup \text{B}}(t) = \mathcal{A}_\text{A} (t) \otimes \mathcal{A}_B(t)$ and, at $t_\ff = t + dt$, in another algebra related by a unitary transformation  $\mathcal{A}_{\text{A} \cup \text{B}}(t+dt) = \mathcal{A}_{\text{A}}(t+dt) \otimes \mathcal{A}_{\text{B}}(t+dt)$, i.e. $\mathcal{A}_{\text{A}}(t+dt) = U_{\g_{\text{A}}(t)} \mathcal{A}_{\text{A}}(t) U^{\dag}_{\g_{\text{A}}(t)}$ and  $\mathcal{A}_{\text{B}}(t+dt) = U_{\g_{\text{B}}(t)} \mathcal{A}_{\text{B}}(t) U^{\dag}_{\g_{\text{B}}(t)}$. Thus, the joint algebra $\mathcal{A}_{\text{A} \cup \text{B}}(t+dt)$ changes between $t$ and $t+dt$ by the unitary operator $U_{\g_{AB}(t)} = U_{\g_{A}(t)} \otimes U_{\g_{B}(t)}$. Then, if one assumes internal equilibrium at $t$, we obtain an infinitesimal version of \eqref{eq : gen bipartite second law}
\be
    \text{d} S_{\mathcal{A}_{\text{A},t}}(\rho) + \text{d} S_{\mathcal{A}_{\text{B},t}}(\rho) = \text{d} S_{\mathcal{A}_{\text{A},t} \otimes \mathcal{A}_{\text{B},t}} (\rho) + \text{d} I_{\mathcal{A}_{\text{A},t} \otimes \mathcal{A}_{\text{B},t}} \geq \text{d} I_{\mathcal{A}_{\text{A},t} \otimes \mathcal{A}_{\text{B},t}}.
\ee
Decomposing the modular Hamiltonian of the coarse-grained state of B, introduced in Section.~\ref{sec: two point m bipartite approach gena}, over an orthogonal operator basis $\{ \Id_{\text{B}}, H_{\mathcal{A}_{\text{B},t}}, O^1_{\mathcal{A}_{\text{B},t}}, \cdots, O^N_{\mathcal{A}_{\text{B},t}} \} \in \mathcal{A}_\text{B}(t)$, where $H_{\mathcal{A}_{\text{B},t}}$ denotes the representative of the microscopic Hamiltonian $H_{\text{B}}(t)$ in the algebra $\mathcal{A}_{\text{B}}(t)$, yields
\be
    \rho_{B, t} = \frac{e^{- \beta(t) (H_{\mathcal{A}_{\text{B},t}} + \sum_k \l_k(t) O^k_{\mathcal{A}_{\text{B},t}} )}}{\tr{(e^{- \beta(t) (H_{\mathcal{A}_{\text{B},t}} + \sum_k \l_k(t) O^k_{\mathcal{A}_{\text{B},t}} )}})} 
\ee
and allows us to write
\bea \label{eq: gen bipartite second law heat}
    \text{d} S_{\mathcal{A}_{\text{A},t}}(\rho) + \beta(t) (\text{d} \langle H_{\mathcal{A}_{\text{B},t}} \rangle_{\rho_{\text{B}}} + \sum_k \l_k(t) \text{d} \langle O^i_{\mathcal{A}_{\text{B},t}} \rangle_{\rho_{\text{B}}}) \nonumber\\
    =\text{d} S_{\mathcal{A}_{\text{A},t} \otimes \mathcal{A}_{\text{B},t}} (\rho) + \text{d} I_{\mathcal{A}_{\text{A},t} \otimes \mathcal{A}_{\text{B},t}} + O(dt^2) \geq \text{d} I_{\mathcal{A}_{\text{A},t} \otimes \mathcal{A}_{\text{B},t}} + O(dt^2)
\eea
(see proof in App.~\ref{app: proofs}). The definition of the heat increment in B follows
\be
    \d \mathcal{Q}_{\text{B}, t, t+dt} := \text{d} \langle H_{\mathcal{A}_{\text{B},t}} \rangle_{\rho_{\text{B}}} + \sum_i \l_i(t) \text{d} \langle O^i_{\mathcal{A}_{\text{B},t}} \rangle_{\rho_{\text{B}}},
\ee
leading to a definition of the work increment in B as
\begin{align}
    \d \mathcal{W}_{\text{B}, t, t+dt} &= \text{d} E_{\text{B}, t, t+dt} - \d \mathcal{Q}_{\text{B}, t, t+dt} \nn \\
    &= \tr{((H_{\mathcal{A}_{\text{B},t+dt}} - H_{\mathcal{A}_{\text{B},t}}) \rho_{\text{B},t})} - \sum_i \l_i(t) \text{d} \langle O^i_{\mathcal{A}_{\text{B},t}} \rangle_{\rho_{\text{B}}}
\end{align}
As for the open-system case, this generalization is necessary to describe
coherent quantum processes without restricting the observer to a single
measurement basis. We illustrate these algebra-dependent thermodynamic laws
in App.~\ref{app: collisional model} for a two-level emitter weakly coupled to a one-dimensional
propagating electromagnetic field. In the weak-coupling regime, the field can 
be decomposed into collision modes, that is, coarse-grained temporal modes
interacting sequentially with the emitter (see Fig.~\ref{fig:collmodel}). This
formulation of the dynamics allows us to easily compare
different levels of access to the field observables, ranging from the full
algebra of individual collision modes to local energy algebras and global
algebras associated with the emitted wave train.
We show in particular that coherence transferred from the emitter to the field
appears as a coherent displacement carrying non equilibrium free energy and
therefore constitutes a resource from which work could be extracted. Restricting the accessible algebra so
that this coherence is no longer resolved increases the assigned entropy and
reduces the accessible free energy, while a further coarse-graining at the
level of the whole wave train additionally discards information about the
localization of the emitted excitation. Conversely, when the accessible
algebra retains the coherences between the vacuum and one-particle sectors,
the coherent field amplitude is explicitly identified as a work contribution.
This example therefore shows how the same microscopic emitter--field dynamics
gives rise to different apparent entropy production, heat, work, and
thermodynamic resources depending on the measurement and control capabilities
encoded in the chosen accessible algebra.

\subsection{Stochastic version}

Finally, we propose a formulation of stochastic quantum thermodynamics within the autonomous viewpoint using our general algebraic framework. Unlike in the previous sections, where the analysis was first developed for two-point measurement protocols and only subsequently extended to the continuous-time limit, we adopt a trajectory-based description from the outset.

In contrast with the average non-commutative analysis of Secs.~\ref{sec: noncomm open}-\ref{sec:algebraic_heat_and_work} which was a direct generalization of the commutative case, defining stochastic trajectories at the subalgebra level requires a slight paradigm shift. While before, single realizations (i.e. trajectories) were naturally related to a single set of measurement outcomes in the coarse-measurement performed at time $t$, the quantum macrostates are by construction obtained from the combination of several measurements which cannot be performed simultaneously, as they involve non commuting observables. 

However, we can identify directly useful approaches inspired by the usual setup of quantum stochastic thermodynamics which is based on a continuous monitoring of the \emph{environment} of the system of interest. The resulting measurement outcomes define the stochastic trajectories on which the stochastic thermodynamic analysis is based. The density operator of the system conditioned to the measurement results is precisely what is called a quantum trajectory \cite{WisemanBook}. The latter can be accessed in two ways: First, it can be inferred by applying a POVM element selected by the outcome of the measurement on the environment, allowing to predict work and heat increments \cite{Elouard17,Manzano18}. This strategy is not included in the framework of Sections.~\ref{sec: two point thermodynamics}-\ref{sec: heat and work}, as the system would have to be described by a full density operator rather than a diagonal matrix (or we would have to base our entropy maximization on a POVM, see App.~\ref{app: POVM}). However, it is naturally included in an algebraic stochastic framework where the full algebra of the system is considered as accessible, and the subalgebra of the environment is taken to be commutative, i.e. a set of projectors whose outcomes define the stochastic trajectory. Interestingly, our approach allows us to straightforwardly authorize coarse measurements on the environment, which in particular encompass the experimentally important case of finite detection efficiency, i.e. imperfect measurements on the bath.  

Second, the quantum trajectory associated with the outcomes of the measurements on the bath can also be directly characterized by a full quantum tomography of the system state at various stopping times, post-selected over a particular sequence of results on the reservoir. This requires to implement enough repetitions of the protocol to be able to observe many iterations of the \emph{same quantum trajectory}, but was nonetheless realized experimentally for simple enough systems, such as open qubits or oscillators, and engineered environments \cite{Murch13,Campagne-Ibarcq16,Rossi19}. Once again, this situation goes beyond the commutative setting of Secs.~\ref{sec: two point thermodynamics}-\ref{sec: heat and work}, as the conditional system state is reconstructed from several incompatible measurements. The framework developed in this section addresses this situation with the crucial extension that the tomography of the system(s) is now allowed to be partial as it can be restricted to a subalgebra. It thereby opens the applicability of quantum stochastic thermodynamics to much larger systems and larger class of environments.

Practically, we will treat those two situations at once by encompassing them in a slightly broader setting where we allow both systems A and B to be described by noncommutative subalgebras, and the stochastic trajectories are defined by the sequence of the sectors $(J_A,J_B)$ in which the systems are found every $dt$ (see Eq.~\eqref{eq: auto AB subalgebra decomp}). The block-diagonal structure of the accessible algebra ensures that we can interpret the evolution as a classical mixture of these sector trajectories. Practically however, the quantum states of A and B conditioned to these trajectory have to be either inferred by a POVM update model, or measured by tomography over multiple repetitions. In the latter case, the coarse-grained nature of the trajectory should significantly increase the probability to reproduce each sector trajectory, and therefore make possible the analysis of a the joint trajectory of two macroscopic quantum systems. The special case of a projective measurement on B and a fine-grained description of A is recovered for $\text{dim}{\cal H}_{J_B} = 1$ and $\text{dim}{\cal H}_{J_A}'=1$, with a single sector $J_A$ when the full algebra $\mathcal{B}(\mathcal{H}_{\mathrm A})$ is accessible.

Throughout this subsection, we assume that deviations from internal equilibrium can be neglected. Accordingly, we disregard the coarse-grained contribution $s_{IF}^{\mathrm{cg}}$ encoding such deviations. Nevertheless, when restricting the analysis to two-point measurement protocols, fluctuation theorems closely analogous to \eqref{eq:FT3} can be derived by following the same line of reasoning as in the previous sections.\\

As in Section.~\ref{sec: stoch thermo comm}, we proceed by first identifying a microstate trajectory. Since all observables in $\mathcal{A}_{\text{A} \cup \text{B}}(t) = \mathcal{A}_{\text{A}}(t) \otimes \mathcal{A}_{\text{B}}(t)$, are accessible at any time $t \in (t_\ii, t_\ff)$, the coarse-grained state of $A \cup B$ at time $t$ can be diagonalized as
\be
    \rho_{\text{A} \text{B}}^{\text{cg}}(t) = \sum_{J_A J_B} \rho_{J_A J_B}(t) \otimes \frac{\Id_{\mathcal{H}_{J_A}^{'} \otimes \mathcal{H}_{J_B}^{'}}}{V_{J_A}^{'} V_{J_B}^{'}}
\ee
in a basis $\{\ket{j_{AB}(t)}\}$ of $ \mathcal{H}_{J_{\text{A}}} \otimes \mathcal{H}_{J_{\text{B}}}$ so that 
\begin{align}
    \rho_{\text{A} \text{B}}^{\text{cg}}(t) &= \sum_{J_A J_B} \sum_{j_{AB} \in J_A J_B} p_{j_{AB}}(t) \ket{j_{AB}(t)} \bra{j_{AB}(t)} \otimes \frac{\Id_{\mathcal{H}_{J_A}^{'} \otimes \mathcal{H}_{J_B}^{'}}}{V_{J_A}^{'} V_{J_B}^{'}} \nn \\
    &= \sum_{J_A J_B, J'_A J'_B} \sum_{j_{AB} \in J_A J_B} \sum_{j'_{AB} \in J'_A J'_B} \frac{p_{j_{AB}}(t)}{V_{J_A}^{'} V_{J_B}^{'}} \ket{j_{AB}   j'_{AB}(t)} \bra{j_{AB} j'_{AB}(t)}
\end{align}
where $\{ \ket{j'_{AB}} \}$ is an arbitrary orthonormal basis of $\mathcal{H}'_{J_A} \otimes \mathcal{H}'_{J_B} $. The joint forward and backward probabilities associated with the microscopic trajectories $\g_{(i_{AB}i_{AB}')(t) (f_{AB} f_{AB}')(t+dt) }$ can thus be formally defined by replacing the map appearing in \eqref{eq: autonompus forward prob} and \eqref{eq: autonompus backward prob} with $$\Lambda_{t, t+dt} := P_{C_{\mathcal{A}_{t+dt, \text{A}} \otimes \mathcal{A}_{t+dt, \text{B}}}, u} \circ C_{\mathcal{A}_{t+dt, \text{A}} \otimes \mathcal{A}_{t+dt, \text{B}}} \circ \mathcal{U}_{t, t+dt}$$. The entropy production increment can then be obtained directly via
\be \label{eq: entropduction auto bipartite alg}
    \delta s_c(\g_{(i_{AB}i_{AB}') (f_{AB} f_{AB}') }) := \ln{\frac{p(\g_{(i_{AB}i_{AB}') (f_{AB} f_{AB}') })}{p(\g_{(i_{AB}i_{AB}') (f_{AB} f_{AB}') }^R)}} = s_{f_{AB}} -s_{i_{AB}}  
\ee
where  $s_{i_{AB}} = - \ln{\frac{p_{i_{AB}}}{V_{I_{AB}}'}}$ and $s_{f_{AB}} = - \ln{\frac{p_{f_{AB}}}{V_{F_{AB}}'}}$. Then, we write 
\be \label{eq: entopyjABcond}
    s_{j_{AB}} = s_{J_{AB}} + s(j_{AB} \lvert J_{AB})
\ee
with $j = i,f$, $J = I,F$ and where
\be \label{def: entropycond}
   s_{J_{AB}} := - \ln{\frac{p_{J_{A} J_B}}{V_{J_A J_B}'}}, \qquad p_{J_{A} J_B} = \sum_{j_{AB} \in J_{AB}} p_{j_{AB}}
\ee
are respectively the stochastic coarse-grained entropy and the probability of the macrostate associated to sector $J_{AB} = (J_A,J_B)$. Moreover,
\be
s(j_{AB} \lvert J_{AB}) := -\ln{\frac{p_{j_{AB}}}{p_{J_{A}J_B}}}.
\ee
 must be interpreted as a stochastic conditional entropy of microstate $j_{AB}$, conditioned on being in the sector $J_{AB}$. Then, we further introduce (classical) stochastic observational mutual information between the trajectories of A and B 
\be \label{def: mutual inf class alg}
    i_{J_{AB}} := s_{J_A} + s_{J_B} - s_{J_{AB}}
\ee
where $s_{J_A} = - \ln{\frac{p_{J_A}}{V_{J_A}'}}$ and $s_{J_B} = - \ln{\frac{p_{J_B}}{V_{J_B}^{'}}}$, and 
\be
    p_{J_A} = \sum_{J_B} p_{J_{A}J_B}, \quad
    p_{J_B} = \sum_{J_A} p_{J_{A}J_B}
\ee
are the individual probabilities for A and B to be in sectors $J_A$ and $J_B$, respectively. From Eqs.~\eqref{eq: entopyjABcond}, \eqref{def: entropycond} and \eqref{def: mutual inf class alg}, one can recast Eq.~\eqref{eq: entropduction auto bipartite alg} as 
\be
     \delta s_c(\g_{(i_{AB}i_{AB}') (f_{AB} f_{AB}') }) = \text{d} s_{J_A} + \text{d} s_{J_B} - \text{d} i_{J_{AB}} + \text{d} s(j_{AB} \lvert J_{AB})
\ee
where the infinitesimal variation $\mathrm{d}$ is evaluated between $t_\ii=t$ and $t_\ff=t+dt$, for the microstates $j_{AB}j_{AB}'=i_{AB}i_{AB}',f_{AB}f_{AB}'$ and their corresponding macrostates. By concatenating the infinitesimal entropy production increments along the total trajectory from $t_\ii$ to $t_\ff$, we finally obtain the general version of our integral fluctuation theorem
\be \label{eq: FTalg00}
    \langle e^{s_{I_A} - s_{F_A} + s_{I_B} - s_{F_B} + i_{F_{AB}} - i_{I_{AB}} - s(f_{AB} \lvert F_{AB}) + s(i_{AB} \lvert I_{AB})} \rangle =1
\ee
where $i_{AB}$ and $f_{AB}$ label the initial and final macrostates of the bipartite system at time $t_\ii$ and $t_\ff$ respectively, and we have set $\xi = 1$ in \eqref{eq: FTalg00} (no absolute irreversibility, see Section.~\ref{sec: stoch open}). Convexity of the exponential implies
\be \label{eq: conv gen}
    \langle s_{F_A} - s_{I_A} + s_{F_B} - s_{I_B} + s(f_{AB} \lvert F_{AB}) -s(i_{AB} \lvert I_{AB})  \rangle \geq \langle i_{F_{AB}} - i_{I_{AB}}\rangle 
\ee
where the difference $s(f_{AB} \lvert F_{AB}) -s(i_{AB} \lvert I_{AB})$ captures the quantum contribution associated with the evolution of the density matrices within each pair of sectors $(I_{AB}, F_{AB})$. 
Notice, however, that the stochastic entropy term appearing in the left-hand side of Eq.~\eqref{eq: conv gen} does not exactly match the variation of coarse-grained entropies $\Delta S_{{\cal A}_A}+\Delta S_{{\cal A}_B}$.
This is because the eigenbases of the macrostates $\rho_{J_A}$ and $\rho_{J_B}$ associated with the sector $J_A$ and $J_B$ is not in general compatible with the $\ket{j_{AB}}$ trajectory basis. As a consequence, the left-hand side of Eq.~\eqref{eq: conv gen} also contains the quantum mutual information between A and B within the sector $J_{AB}$. Explicitly, using that
\bea\label{eq: other gen ave sec law}
    \langle s(j_{AB} \lvert J_{AB}) \rangle &=& \sum_{J_A J_B} p_{J_A J_B} S(\tilde{\rho}_{J_{AB}})\nn\\
    &=& \sum_{J_A } p_{J_A} S_\text{v.N}(\tilde{\rho}_{J_{A}})+\sum_{J_B} p_{J_B} S_\text{v.N}(\tilde{\rho}_{J_{B}})-\sum_{J_A J_B} p_{J_A J_B} I_{J_A J_B}
\eea
with $J=I,F$ and where $I_{J_A J_B} = S(\tilde{\rho}_{J_A}) + S(\tilde{\rho}_{J_B}) - S(\tilde{\rho}_{J_{AB}})$ is the quantum mutual information attached to sector $J_{AB}$, one can rewrite  Eq.~\eqref{eq: conv gen} under the form of Eq.~\eqref{eq : gen bipartite second law} (see App.~\ref{app: proofs} for a proof).

Finally, we specialize to the practically relevant case in which the environment B is continuously monitored in a single basis at each time $t$, that is, we represent it via the family of (unitarily equivalent) commutative subalgebras $\{ \mathcal{A}_\text{B}^c(t) \lvert t \in (t_\ii, t_\ff) \}$ and where A is measured at $t_\ii$ and $t_\ff$ in some basis. This corresponds in particular to the usual setup of quantum stochastic thermodynamics, but now allowing for coarse (e.g. noisy) measurements on the environment.
In this case, the entropy increment $s_{B,t}$ can be decomposed at each time $t$ according to \eqref{eq: entropy B element}. Using the results of Section \eqref{sec: autonomou sapproach continuous stoch}, we obtain the integral fluctuation theorem
\be\label{eq: alg stoch th heat like}
    \langle e^{s_{I_A} - s_{F_A} - \int_{\gamma_B} (\beta(t) \dot{Q}_B + D_B(t)) dt  + i_{F_{AB}} - i_{I_{AB}}} \rangle = 1
\ee
for the full trajectory from $t_\ii$ to $t_\ff$ which naturally identifies a stochastic heat contribution.
Remarkably, for the system, only the difference between the initial and final stochastic entropy terms appears. This means that the system may only be measured at times $t_\ii$ and $t_\ff$ (in its instantaneous state eigenbasis), rather than throughout the evolution.
By contrast, environment B must be continuously monitored in some (possibly coarse-grained) basis for $t \in (t_\ii, t_\ff)$. 
Note that we have used in Eq.~\eqref{eq: alg stoch th heat like} that $s(f_{AB} \lvert F_{AB}) = s(i_{AB} \lvert I_{AB}) = 0$ because of the measurements on A and B at the endpoints $t_\ii$ and $t_\ff$ (but not between $t_\ii$ and $t_\ff$).  
At first sight, this relation appears very close to \eqref{eq: FTinfbob}. However, the underlying dynamics are fundamentally different. In the present construction, since no measurement is performed on system A except at the initial and final times $t_\ii$ and $t_\ff$, system A evolves ``freely'' (or more precisely, according to the dynamics induced by B) on the interval $(t_\ii,t_\ff)$ and the fluctuation theorem Eq.~\eqref{eq: alg stoch th heat like} brings information about this spontaneous process. By contrast, the derivation of \eqref{eq: FTinfbob} was relying on measurements of A at all intermediate times $t\in (t_{\mathrm i},t_{\mathrm f})$, which continuously perturb the dynamics through measurement back-action. Eq.~\eqref{eq: alg stoch th heat like} is well suited to express thermodynamic constraints, beyond the average for large quantum open systems via the continuous monitoring of their environments.

\section{Conclusion}

We have proposed a unified framework to formulate the laws of thermodynamics for possibly many-body quantum systems, both at the average and stochastic level. Our operational approach is based on a flexible notion of quantum coarse-graining, setting a distinction between accessible observables, whose value is assumed to be tracked, and internal observables which are ignored. This distinction may be done by specifying a choice of projective measurement scheme, that is, a commutative algebra of accessible observables, in which case one obtains an essentially classical macrostate description of the system. Instead, in cases where several measurements in different bases can be performed, one can specify the coarse-graining via the von Neumann subalgebra of the accessible observables, whose expectation values characterize a reduced density operator, that is, a fully quantum macrostate. 
Applying the Jaynes principle, we have defined an entropy function that depends on both the coarse-grained state and the accessible algebra. It exactly corresponds to the observational entropy in the commutative case, and otherwise extends it to non-classical coarse-grained descriptions. From a two-point measurement procedure, we have shown that this entropy fulfills a generalized version of the second law and fluctuation theorems. In particular, 
we have identified new contributions arising when the ignored (internal) degrees of freedom are capable of affecting the macroscopic dynamics, causing apparent violations of the coarse-grained second law. These contributions vanish if the coarse-graining is chosen such that the internal (inaccessible) degrees of freedom are in a micro-canonical distribution --which can be understood as a condition of internal equilibrium expected to be fulfilled for a fast and ergodic internal dynamics-- or if the internal dynamics are effectively decoupled from the accessible observables. By considering quasi-static processes, in which internal equilibrium (or decoupling between internal and accessible dynamics) is maintained at all times, we identified emergent notions of work and heat that bridge the gap between quantum-thermodynamic and macroscopic definitions, thereby connecting our informational second law and fluctuation theorems to an operational formulation of the first law and fluctuating energy currents.
In particular, we have derived a fluctuation theorem involving the heat fluctuations between a quantum system and its environment, which only requires monitoring the environment with coarse (imperfect) measurements, henceforth providing an experimental-friendly access to quantum stochastic thermodynamics analysis of large quantum systems.
We have illustrated those concepts on basic examples such as the spontaneous emission of a two-level atom, the thermalization of a qubit probed in a time-dependent basis, and finally, a qubit coupled to a waveguide where the impact of accessing different algebras of the emitted light was discussed. 

The flexibility of our framework in defining the set of accessible observables opens new perspective to analyze genuinely many-body situations, both theoretically and experimentally.
An immediate step is therefore to apply it on more complex many-body systems, using numerically solvable and analytically integrable models as benchmarks before testing it experimentally. 
One of the natural lines of investigation could be to characterize under which circumstances the apparent violations of the coarse-grained second law arise and vanish.

Another natural follow-up would be the analysis of quantum engines of batteries based on many-body systems, in an experimental-friendly formulation. Importantly, all components of the machines (work loads, energy sources, and working media) may themselves be many-body systems and addressed in an appropriately coarse-grained way, which can be tuned to balance operationality, accuracy of the thermodynamic laws, and invasiveness of the probing scheme.
 
Finally, owing to its algebraic formulation, our formalism is well suited to analyze the thermodynamics of truly macroscopic quantum systems -- composed of infinitely many degrees of freedom like fields -- and therefore investigate the quantum-to-classical transition and the quantum measurement process with a rigorous thermodynamic approach \cite{VanDenBossche:2022clz, Bossche:2023sfr,van2023revisiting, VanDenBossche2024}. A technical extension to allow the accessible subalgebra to be a type II or III von Neumann algebra would be interesting as it would offer the perspective of a quantum thermodynamic standpoint on quantum field theory problems such as the thermodynamics of quantum black holes/horizons \cite{Wall:2009wm, Wall:2011hj, Chandrasekaran:2022cip, Chandrasekaran:2022eqq, Kudler-Flam:2023qfl2, Jensen:2023yxy, Faulkner:2024gst2, Rignon-Bret:2024zhj, Rignon-Bret:2026ttb, Rignon-Bret:2026jzt} or defining and quantifying the available resources in a quantum field in a compact region of spacetime. In the same line, the emergence of classicality can be related to the emergent commutative behavior of sufficiently coarse-grained subalgebras \cite{Bibak25}. Our framework sets a natural stage to investigate how the thermodynamic behavior, or the expressions for the work and heat, increments are affected by this transition.

\section*{Acknowledgments}

The authors thank warmly Mohamed Boubakour and Dragi Karevski for scientific discussions.
This work is funded by the European Union. Views and opinions expressed are however those of the authors only and do not necessarily reflect those of the European Union or the European Research Council. Neither the European Union nor the granting authority can be held responsible for them. This work is supported by ERC grant QARNOT, project number 101163469.

\appendix

\section{Proofs}

\begin{itemize}

    \item Proof of \eqref{eq: coarse-grained state ga}: One uses first \eqref{eq: algIdecompo} to decompose any operator $a \in \mathcal{A}$ as 
    \be
        a = U \big( \bigoplus_J a_J \otimes \mathbb{I}_{\mathcal{H}_J^{'}}  \big) U^{-1}, \qquad \forall J, \quad a_J \in \mathcal{B}(\mathcal{H}_J) 
    \ee
    for some unitary operator $U$. Let us now characterize the state $\rho$ associated to a given value of the accessible observables $\mean{a}_\rho$. If $\{ \Pi_J \}$ is the set of orthogonal projectors on the subspaces $J$, then one has 
    \begin{align}
        \tr{(a\rho)} &= \sum_J \tr_{\mathcal{H}_J}\tr_{\mathcal{H}_J^{'}}{\big( a_J \Pi_J U^{-1} \rho U \Pi_J \big)} \nn \\
        &= \sum_J \tr_{\mathcal{H}_J}{(a_J \rho_J)}
    \end{align}
    with
    \be \label{eq: rhoJ from rho}
        \tr_{\mathcal{H}_J^{'}}{ (\Pi_J U^{-1} \rho U \Pi_J)} = \rho_J \in \mathcal{B}(\mathcal{H}_J).
    \ee
    and
     \be \label{eq: aJ}
        \tr_{\mathcal{H}_J}{(a_J \rho_J)} = \mean{a}_{\rho_J}.
    \ee
   Therefore, since $a_J$ is an arbitrary operator in $\mathcal{B}(\mathcal{H}_J)$, a state $\rho$ satisfies the constraints
    \be \label{eq: app constraints}
        \forall a \in \mathcal{A}, \qquad \tr{(a \rho)} = \langle a \rangle_\rho.
    \ee
    if it only if for any $J$ it also satisfies Eqs.~\eqref{eq: rhoJ from rho}-\eqref{eq: aJ}.
   In particular, the coarse-grained state
   \be \label{eq: coarse-grained state 2}
        \rho_{\mathcal{A}}^{\text{cg}} = 
         U \big( \sum_J   \rho_J \otimes \frac{\Id_{\mathcal{H}_J'}}{V_{J'}} \big) U^{-1}
    \ee
    satisfies \eqref{eq: rhoJ from rho}. Hence the only remaining point needed to be proven is that among all the states satisfying \eqref{eq: rhoJ from rho}, the coarse-grained state \eqref{eq: coarse-grained state 2} on the algebra $\mathcal{A}$ is the state with maximum entropy.
    One starts with
    \be \label{eq: positivit of rela rho rhoacg}
        S(\rho \lvert \lvert \rho^{\text{cg}}_{\mathcal{A}}) \geq 0
    \ee
    from the positivity of the relative entropy. Then, Eq.~\eqref{eq: positivit of rela rho rhoacg} can be recast as
    \begin{align}
        S(\rho \lvert \lvert \rho^{\text{cg}}_{\mathcal{A}}) &= S(\rho \lvert \lvert U \big( \sum_J   \rho_J \otimes \frac{\mathcal{I}_{\mathcal{H}_J^{'}}}{V_{J'}} \big) U^{-1}) \nn \\
        &= \tr{(\rho \ln{\rho})} - \sum_J \tr{\big( \rho U \ln{(\rho_J \otimes \frac{\mathcal{I}_{\mathcal{H}_J^{'}}}{V_{J'}}) U^{-1}  \big) }} \nn \\
        &= \tr{(\rho \ln{\rho})} - \sum_J \tr{(\rho_J \ln{\rho_J})} + \tr{\rho_J \ln{V_{J'}}} \nn \\
        &= -S_{\text{v.N}}(\rho) + S_{\mathcal{A}}(\rho)
    \end{align}
    so that \eqref{eq: positivit of rela rho rhoacg} allows us to conclude that for any $\rho$ satisfying the constraints \eqref{eq: app constraints}
    \be
        S_{\mathcal{A}}(\rho) \geq S_{\text{v.N}}(\rho).
    \ee

    \item Proof of \eqref{eq: petzmapcgena}: One remembers that the map $C_\mathcal{A}$ is the composition of a unitary, a projective decomposition and a partial trace. In order to compute the Petz map attached to the map $C_\mathcal{A}$, one needs to compute the adjoint of each individual maps composing it, i.e. the adjoint map of the partial trace, the projective decomposition and the unitary. Since 
    \be
        \tr_{\mathcal{H}_J^{'}}^\dag{\rho_J} = \rho_J \otimes \Id_{\mathcal{H}_J^{'}},
    \ee
    and that the projective decomposition $\cdot \rightarrow \sum_J \Pi_J \cdot \Pi_J$ is self-adjoint, $U^\dag = U^{-1}$, and that 
    \be
        C_\mathcal{A}(u = \frac{\Id_{\mathcal{H}}}{D}) = \sum_J \frac{V_J^{'}}{D} \Id_{\mathcal{H}_J}
    \ee
    where $D := \dim{\mathcal{H}}$ and $V_J^{'}:= \dim{\mathcal{H}_J^{'}}$ so that since $D = \sum_J D_J V_J^{'}$, where $D_J = \dim{\mathcal{H}_J}$, one recovers immediately \eqref{eq: petzmapcgena} using the definition of the Petz map.
    \item Proofs of \eqref{eq: entropy hierarchy} and \eqref{eq: entropy hierarchy 2}: Since the map \eqref{eq: petzmapcgena} is unital, Equations \eqref{eq: entropy hierarchy} and \eqref{eq: entropy hierarchy 2} directly follow from monotonicity of the relative entropy. Of course, the reference state in the relative entropy is chosen to be the maximally mixed state. 

    \item Proof of \eqref{eq: entropy prod rate}: First, we take $\rho_B(t_\ii) = \tr_{A} \rho_{AB}(t_\ii)$ to be the state of the environment at the beginning of the protocol. If the system A and the environment B are assumed to be isolated, then there exists a generic CPTP map
    \begin{align} \label{eq: genev}
        \mathcal{N}_{t_\mathrm{i}, t_\mathrm{f}}^{(AB)}: \mathcal{B}(\mathcal{H}_\text{A}) \otimes \mathcal{B}(\mathcal{H}_\text{B}) &\longrightarrow \mathcal{B}(\mathcal{H}_\text{A}) \nn \\
        \rho_{AB}(t_\mathrm{i}) &\longrightarrow \rho_{A}(t_\mathrm{f}) = \tr_B(U_{t_\mathrm{i}, t_\mathrm{f}} \rho_{AB}(t_\mathrm{i}) U_{t_\mathrm{i}, t_\mathrm{f}}^\dag )
    \end{align}
    giving the final state $\rho_A(t_\ff)$ of the system A, where $U_{t_\mathrm{i}, t_\mathrm{f}}$ denotes the unitary operator governing the time evolution of the joint system-environment. When applied on a factorized initial state $\rho_A(t_\mathrm{i})\otimes\rho_B(t_\mathrm{i})$, this map can be used to generate a CPTP map for the system only, i.e. $\mathcal{N}^{(A)}_{t_\mathrm{i},t_\mathrm{f}}: \rho_A(t_\mathrm{i}) \rightarrow \tr_B{(U_{t_\mathrm{i},t_\mathrm{f}} \rho_A(t_\mathrm{i}) \otimes \rho_B(t_\mathrm{i}) U_{t_\mathrm{i},t_\mathrm{f}}^\dag})$. Let $\s_A \in \mathcal{B}(\mathcal{H}_\text{A})$ be a fixed point of the map $\mathcal{N}^{(A)}_{t_\mathrm{i},t_\mathrm{f}}$. Notice that since $\mathcal{N}^{(A)}_{t_\mathrm{i},t_\mathrm{f}}$ is CPTP, a fixed point $\s_A$ always exists. In addition, let us assume that $\s_A$ is full rank.
    Then, we use the monotonicity of the relative entropy for the CPTP map \eqref{eq: genev} between an arbitrary state $\rho_{AB}(t_\mathrm{i}) \in \mathcal{B}(\mathcal{H}_\text{A}) \otimes \mathcal{B}(\mathcal{H}_\text{B})$ and a reference state $\s_A \otimes \rho_{B}(t_\mathrm{i}) \in \mathcal{B}(\mathcal{H}_\text{A}) \otimes \mathcal{B}(\mathcal{H}_\text{B})$, and we obtain 
    \be
    S(\mathcal{N}_{t_\mathrm{i}, t_\mathrm{f}}^{(AB)}(\rho_{AB}(t_\mathrm{i}))||\underbrace{\mathcal{N}_{t_\mathrm{i}, t_\mathrm{f}}^{(AB)}(\s_{A}\otimes\rho_B(t_\mathrm{i})))}_{{\cal N}^{(A)}_{t_\mathrm{i}, t_\mathrm{f}}(\s_{A})=\s_A}\leq S(\rho_{AB}(t_\mathrm{i})||\s_{A}\otimes\rho_B(t_\mathrm{i})).
    \ee
    Hence, after expanding the relative entropies
    \be
        S_{\text{v.N}}(\rho_A(t_\mathrm{f})) - S_{\text{v.N}}(\rho_{AB}(t_\mathrm{i})) - (\langle K_A \rangle_{\rho(t_\mathrm{f})} - \langle K_A \rangle_{\rho(t_\mathrm{i})}) - \tr{(\rho_B(t_\mathrm{i}) \ln{\rho_B(t_\mathrm{i}))}}  \geq 0
    \ee
    that can be rearranged into 
    \be
        \Delta S_{\text{v.N}}(\rho_A) - \Delta \langle K_A \rangle \geq - I_{AB}(t_\mathrm{i})
    \ee
    where $I_{AB}(t_\mathrm{i}) = S(\rho_{AB}(t_\mathrm{i}) \lvert \lvert \rho_A(t_\mathrm{i}) \otimes \rho_B(t_\mathrm{i}))$ is the initial quantum mutual information between A and B.

    \item Proof of \eqref{eq: generic second law commalg opens}: Let us work out first the case without initial correlations between the system A and the environment B. We use the monotonicity of relative entropy for the CPTP map $\Lambda_{t_\mathrm{i}, t_\mathrm{f}}$ \eqref{eq: evolutionmap gena} to get 
    \be \label{monotnot2map}
        S(\rho(t_\mathrm{i}) \lvert \lvert \s_\ii) \geq S(\rho^{\text{cg}}_\ff(t_\ff) \lvert \lvert \s_\ff)
    \ee
    where $\rho^{\text{cg}}_\ff := \Lambda_{t_\mathrm{i}, t_\mathrm{f}}(\rho(t_\ii)) $ is the coarse-grained state associated to the final macrostate and $\s_\ff$ is the image of the fixed point $\s_\ii$ of the CPTP map $\tilde{\Lambda}_{t_\mathrm{i},t_\mathrm{f}}$ \eqref{eq: tildelambda}. Then, we use the fact that $\s_\ii \in \mathcal{A}^{c}_\ii$ so that 
    \be
        \tr{(\rho(t_\mathrm{i}) \ln{\s_\ii})} = \tr{(\rho^{\text{cg}}_\ii(t_\ii) \ln{\s_\ii})}  
    \ee
    where $\rho^{\text{cg}}_\ii(t_\ii) = (P_{C_\ii, u} \circ C_\ii)(\rho(t_\mathrm{i}))$ is the initial coarse-grained state. Likewise, since $\rho^{\text{cg}}_\ii \in \mathcal{A}^c_\ii$ as well, one also has
    \be
        \tr{(\rho(t_\mathrm{i}) \ln{\rho_\ii^{\text{cg}}(t_\ii)})} = \tr{(\rho_\ii^{\text{cg}}(t_\ii) \ln{\rho_\ii^{\text{cg}}}(t_\ii))}
    \ee
    so that \eqref{monotnot2map} becomes 
    \be \label{step1proofmonoto}
        S(\rho^{\text{cg}}_\ii(t_\ii) \lvert \lvert \s_\ii) - S(\rho^{\text{cg}}_\ff(t_\ff) \lvert \lvert \s_\ff) \geq - S(\rho(t_\mathrm{i}) \lvert \lvert \rho^{\text{cg}}_\ii(t_\ii))
    \ee
    which directly implies \eqref{eq: generic second law commalg opens}.

    In the presence of initial correlations, one proceeds similarly to the previous point. One uses monotonicity of relative entropy between an arbitrary initial state $\rho_{AB}(t_\ii) \in \mathcal{B}(\mathcal{H}_{\text{A}} \otimes \mathcal{H}_{\text{B}})$ and the state $\s_\ii \otimes \rho_B(t_\ii)$, where $\s_\ii$ is a fixed point of the map $\tilde{\Lambda}_{t_\mathrm{i},t_\mathrm{f}}$ \eqref{eq: tildelambda}, under the CPTP map
    \begin{align}
        \mathcal{B}(\mathcal{H}_{\text{A}} \otimes \mathcal{H}_{\text{B}}) &\longrightarrow \mathcal{A}^c_{\text{A},\ff} \nn \\
        \Lambda^{(AB)}_{t_\ii, t_\ff} &= \mathcal{P}_{C_\ff, u} \circ C_\ff \circ \mathcal{N}^{(AB)}_{t_\ii, t_\ff}
    \end{align}
    where $\mathcal{N}^{(AB)}_{t_\ii, t_\ff}$ is given by \eqref{eq: genev}. Then, from monotonicity of relative entropy, we obtain
    \be \label{eq: app proof general mut inf}
    S(\Lambda_{t_\mathrm{i}, t_\mathrm{f}}^{(AB)}(\rho_{AB}(t_\mathrm{i}))||\underbrace{\Lambda_{t_\mathrm{i}, t_\mathrm{f}}^{(AB)}(\s_{\ii}\otimes\rho_B(t_\mathrm{i})))}_{{\Lambda}_{t_\mathrm{i}, t_\mathrm{f}}(\s_{\ii})=\s_\ff}\leq S(\rho_{AB}(t_\mathrm{i})||\s_{\ii}\otimes\rho_B(t_\mathrm{i})).
    \ee
    from which Eq.~\eqref{eq: generic second law commalg opens} follows directly from the preceding results, upon first identifying the initial mutual-information contribution using Eq.~\eqref{eq: app proof general mut inf}, and then recovering the initial observational entropy exactly as in the case without initial correlations.

    \item Proof of \eqref{eq: lower boundpivi}: Remember that $\rho^{\text{cg}}(t_\ii) := \sum_I \frac{p_I}{V_I} \Pi_I$ and $\rho(t_\mathrm{i}^+) = \sum_I p_I \rho_I^+$ where $\rho_I^+ = \frac{\Pi_I \rho(t_\mathrm{i}) \Pi_I}{p_I}$ so that:
    \bea
        S(\rho(t_\mathrm{i}^+) \lvert \lvert \rho^{\text{cg}}(t_\ii))) &=& S\left(\sum_I p_I \rho_I^+ \Big\vert\Big\vert \sum_I \frac{p_I}{V_I} \Pi_I\right) \leq \sum_I p_I S\left(\rho_I^+ \Big\vert \Big\vert \frac{\Pi_I}{V_I}\right)\nonumber\\
        &=& \sum_I p_I (\ln{V_I}- S_{\text{vN}}(\rho_I^+)) \leq \sum_I p_I \ln{V_I}
    \eea
    where we used the joint convexity of quantum relative entropy to get the first inequality and the fact that the von Neumann entropy is positive to obtain the second inequality.

    \item Proof of \eqref{eq: average coarse grained stat}, left inequality: One has
    \bea
    p(\gamma_{IF}) &=& \sum_{i\in I, f\in F} p_i \bra{f} P_{C_\ff,u}\circ C_\ff\circ{\cal N}_{t_\ii,t_\ff}[\ket{i} \bra{i}] \ket{f}  \nonumber\\
    &=&\sum_{i\in I} p_i \bra{F}C_\ff\circ{\cal N}_{t_\ii,t_\ff}[\ket{i} \bra{i}] \ket{F},
    \eea
    such that
\bea
s_{IF}^\text{cg} &=& -\ln \sum_{i\in I}\frac{p_I/V_I}{p_i}\frac{p_i\bra{F}C_\ff\circ {\cal N}_{t_\ii,t_\ff}[\ket{i}\bra{i}]\ket{F}}{\sum_{i\in I}p_i\bra{F}C_\ff\circ {\cal N}_{t_\ii,t_\ff}[\ket{i}\bra{i}]\ket{F}}\nonumber \\
&=& -\ln \frac{\sum_{i\in I}\frac{p_I}{V_I}\bra{F}C_\ff\circ {\cal N}_{t_\ii,t_\ff}[\ket{i}\bra{i}]\ket{F}}{\sum_{i\in I}p_i\bra{F}C_\ff\circ {\cal N}_{t_\ii,t_\ff}[\ket{i}\bra{i}]\ket{F}}, 
\eea
and 
\begin{align} \label{eq: average sif stoch}
\langle s_{IF}^\text{cg}\rangle &= \sum_{IF} \sum_{i\in I, f \in F} p(\gamma_{if}) s_{IF}^\text{cg} = \sum_{IF}  \sum_{i\in I}p_i\bra{F}C_F\circ {\cal N}_{t_\mathrm{i},t_\mathrm{f}}[\ket{i}\bra{i}]\ket{F} s_{IF}^\text{cg} \nn \\
&=  S\left(\sum_I \ket{I}\bra{I}\otimes\sum_{i\in I}p_i C_\ff\circ {\cal N}_{t_\ii,t_\ff}[\ket{i}\bra{i}]\bigg\| \sum_I \ket{I}\bra{I}\otimes\sum_{i\in I}\frac{p_I}{V_I}C_\ff\circ {\cal N}_{t_\ii,t_\ff}[\ket{i}\bra{i}]\right) \geq 0.
\end{align}
where in the last line, one could alternatively define the probability distributions 
\bea
    p_{IF} &=& \sum_{i\in I} p_i \bra{F} C_\ff\circ {\cal N}_{t_\ii,t_\ff}[\ket{i}\bra{i}] \ket{F}, \quad \sum_{IF} p_{IF} = 1\nonumber\\
    q_{IF} &=& \sum_{i\in I}\frac{p_I}{V_I} \bra{F} C_\ff\circ {\cal N}_{t_\ii,t_\ff}[\ket{i}\bra{i}]\ket{F}, \quad \sum_{IF} q_{IF} = 1
\eea
and obtain the inequality from the positivity of the classical relative entropy between the distributions $\{ p_{IF}\}$ and $\{ q_{IF}\}$. 

    \item Proof of \eqref{eq: average coarse grained stat}, right inequality: We first introduce the CPTP map ${\cal V}$ acting on any state $\rho(t_\mathrm{i}^+) = \sum_I \Pi_I \rho(t_\mathrm{i}) \Pi_I$
    as
\be
    {\cal V}\left[\sum_I \Pi_I \rho(t_\mathrm{i}) \Pi_I\right] := \sum_I \ket{I} \bra{I} \otimes \Pi_I \rho(t_\mathrm{i}) \Pi_I
\ee
and then use the monotonicity of relative entropy under the map ${\cal E}=\left[\Id\otimes (C_F \circ \mathcal{N}_{t_\mathrm{i},t_\mathrm{f}})\right] \circ {\cal V}$
to get 
\begin{align}
    S(\rho(t_\ii^+) \lvert \lvert \rho^\text{cg}_\ii) &\geq S({\cal E} (\rho(t_\ii^+)) \lvert \lvert {\cal E} (\rho^\text{cg}_\ii)) \nn \\
    = S\bigg((\Id\otimes (C_F \circ \mathcal{N}_{t_\mathrm{i},t_\mathrm{f}}))&\left[\sum_I \ket{I} \bra{I} \otimes \sum_{i \in I} p_i \ket{i} \bra{i}\right] \Big\lvert \Big\lvert (\Id\otimes (C_F \circ \mathcal{N}_{t_\mathrm{i},t_\mathrm{f}})) \left[\sum_I \ket{I} \bra{I} \otimes \sum_{i \in I} \frac{p_I}{V_I} \ket{i} \bra{i})\right]\bigg) \nn \\
    &=  \langle s_{IF}^\text{cg}\rangle 
\end{align}
where we used \eqref{eq: average sif stoch} to obtain the last equality.

\item Proof of \eqref{eq: first order fixed point projector}: We have $\tilde{\Lambda}_{t,t+dt}=\text{A}_{t, t+dt}\circ\Lambda_{t,t+dt}$, where 
\bea
\text{A}_{t,t+dt}[\rho] &=& P_{C_t,u}\circ C_t[\rho]\nn\\
&=& \sum_{J} \tr{\left(\rho \frac{\Pi_J(t)}{V_J}\right)} \Pi_J(t),
\eea
is the connection mapping back the final state of the interval $(t,t+dt)$ onto the algebra of time $t$. The fixed point $\sigma_t$ of $\tilde{\Lambda}_{t,t+dt}$ verifies by:
\bea \label{eq: sigma t Lambda sigma t }
  \sigma_t &=& (\text{A}_{t,t+dt}\circ\Lambda_{t,t+dt})[\sigma_t]\nonumber\\
  &=&\sum_{J} \frac{\tr{\left(\Lambda_{t,t+dt}[\sigma_t] \Pi_J(t)\right)}}{V_J} \Pi_J(t) \nn \\
    &=& \sum_{J} \frac{\tr{\left(\Lambda_{t,t+dt}[\sigma_t] \Pi_J(t+dt)\right)}}{V_J} \Pi_J(t) + O(dt^2)\nn \\
    &=& \sum_{J} \frac{\sigma_J^{\Lambda}}{V_J} \Pi_J(t) + O(dt^2),
\eea
with $\sigma_J^{\Lambda}:=\tr{\left(\Lambda_{t,t+dt}[\sigma_t] \Pi_J(t+dt)\right)}$. In the third line, we have used property \eqref{eq: projectorproductorder2}. We also have to use that, since the measurement is performed at $t+dt$, we also have
\be
\rho_{t+dt} = \sum_J p_J(t+dt)\frac{\Pi_J(t+dt)}{V_J}.
\ee
Therefore,
\bea
\tr{(\rho_{t+dt} \ln{\s_t})} &=&  \tr{\left[\sum_K p_{K}(t+dt) \frac{\Pi_K(t + dt)}{V_K} \sum_{J} \ln{\left(\frac{\s_{J}^\Lambda}{V_J}\right) \Pi_{J}(t)}\right]} + O(dt^2) \nn \\
    &=& \tr{(\rho_{t+dt} \ln{\Lambda_{t,t+dt}(\s_t)})} + O(dt^2),
\eea
where Eq.~\eqref{eq: projectorproductorder2} was used once more to obtain the desired relation.

\item Proof of \eqref{eq: relative entropy odt2}: The relative entropy
\begin{align}
    \mathcal{B}(\mathcal{H}) &\longrightarrow \mathbb{R}_+ \nn \\
    \rho &\longrightarrow S(\rho \lvert \lvert \rho^{\text{cg}}_{B,t})
\end{align}
vanishes and therefore reaches its minimum for $\rho = \rho^{ \text{cg}}_{B,t}$, and since $\rho^{\text{cg}}_{B,t+dt} = \rho^{ \text{cg}}_{B,t} + O(dt)$, it implies that 
\be
    S(\rho^{\text{cg}}_{B,t+dt} \lvert \lvert \rho^{ \text{cg}}_{B,t}) = O(dt^2)
\ee
since the first functional derivative of the relative entropy vanishes for $\rho = \rho^{\text{cg}}_{B,t}$.

    \item Proof of \eqref{eq: 3 identities mod proj ttdt}. One starts from \eqref{eq: sigma t Lambda sigma t }, which allows us to write
    \begin{align} \label{3 id ttdt proof}
        \text{Tr}\{\Pi_J(t)\ln{\sigma_t} \} &=  \text{Tr}\{\Pi_J(t)\ln{(\sum_K \frac{\s_K^\Lambda}{V_K} \Pi_K(t) + O(dt^2))} \} \nn \\
        &= \text{Tr}\{\Pi_J(t)\ln{(\sum_K \frac{\s_K^\Lambda}{V_K} \Pi_K(t))} \} + O(dt^2) \nn \\
        &= \text{Tr}\{\Pi_J(t)\ln{(\sum_K \frac{\s_K^\Lambda}{V_K} \Pi_K(t+dt))} \} + O(dt^2) \nn \\
        &= \text{Tr}\{ \Pi_J(t)\ln{\Lambda_{t,t+dt}(\sigma_t)\}} + O(dt^2)
    \end{align}
    where we used property \eqref{eq: projectorproductorder2} again to go from the second line to the third line and prove the desired relation. In order to get the last equality in \eqref{eq: 3 identities mod proj ttdt}, one just starts from the second line in the right hand side of \eqref{3 id ttdt proof} and notice that we can replace for free $\Pi_J(t)$ and $\Pi_K(t)$ by $\Pi_J(t+dt)$ and $\Pi_K(t+dt)$ respectively. Then, we get directly
    \be
        \text{Tr}\{\Pi_J(t)\ln{\sigma_t} \} = \text{Tr}\{\Pi_J(t+dt)\ln{\Lambda_{t,t+dt}(\sigma_t)} \} + O(dt^2)
    \ee
    which completes the proof.
    \item Proof of \eqref{eq: mean value of D}: One has 
    \be
        D_{t,F_B} = \p_t K_{t,F_B} = - \frac{\p_t p_{F_B}}{p_{F_B}}
    \ee
    Therefore, one has 
    \be
        \langle D_{t,F_B} \rangle = \sum_{F_B} p_{F_B} D_{t,F_B} = - \sum_{F_B} \p_t p_{F_B} = 0 
    \ee
    because of conservation of probability.

    \item Proof of \eqref{eq: secondlaw mean value genalg}: One proceed exactly as the for the proof of \eqref{eq: generic second law commalg opens} detailed above, with the appropriate modifications.
     
    \item Proof of \eqref{eq: trivial connection geenral setting}: The proof is based on the fact that the coarse-graining maps  $P_{C_{\mathcal{A}_t, u}} \circ C_{\mathcal{A}_t} $ are projectors. It directly implies that 
    \be \label{eq: doubleprojec}
        ( P_{C_{\mathcal{A}_{t+dt}}, u} \circ C_{\mathcal{A}_{t+dt}} ) \circ (P_{C_{\mathcal{A}_t}, u} \circ C_{\mathcal{A}_t} )(\Lambda_{t,t+dt}(\s_{t})) = \Lambda_{t,t+dt}(\s_{t}) + O(dt^2) 
    \ee
    and therefore 
    \be \label{eq: projector is lambda}
        ( P_{C_{\mathcal{A}_{t+dt}}, u} \circ C_{\mathcal{A}_{t+dt}} ) (\s_t) = \Lambda_{t,t+dt}(\s_{t}) + O(dt^2)
    \ee
    since $P_{C_{\mathcal{A}_t, u}} \circ C_{\mathcal{A}_t} = A_{t,t+dt}$ is the connection and $\s_t$ is the fixed point of $\tilde{\Lambda}_{t,t+dt} = A_{t,t+dt} \circ \Lambda_{t,t+dt}$.
    To prove \eqref{eq: doubleprojec}, we start from 
    \be
        \Lambda_{t,t+dt}(\s_{t}) = U_{t+dt} \big( \sum_J \s_J^{\Lambda}(t+dt) \otimes \frac{\Id_{\mathcal{H}_{J'}}}{V_J^{'}} \big) U_{t+dt}^\dag.
    \ee
    Introducing $U_{\gamma(t)} = U^{- 1}_t U_{t+dt} := e^{i G_{\gamma(t)} dt}$, we get
    \begin{align}
        \s_t &= (P_{C_{\mathcal{A}_t}, u} \circ C_{\mathcal{A}_t} ) (\Lambda_{t,t+dt}(\s_{t})) \nn \\
        &= U_t \sum_J \left\{\s_J^\Lambda(t + dt) + i dt \tr_{\mathcal{H}_{J'}}\left[G_{\g(t),J}, \s_J^{\Lambda}(t + dt) \otimes \frac{\Id_{\mathcal{H}_{J'}}}{V_J'}\right]\right\} \otimes \frac{\Id_{\mathcal{H}_{J'}}}{\bar{V}_J'} U_t^\dag + O(dt^2) \nn \\
        &:= U_t \sum_J \s_J^{\Lambda}(t) \otimes \frac{\Id_{\mathcal{H}_{J'}}}{V_J'} U_t^\dag + O(dt^2).
    \end{align}
   In the second line, we have defined $G_{\g(t),J} := \Pi_J A \Pi_J$, $\Pi_J$ being projectors on the subspace $J$, and in the thrid line  $\s_J^{\Lambda}(t) := \s_J^\Lambda(t + dt) + i dt \tr_{\mathcal{H}_{J'}}[G_{\g(t),J}, \s_J^{\Lambda}(t + dt) \otimes \frac{\Id_{\mathcal{H}_{J'}}}{V_J'}] + O(dt^2)$. By, symmetry, one has:
    \begin{align}
        (P_{C_{\mathcal{A}_{t+dt}}, u} \circ C_{\mathcal{A}_{t+dt}} ) (\s_{t}) &= U_{t+dt} \sum_J (\s_J^{\Lambda}(t) - i dt \tr_{\mathcal{H}_{J'}}[G_{\g(t),J}, \s_J^{\Lambda}(t) \otimes \frac{\Id_{\mathcal{H}_{J'}}}{V_J'}]) \otimes \frac{\Id_{\mathcal{H}_{J'}}}{\bar{V}_J'}) U_{t+dt}^\dag + O(dt^2) \nn \\
        &= \Lambda_{t,t+dt}(\s_{t}) + O(dt^2)
    \end{align}
    and therefore we proved \eqref{eq: doubleprojec}.
    Then, since $\rho_{t+dt} \in \mathcal{A}_{t+dt}$, $\rho_{t+dt}$ is stable by the application of the projector $P_{C_{\mathcal{A}_{t+dt}}, u} \circ C_{\mathcal{A}_{t+dt}}$, one has 
    \begin{align} \label{eq: analyticdecomp}
        \tr{(\rho_{t+dt} \ln{\s_t})} &= \tr{\left[\rho_{t+dt} ( P_{C_{\mathcal{A}_{t+dt}}, u} \circ C_{\mathcal{A}_{t+dt}} )(\ln{\s_t})\right]} 
    \end{align}
    Moreover, $\s_t$ being a full rank density operator, its eigenvalues lie in the interval $(0,1)$, so that one can write
    \begin{align}
        \ln{\s_t} &= \ln{(\Id - (\Id-\s_t))}  =-\sum_{n = 1}^{+ \infty} \frac{1}{n} (\Id-\s_t)^n = -\sum_{n = 1}^{+ \infty} \frac{1}{n} ((\Id -\Lambda_{t,t+dt}(\s_{t})) + (\Lambda_{t,t+dt}(\s_{t}) - \s_t))^n \nn \\
        &= -\sum_{n = 1}^{+ \infty} \frac{1}{n} \{ (\Id -\Lambda_{t,t+dt}(\s_{t}))^{n} +\sum_{k= 0}^{n} [(\Lambda_{t,t+dt}(\s_{t}) - \s_t) (\Id -\Lambda_{t,t+dt}(\s_{t}))^{n-1} \nn \\
        &+ (\Id -\Lambda_{t,t+dt}(\s_{t})) (\Lambda_{t,t+dt}(\s_{t}) - \s_t) (\Id -\Lambda_{t,t+dt}(\s_{t}))^{n-2} + \cdots \nn \\
        &+ (\Id -\Lambda_{t,t+dt}(\s_{t}))^{n-1} (\Lambda_{t,t+dt}(\s_{t}) - \s_t) ] \} + O(dt^2).
    \end{align}
    Then, one has
    \begin{align} \label{eq: gen alg connect wwwtshow}
        \tr{(\rho_{t+dt} \ln{\s_t})} =  \tr{(\rho_{t+dt} \ln{\Lambda_{t,t+dt}(\s_t)})} + \sum_{n = 1}^{+ \infty} \frac{1}{n} \tr{(\mathcal{X}_{t+dt} (\Lambda_{t,t+dt}(\s_{t}) - \s_t)  )} + O(dt^2)
    \end{align}
    where $\mathcal{X}_{t+dt}$ represents an operator in $\mathcal{A}_{t+dt}$ since obtained as the sum and the product of $(\rho_{t+dt}, \Id - \Lambda_{t,t+dt}(\s_t)) \in \mathcal{A}_{t,t+dt}$. Therefore, since $\mathcal{X}_{t+dt}$ is stable through $P_{C_{\mathcal{A}_{t+dt}}, u} \circ C_{\mathcal{A}_{t+dt}}$, one has 
    \begin{align}
        \tr{(\mathcal{X}_{t+dt} (\Lambda_{t,t+dt}(\s_{t}) - \s_t)  )} &= \tr{(\mathcal{X}_{t+dt} (P_{C_{\mathcal{A}_{t+dt}},u} \circ C_{\mathcal{A}_{t+dt}})(\Lambda_{t,t+dt}(\s_{t}) - \s_t)  )}  \nn \\
        &= O(dt^2)
    \end{align}
    where the last line follows immediately for \eqref{eq: projector is lambda}. Then, from \eqref{eq: gen alg connect wwwtshow}, we directly get 
    \be
        \tr{(\rho_{t+dt} \ln{\s_t})} =  \tr{(\rho_{t+dt} \ln{\Lambda_{t,t+dt}(\s_t)})}  + O(dt^2)
    \ee
    which is the desired result.
    
    \item Proof of \eqref{eq: gen bipartite second law heat}: One has 
    \begin{align}
    d S_{\mathcal{A}_B}(\rho) &= - \tr{\rho_{B,t+dt} \ln{\rho_{B,t+dt}}} + \tr{\rho_{B,t} \ln{\rho_{B,t}}} \nn \\
    &= - \tr{((\rho_{B,t+dt} - \rho_{B,t}) \ln{\rho_{B,t}})} - S(\rho_{B,t+dt} \lvert \lvert \rho_{B,t}) \nn \\
    &= \beta_t \tr{((\rho_{B,t+dt} - \rho_{B,t}) (H_{B,t} + \sum_i \l_i(t) O^i_{B,t} ))} + O(dt^2)
    \end{align}
    since $S(\rho_{B,t+dt} \lvert \lvert \rho_{B,t}) = O(dt^2)$ which is due to the fact that $S(\rho_B \lvert \lvert \rho_{B,t})$ is minimal and vanishes for $\rho_B = \rho_{B,t}$ so that the term of order $O(dt)$ vanishes in the Taylor expansion (see the proof of \eqref{eq: relative entropy odt2} discussed above).

    \item Proof of \eqref{eq: other gen ave sec law}: To go from the first line to the second line of \eqref{eq: other gen ave sec law}, it is enough to prove that 
    \be \label{eq: equality mutual information}
        I_{\mathcal{A}_\text{A} \otimes \mathcal{A}_\text{B}} = I_{\mathcal{A}_\text{A}^c \otimes \mathcal{A}_\text{B}^c} + \sum_{J_A J_B} p_{J_A J_B} I_{J_A J_B}
    \ee
    where $I_{\mathcal{A}_\text{A}^c \otimes \mathcal{A}_\text{B}^c} = - \sum_{J_A J_B} p_{J_A J_B} \ln{\frac{p_{J_A J_B}}{p_{J_A} p_{J_B}}}$ is the classical mutual information attached to the center $\mathcal{A}^c_{\text{A}} \otimes \mathcal{A}^c_{\text{B}}$ of the algebra $\mathcal{A}_{\text{A}} \otimes \mathcal{A}_{\text{B}}$ and $I_{J_A J_B} = S(\tilde{\rho}_{J_A}) + S(\tilde{\rho}_{J_B}) - S(\tilde{\rho}_{J_{AB}})$.
    Since the state $\rho_{\text{A} \cup \text{B}}$ is assumed to be a coarse-grained state, it can be written as 
    \be
        \rho_{\text{A} \cup \text{B}} = \sum_{J_A J_B} p_{J_A J_B} (\tilde{\rho}_{J_{AB}} \otimes \frac{\Id_{J_A' J_B'}}{V_{J_A'} V_{J_B'}} )
    \ee
    so that 
    \be
        S(\rho_{\text{A} \cup \text{B}}) = - \sum_{J_A J_B} p_{J_A J_B} \ln{\frac{p_{J_A J_B}}{V_{J_A'} V_{J_B'}}} + \sum_{J_A J_B} p_{J_A J_B} S(\tilde{\rho}_{J_{AB}})
    \ee
    while orthogonality of the sectors $J_A$ and $J_B$ implies
    \begin{align}
        S_{\mathcal{A}_{\text{A}}}(\rho_{\text{A}}) &= - \sum_{J_A} p_{J_A} \ln{\frac{p_{J_A}}{V_{J_{A}'}}} + \sum_{J_A} p_{J_A} S(\tilde{\rho}_{J_A}) = - \sum_{J_A} p_{J_A} \ln{\frac{p_{J_A}}{V_{J_A'}}} + \sum_{J_A J_B} p_{J_A J_B} S(\tilde{\rho}_{J_A}) \nn \\
        S_{\mathcal{A}_{\text{A}}}(\rho_{\text{B}}) &= - \sum_{J_B} p_{J_B} \ln{\frac{p_{J_B}}{V_{J_B'}}} + \sum_{J_B} p_{J_B} S(\tilde{\rho}_{J_B}) = - \sum_{J_B} p_{J_B} \ln{\frac{p_{J_B}}{V_{J_B'}}} + \sum_{J_A J_B} p_{J_A J_B} S(\tilde{\rho}_{J_B})
    \end{align}
    from which one directly reads \eqref{eq: equality mutual information}
\end{itemize}
\label{app: proofs}

\section{A positive-temperature parametrization}
\label{app: negativetemperature}
Effective negative temperatures may naturally arise in systems with bounded energy spectra and population inversion. More precisely, such systems can reach states of the form $\rho \propto e^{-\beta_\text{eff}H}$ with $\beta_\text{eff}<0$. Although interpreting $\beta_\text{eff}$ as a negative temperature in this case is interesting in some cases, in the context of deriving a systematic expression for the 2nd law, it is convenient to treat them in a way acknowledging their nonequilibrium nature and the extra thermodynamic resources they contain with respect to a thermal state with positive temperature. 
This approach has the advantage of preserving the usual interpretation of heat while separately accounting for the work potential associated with population inversions.
In this spirit, we introduce a positive nonequilibrium temperature 
that coincides with the conventional thermodynamical
temperature whenever the latter is positive. Otherwise, it characterizes the thermal state obtained when extracting from the system all the work associated with non-thermal populations. To achieve
this, we first isolate the component of the
coarse-grained state which is generated by the coarse-grained Hamiltonian. When a single-parameter parametrization of the nonequilibrium state would yield a negative temperature, we instead introduce a two-parameter parametrization of the same state that allows for a positive temperature, with the second parameter playing the role of a chemical potential.

We start from the coarse-grained state
$\sigma_t=e^{-K_t}$ and decompose its modular Hamiltonian
over the orthogonal operator basis
$\{\Id,H_t,O_t^1,\ldots,O_t^N\}$ of the accessible algebra
$\mathcal{A}(t)$, as introduced in
Eq.~\eqref{eq: Kt in op basis}:
\be \label{eq: local Kt appendix}
    K_t
    =\beta(t)H_t
    +\sum_{k\geq 1}\tilde{\lambda}_k(t)O_t^k
    +\ln Z_t\,\Id,
\ee
where
\be
    Z_t
    =\tr{(e^{
        -\beta(t)H_t
        -\sum_{k\geq 1}\tilde{\lambda}_k(t)O_t^k})}
\ee
ensures the normalization of $\sigma_t$.

When $\beta(t)\geq 0$, the coefficient $\beta(t)$ is the conventional inverse
thermodynamic temperature associated with this generalized
Gibbs ensemble. Indeed, the entropy is the mean value of the modular Hamiltonian $K_t$ in the state $\s_t$ and therefore satisfies
\be \label{eq: thermo entropy}
    S = \langle K_t \rangle
    =\beta(t) E
    +\sum_{k\geq 1}\tilde{\lambda}_k(t)
        \langle O_t^k\rangle
    +\ln Z_t,
\ee
with $E=\langle H_t\rangle$. The conventional thermodynamic inverse
temperature can then be computed from \eqref{eq: thermo entropy} and is given by  
\be \label{def: thermotemperature}
    \frac{1}{T}
    :=\left.
    \frac{\partial S}{\partial E}
    \right|_{\{\langle O_t^k\rangle\},t}
    =\beta(t).
\ee

We now allow $\beta(t)$ to have any sign and isolate the thermal component of $\sigma_t$
by setting all additional parameters
$\tilde{\lambda}_k$ to zero and renormalizing the
resulting state. This yields
\be
    \omega_\beta(t)
    =\frac{e^{-\beta(t)H_t}}
    {\tr{e^{-\beta(t)H_t}}}.
\ee
Equivalently, this construction amounts to projecting
the modular Hamiltonian onto the subspace spanned
by the identity and $H_t$, while retaining the
 original coefficient $\beta(t)$. 

We now build the positive nonequilibrium temperature $\beta_0(t)$
following the construction of~\cite{Elouard23}. Namely, we define $\beta_0(t)$ as the positive  solution of the equation 
\be \label{eq: positive beta0}
    S^{\mathrm{ob}}(\omega_{\beta_0}(t))
    =
    S^{\mathrm{ob}}(\omega_\beta(t)),
\ee
where 
\be
    \omega_{\beta_0}(t)
    =\frac{e^{-\beta_0(t)H_t}}
    {\tr{e^{-\beta_0(t)H_t}}}.
\ee
Since the observational entropy is the von Neumann
entropy of the coarse-grained state, and the
thermal states considered here are constructed
directly from the coarse-grained Hamiltonian,
Eq.~\eqref{eq: positive beta0} is an equality
of von Neumann entropies. Clearly, whenever $\beta(t)\geq 0$, the solution is simply given by $\beta_0(t)= \beta(t)$. In general, for a fixed Hamiltonian $H_t$, the entropy of
a thermal state satisfies
\be
    \frac{\partial S_{\mathrm{v.N.}}(\omega_{\beta_0})}{\partial\beta_0}
    =-\beta_0 \langle (H_t - \langle H_t \rangle)^2 \rangle \leq 0
\ee
It is therefore monotonically decreasing with
$\beta_0$ as the latter is defined to be positive.
For a nontrivial finite-dimensional Hamiltonian
with a nondegenerate ground state, this entropy
ranges from $\ln{D}$ at infinite temperature
to $0$ at zero temperature, where $D$ denotes
the dimension of the Hilbert space \footnote{The procedure can be straightforwardly extended to the case of non degenerate ground states, see \cite{Elouard:2023kiu}.} Thus, for any prescribed entropy between these
two values, there exists a unique positive-temperature
Gibbs state with that entropy. 
We have therefore defined a positive reference
inverse temperature $\beta_0(t)$ that coincides
with $\beta(t)$ whenever the latter is nonnegative,
while for negative $\beta(t)$ it is determined
by the entropy of the isolated thermal component.

\vspace{0.3 cm} 

The motivation behind this entropy-based definition of temperature is its ability to yield correction to work and heat flow which still comply with the second law. To see this, we denote
$E_\beta=\tr{(\omega_\beta(t)H_t)}$
and
$E_{\beta_0}=\tr{(\omega_{\beta_0}(t)H_t)}$,
and write the relative entropy between $\omega_\beta(t)$ and $\omega_{\beta_0}(t)$:
\be S(\omega_\beta(t)\Vert\omega_{\beta_0}(t))
    =\beta_0(t)(E_\beta-E_{\beta_0})
    \geq 0.
\ee
Consequently,
\be
    E_\beta-E_{\beta_0}\geq 0.
\ee
When $\beta(t)\leq 0$, the nonequilibrium state $\omega_{\beta(t)}$ then contains an extra amount of energy with respect to $E_{\beta_0}$ which could be totally extracted as work under a reversible isentropic transformation. The actual variation of $E_\beta-E_{\beta_0}$ happening during the dynamics corresponds to the contribution to the work associated with the system having non-thermal populations. \\

We formalize this point by rewriting Eq.~\eqref{eq: local Kt appendix}
as
\begin{align}
    K_t
    &=\beta_0(t)H_t
    +[\beta(t)-\beta_0(t)]H_t
    \nonumber\\
    &\quad
    +\sum_{k\geq 1}\tilde{\lambda}_k(t)O_t^k
    +\ln Z_t\,\Id.
\end{align}
For $\beta_0(t)>0$, we introduce the generalized
potential
\be
    \mu_0(t)
    =\frac{\beta_0(t)-\beta(t)}{\beta_0(t)}
\ee
and define
$\lambda_k(t)=\tilde{\lambda}_k(t)/\beta_0(t)$.
The modular Hamiltonian becomes
\begin{align}
    K_t
    &=\beta_0(t)\left[
        H_t-\mu_0(t)H_t
        +\sum_{k\geq 1}\lambda_k(t)O_t^k
    \right]
    +\ln Z_t\,\Id
    \nonumber\\
    &=\beta_0(t)\left[
    H_t+\sum_{k\geq 0}\lambda_k(t)O_t^k
    \right]
    +\ln Z_t\,\Id,
    \label{eq: Kt in op basis 2}
\end{align}
where $O_t^0=H_t$ and
$\lambda_0(t)=-\mu_0(t)$. For $\beta(t)\geq 0$, we have
$\beta_0(t)=\beta(t)$ and $\mu_0(t)=0$,
so that the conventional thermal contribution
is recovered. For $\beta(t)<0$, the generalized
potential $\mu_0(t)$ is strictly positive and
encodes the additional contribution associated
with population inversion. In this way, we recover the decomposition
\eqref{eq: KBt in op basis}, now expressed
in terms of a positive reference temperature.
The multiplicative parametrization above applies
when $\beta_0(t)>0$. At $\beta_0(t)=0$, the
preceding additive decomposition of $K_t$
remains well defined, without requiring a
division by $\beta_0(t)$.
Importantly, this reparametrization modifies
neither the modular Hamiltonian nor the
coarse-grained state. It only changes the
identification of the thermal contribution
and the additional resources entering our
definitions of heat and work, see Eqs.~\eqref{eq: heat def sec 3}-\eqref{eq: work def sec 3}, while ensuring that those quantites comply with the second law under the form of Clausius inequality Eq.~\eqref{eq: Clausius sec 3}.

\begin{remark}
    Our prescription is motivated by the
    distinction between the thermal contribution
    and the additional thermodynamic resources
    encoded in the modular Hamiltonian.
    Nevertheless, we do not claim that this
    choice of reference temperature is unique
    or necessarily preferable in every
    physical situation.
    An alternative is the \emph{isentropic
    temperature} introduced in Ref.~\cite{Elouard23}, defined as the
    temperature of the positive-temperature
    Gibbs state having the same entropy
    as the original coarse-grained state:
    \be
    S_{\mathrm{v.N}}(\sigma_t) = S_{\mathrm{v.N}}(\omega_{\beta_{\mathrm{is}}}(t)).
    \ee
    This prescription differs from ours
    because it applies the entropy-matching
    condition directly to the full
    coarse-grained state, rather than
    to its isolated thermal component.
    Consequently, the resulting temperature $T_{\mathrm{is}} = \frac{1}{\beta_{\mathrm{is}}}$
    generally differs from the conventional
    thermodynamic temperature defined in
    Eq.~\eqref{def: thermotemperature}, even
    when $\beta>0$.
    By contrast, our two-step construction
    preserves the conventional thermodynamic
    temperature whenever the latter is positive and applies the isentropic
    prescription only to the isolated thermal
    component when its temperature is negative.
    More generally, the techniques developed
    in this paper do not rely on a particular
   prescription for a nonequilibrium temperature. For any
    physically motivated inverse temperature
    $\beta_0(t)$, we can decompose the modular
    Hamiltonian as
    \be
        K_t
        =\beta_0(t)H_t
        +[K_t-\beta_0(t)H_t].
    \ee
    The same procedure as in the main text
    can then be used to identify heat and
    work, with the residual operator
    $K_t-\beta_0(t)H_t$ encoding the
    additional thermodynamic resources,
    up to terms proportional to the identity.
    Thus, while the identification of heat
    and work depends on the chosen reference
    temperature, the modular Hamiltonian,
    the underlying dynamics, and the
    thermodynamic identities established
    in the main text remain unchanged.
\end{remark}

\section{Internal Maxwell demons}
\label{app: Maxwell demons}

In this appendix, we study a toy model for a $N\gg 1$ particle gas in a box designed to emphasize how the lower bound on the right hand side of \eqref{eq: lower boundpivi} can be reached.
In this model, the Hilbert space $\mathcal{H}_i$ attached to each particle is two dimensional, with state $\ket{L}_i$ ($\ket{R}_i$) corresponding to the particle $i$ being on the left side (right side). Therefore, the particles can freely move from one compartment of the box to another, which changes the microstate of the system. Therefore, the microstates
\be \label{eq: ketsigma}
    \ket{\sigma} = \bigotimes_{i = 1}^N \ket{\sigma_i}, \qquad \sigma_i \in\{ L_i, R_i\}^N
\ee
generate the $2^N$-dimensional Hilbert space $\mathcal{H} = \otimes_i \mathcal{H}_i$ of the gas. Moreover, we assume that the evolution of the gas is given by the repeated action of the unitary operator
\be
    \ket{\pi(\sigma)} = U \ket{\sigma} = \bigotimes_{i = 1}^N \ket{\pi(\sigma_i)}, \qquad \pi(\sigma_i) \in \{ L_i, R_i \}^N
\ee
where the map $\pi$ cyclically permutes the $2^N$ elements of the basis \eqref{eq: ketsigma}, so that each particle either remains in its compartment or moves to the other one.  
Therefore, after the process, the particles adopt a new configuration within the box, corresponding to a new microstate. 
Now, assume that we are not able to measure the microstate of the system and that the only information we can have access to is the presence or the absence of at least one particle in the right compartment. Therefore, the space of macrostates is generated by the two macrostates
\begin{align} \label{eq: macrostates examples}
    \ket{\text{Yes}} &= \text{``At least one particle is in the right compartment " } \nn \\
    \ket{\text{No}} &= \text{``No particle is in the right compartment " }
\end{align}
and $C$ is the coarse-graining map attached to the macrostates \ref{eq: macrostates examples}.
Initially, one assumes  
that the initial macrostate $\bar{\rho}(t_\ii) = C(\rho(t_\ii))$ is 
\be \label{macrosl}
    \bar{\rho}(t_\ii) = (1 - \frac{1}{2^N}) \ket{\text{Yes}} \bra{\text{Yes}} + \frac{1}{2^N} \ket{\text{No}} \bra{\text{No}}
\ee
and $V_{\text{Yes}} = 2^N - 1$ while $V_{\text{No}} = 1$. Notice that the configuration \eqref{macrosl} may a priori seem to be in internal equilibrium, but 
the initial microstate $\rho(t_\ii)$ does not necessarily correspond to the coarse-grained state compatible with the constraints, which is
\be
    \rho^{\text{cg}}(t_\ii) = P_{C,u}( \bar{\rho}(t_\ii)) = \frac{\Id_{\mathcal{H}}}{2^N}.
\ee
 The initial observational entropy is, therefore, given by
\be
    S^{\text{ob}}(t_\ii) = N \ln{2} 
\ee
while we also have that 
\be
    \sum_I p_I \ln{V_I} \underset{N \rightarrow + \infty}{\sim} N \ln{2} 
\ee
Moreover, only one microstate is associated with the macrostate $\ket{\text{No}}$, that is $\ket{\s_L} = \otimes_{i=1}^N \ket{L_i}$. We define $\ket{\s_L^{-1}} = U^{-1} \ket{\s_L}$ and we notice that $\bra{\s_L^{-1}} \ket{\s_L} = 0$, since the unitary $U$ cyclically permutes the basis elements. Then, the initial microstate 
\be \label{badrho+}
    \rho(t_\ii) = (1 - \frac{1}{2^N}) \ket{\s_L^{-1}} \bra{\s_L^{-1}} + \frac{1}{2^N} \ket{\s_L} \bra{\s_L}
\ee
admits \eqref{macrosl} as a macrostate. Thus, the final microstate obtained through microscopic dynamics is $\rho(t_\ff) = U \rho(t_\ii) U^{-1}$. Then, the corresponding macrostate is
\be
    \bar{\rho}(t_\ff) = C \circ (U \rho(t_\ii) U^{-1})  =  \frac{1}{2^N} \ket{\text{Yes}} \bra{\text{Yes}} + (1 - \frac{1}{2^N}) \ket{\text{No}} \bra{\text{No}},
\ee
which directly implies that
\be
    S^{\text{ob}}(t_\ff) \underset{N \rightarrow + \infty}{\sim} \frac{N \ln{2}}{2^N}  \underset{N \rightarrow + \infty}{\longrightarrow} 0
\ee
so that, when the initial microstate is given by \eqref{badrho+}, the observational entropy of the isolated system decreases under unitary evolution
\be
    \Delta S^{\text{ob}} = S^{\text{ob}}(t_\ff) - S^{\text{ob}}(t_\ii) \underset{N \rightarrow + \infty}{\sim} -N \ln{2} \underset{N \rightarrow + \infty}{\sim}  -\sum_I p_I \ln{V_I},
\ee
leading to an apparent violation of the second law of thermodynamics. This violation, however, occurs only for an extremely atypical microstate belonging to a set of negligible measure in Hilbert space. In this highly fine-tuned configuration, all particles are found in the left compartment of the box at the final time.

This gedankenexperiment can be interpreted as a variation of the Maxwell demon scenario. Indeed, if the observer has prior knowledge that the system is initially prepared in the microstate \eqref{badrho+}, then this information can be exploited to extract work while the system evolves toward the configuration in which all particles occupy the left compartment (e.g. by moving to the right a piston initially placed at the interface between the compartments). In this sense, the apparent violation of the second law stems from the availability of highly specific information about the initial state rather than from any genuine breakdown of thermodynamic principles.

\section{Appendix: Spontaneous emission}
\label{app: spontaneous emission}

In this appendix, we consider a two-level atom emitting in an electromagnetic field quantized in a 1D box with perfectly reflective walls. The atom is initially in the excited state $\ket{e}$ and the electromagnetic field is in the ground (vacuum) state $\ket{0}$. The interaction is switched on at $t=0$.
It is clear that the joint evolution of the atom and in the electromagnetic field is unitary. Therefore, if one does not coarse-grain, the total (von Neumann) entropy will remain constant, i.e. there will be no entropy production. In addition, if the size of the box $L$ is sufficiently large, the time needed for the photon to travel that distance will be much longer than the typical time $\tau \sim \frac{1}{\gamma}$ for the atom to decay. In this case, for times
\be \label{timeint}
\frac{L}{c} \gg t \gg \frac{1}{\gamma},
\ee
we can assume that the atom and the field are in a factorized state $\ket{g}\otimes\ket{\psi}$ where the atom has relaxed to its ground state $\ket{g}$ and $\ket{\psi}$ is 1-photon state of the field. This pure, factorized joint state, as well as the atom and field reduced states, have vanishing von Neumann entropy, just as in the initial situation. 

Now, imagine that the observer places a detector in the box to measure the photon energy at time $t$. Denoting the time-resolution of the detector (i.e. the integration time of the measurement) by $\Delta t \ll t \ll \frac{L}{c}$, the energy resolution of the apparatus will be given by 
\be \label{eq: Heisin}
    \Delta E \sim \frac{1}{\Delta t}
\ee
from the time-energy uncertainty relation. In contrast, the energy resolution of the space of microstates of the electromagnetic field, i.e. the energy spacing between two modes of the box is $\delta E\sim c/L$, with $c$ the light velocity. Since $\Delta t \ll \frac{L}{c}$, one has 
\be
    \Delta E \gg \delta E,
\ee
i.e., the finite time resolution of the detector implies a coarse-graining of the energy measurement.

\vspace{0.3 cm}

The most interesting case corresponds to $\Delta E \ll \gamma$, i.e.
\be
    \frac{L}{c} \gg t \gg \Delta t \gg \frac{1}{\gamma} .
\ee
This is the regime one expects for a good apparatus, able to measure precisely the energy of the electromagnetic field. In order to do this, $\Delta t$ must be large enough \eqref{eq: Heisin}. However, one has to keep $\Delta t \ll t$, otherwise the time uncertainty is longer than the detection time $t$ and the latter becomes meaningless. In addition, if $t \gg \frac{L}{c}$ the photon has reached the end of the box and has already bounced back, so that the dynamics cannot be considered Markovian anymore.
The measurement performed by the detector is associated with the projectors $\{ \Pi_{E, E + \Delta E} \}$ each of rank $V_{E, E+ \Delta E} = \frac{L}{c \Delta t} \gg 1$. Assuming that the electromagnetic field was in the vacuum $\ket{0}$ before the interaction with the atom is switched on (at $t=0$), the measurement outcome distribution is
\be \label{eq: energy dist}
    p_{E, E + \Delta E} = \frac{1}{\pi} \frac{\frac{\g}{2} \frac{1}{\Delta t}}{(E - E_0)^2 + \frac{\g^2}{4}} \Delta E,
\ee
i.e., the probability distribution for the electromagnetic field of having energy in the interval $(E, E + \Delta E)$ once the atom emitted a photon.

Initially, the field and the atom were not coupled. As a consequence, 
the observer could use the time interval $(- \infty, 0)$ to prepare the electromagnetic field to ground state $\ket{0}$ with arbitrary high precision, which is equivalent to saying that the integration time for the preparation of the detector is
\be
    \Delta t_{\text{prep}} = +\infty \gg \frac{L}{c}.
\ee
Therefore, it is natural to assume that all the energy eigenvalues $E_i$ of the box can be resolved in this initial measurement, which is captured by projectors $\{\ket{E_i}\bra{E_i}\}$ of rank $V_{E_i} = 1$. The initial outcome statistics is simply $p_{E_i}=\delta_{i,0}$, with $E_0$ the ground state energy.

We can now compute the initial and final observational entropies corresponding to those measurements.
Initially, as the projectors are rank-1, it matches the initial von Neumann entropy
\be
    S^{\text{ob}}_E(t = 0) = S_{\text{v.N}}(t = 0) = 0.
\ee 
At time $t\gg \gamma$, conversely, the observational entropy $S^{\text{ob}}_E(t \gg \frac{1}{\g})$ attached to the energy of the electromagnetic field measured during the time interval $\Delta t$ does not vanish, even though the final field state $\ket{\psi}$ is pure and its von Neumann does vanish. Indeed, we have:
\be \label{deltasobstep1}
    S^{\text{ob}}_E(t \gg \frac{1}{\g}) = \Delta S^{\text{ob}}_E = -\int_{0}^{+ \infty}  \frac{1}{\pi} \frac{\frac{\gamma}{2}}{(E - E_0)^2 + \frac{\gamma^2}{4}} \ln{  \frac{1}{\pi} \frac{\frac{\gamma}{2} \frac{1}{\Delta t}}{(E - E_0)^2 + \frac{\gamma^2}{4}}} dE  + \int_{0}^{+ \infty} \frac{1}{\pi}\frac{\frac{\gamma}{2}}{(E - E_0)^2 + \frac{\gamma^2}{4}} \ln{\frac{L}{c \Delta t}} dE,  
\ee
where we have to understand $\frac{1}{\Delta t} \sim \Delta E = d E$. Notice that one can also write \eqref{deltasobstep1} as
\begin{align} \label{deltasobstep12}
    \Delta S^{\text{ob}}_E &= \int_{0}^{+ \infty} \frac{1}{\pi}\frac{\frac{\gamma}{2}}{(E - E_0)^2 + \frac{\gamma^2}{4}} \ln{ \big( \frac{2 \pi L}{c \gamma} [(E - E_0)^2 + \frac{\gamma^2}{4}] \big) } dE  \nn \\
    &\sim \ln{\frac{L \gamma}{c}}
\end{align}
since the Lorentzian is sharply centered around $E = E_0$ and $\frac{L \g}{c}  \gg 1$. Therefore, the variation of observational entropy is mostly contained in the Boltzmann weight.

\vspace{0.3 cm}

It is enlightening to compare with the case of a detector with instead a short integration time
\be
    \frac{L}{c} \gg t \gg \frac{1}{\gamma} \gg \Delta t
\ee
such that it becomes impossible for the apparatus to distinguish two different frequencies of the photon. In this case, we simply get
\be
    \Delta S^{\text{ob}}_E = \ln{\frac{L}{c \Delta t}}.
\ee

In both cases, we see that coarse-graining leads to a nonzero entropy production, revealing the irreversibility of the spontaneous emission event. However, in the first case the increase of the observational entropy has mainly an informational origin, while in the second case it comes entirely from the coarse-graining. It is very interesting to notice the analogy with the Joule Gay-Lussac expansion. Initially, the atom is excited so that that the electromagnetic field is constrained to be in the ground state, exactly as the atom is constrained to be in one small compartment in the box. Then, the interaction with the electromagnetic field is turned on (the constraint on the particle in the box is relaxed) and the electromagnetic field explores a larger phase space, exactly as the particle which can now move freely in the box once the wall has been removed. Finally, in both cases, the entropy production scales like the logarithm of the volume of the box (replaced here by the length of the cavity), which is typically the entropy production term involved in a gas expansion. Therefore, the particle interpretation of the electromagnetic field in terms of photons can simply be recovered thermodynamic quantities attached to the electromagnetic field.

\section{Example: Thermodynamics of the qubit}
\label{app: qubit}

One can apply our framework to the simplest non trivial example, a two-level quantum system (qubit). In this case, the set of operators $\mathcal{B}(\mathcal{H}) = M_2 (\mathbb{C})$ has, from the representation theorem, only three von Neumann subalgebras (up to unitary equivalence): $ M_2 (\mathbb{C})$, $\mathbb{C} \oplus \mathbb{C}$ and $\mathbb{C} \Id_2$. The algebra $\mathbb{C} \Id_2$ corresponds to scalar operators, proportional to the identity and is then trivial. Moreover, if we take $\mathcal{A} = M_2 (\mathbb{C})$, we recover the full description of the qubit state, and therefore a quantum thermodynamic description based on the full system density operator.  We therefore focus on subalgebras equivalent to $\mathbb{C} \oplus \mathbb{C}$,
which are each obtained by projecting the qubit into a given measurement basis. We then apply the framework of Section \ref{sec: heat and work} where two measurements $C_\ii$ and $C_\ff$ are performed on the qubit at times $t$ and $t + dt$. Of course, for a qubit, the projectors associated with the measurement are necessarily of rank $1$, such that there is no coarse-graining strictly speaking ($V_J = 1$). As a consequence, the condition of internal equilibrium is trivially satisfied. This example is however enlightening to understand the notions of work and heat emerging from time-dependent algebras.

\subsection{No free dynamics}

We first focus on the case where the qubit has no non trivial free dynamics between $t$ and $t+dt$. Of course, since the qubit Hamiltonian is non-vanishing, the qubit undergoes an internal dynamics even in the absence of projective measurements. However, the qubit is measured projectively at time $t$ along the axis $\vec{u}(t)$ and at time $t+dt$ along the neighboring axis $\vec{u}(t+dt)=\vec{u}(t)+O(dt)$. Immediately after the first measurement, the state is diagonal in the corresponding measurement basis. As a consequence, the unitary evolution generated by the qubit Hamiltonian does not modify the populations at first order in $dt$, and is therefore invisible in the measurement statistics at this order. Consequently, the map ${\cal N}_{t,t+dt}$ may be taken either as the free unitary channel $\mathcal{U}_{t,t+dt}(\cdot)=e^{-iHdt}(\cdot)e^{iHdt}$ or, to first order in $dt$, as the identity channel, without affecting any of the results of this section. Alternatively, one may work in the interaction picture, provided that the measurement operators are transformed consistently, which leads to the same physical energy transfers. 
The variation of the qubit state is therefore only due to the quantum backaction of the measurements.

Imagine that at $t$ the qubit is diagonal in the basis  $\{ \ket{+}_{\vec{u}(t)},  \ket{-}_{\vec{u}(t)}\}$ of the eigenstates of $\s_{\vec{u}(t)}$, where $\vec{u}(t) = (\sin{\theta(t)} \cos{\phi(t)}, \sin{\theta(t)} \sin{\phi(t)}, \cos{\theta(t)})$ is a unit vector on the Bloch sphere. In terms of the eigenstates $\ket{\pm} \equiv \ket{\pm}_{(0,0,1)}$ of the Hamiltonian $H = \frac{\om_0}{2}\s_z$, we have
\begin{align}
    \ket{+}_{\vec{u}(t)} &= \cos{\frac{\theta}{2}} \ket{+} + e^{i \phi} \sin{\frac{\theta}{2}} \ket{-} \nn \\
    \ket{-}_{\vec{u}(t)} &= -e^{- i \phi} \sin{\frac{\theta}{2}} \ket{+} + \cos{\frac{\theta}{2}} \ket{-}.
\end{align}
Assuming that the initial measurement is performed on the basis $\{ \ket{+}_{\vec{u}(t)},  \ket{-}_{\vec{u}(t)}\}$, it follows that the representative of the initial state in the initial subalgebra is nothing but the initial state itself, i.e.
\be 
    \rho_t = p_+(t) \ket{+}_{\vec{u}(t)} \bra{+}_{\vec{u}(t)} + p_-(t) \ket{-}_{\vec{u}(t)} \bra{-}_{\vec{u}(t)}.
\ee

Now, the final measurement is performed on the new basis $\{ \ket{+}_{\vec{u}(t+dt)},  \ket{-}_{\vec{u}(t+dt)}\}$ at time $t+dt$. This forces the qubit to reach a diagonal state in this new basis, and can therefore be seen as a \emph{variation of the external constraints the observer is imposing on the system}, in other words, the direct analogue of moving a piston constraining the volume of a gas. 
To make this variation of the constraints quasistatic, we further consider an infinitesimal basis change, given by unitary $U_{t,t+dt} = \Id + O(dt)$ so that  
\be
    \ket{\pm}_{\vec{u}(t+dt)} \bra{\pm}_{\vec{u}(t+dt)} = U_{t,t+dt} \ket{\pm}_{\vec{u}(t)} \bra{\pm}_{\vec{u}(t)} U_{t,t+dt}^\dag.
\ee
After the second measurement, the qubit state is then:
\be
    \rho_{t+dt} = p_+ (t) \ket{+}_{\vec{u}(t+dt)} \bra{+}_{\vec{u}(t+dt)} + p_-(t) \ket{-}_{\vec{u}(t+dt)} \bra{-}_{\vec{u}(t+dt)}  + O(dt^2)
\ee
because of property \eqref{eq: projectorproductorder2}, and once again, the representative of the state in the algebra is equal to the state itself. In the limit $dt\to 0$, the evolution of the system is therefore unitary. One finds that the state of the qubit can be perfectly controlled by a series of measurements in a slowly varying bases, a process sometimes called ``Zeno rotation'' \cite{Itano1990Quantum, Facchi2008Quantum}.

We can now evaluate the work increment associated with this variation of constraints in two ways. First, we could argue that the energy variation of the qubit can only come from the energy provided by the measurement. As the latter induces a transformation that is entropy-preserving (as unitary) and reversible (it is reversed by simply measuring on basis $\{ \ket{+}_{\vec{u}(t)},  \ket{-}_{\vec{u}(t)}\}$), the latter must correspond entirely to work:
\be
\delta W_\text{ext} = \text{Tr}\{H(\rho_{t+dt}-\rho_t)\}.
\ee

However this approach is based on the knowledge of the full qubit algebra (to which $H$ belongs) rather than on the accessible observables only, which we want to avoid in order to enable the coarse-grained analysis. Let us now use the machinery introduced in the present article. We compute the representative of the Hamiltonian $H$ in the accessible algebra (Notice that all $V_J=1$ here):
\begin{align}
    H_{\vec{u}(t)} &= \bra{+}_{\vec{u}(t)} H \ket{+}_{\vec{u}(t)} \ket{+}_{\vec{u}(t)} \bra{+}_{\vec{u}(t)}  + \bra{-}_{\vec{u}(t)} H \ket{-}_{\vec{u}(t)} \ket{-}_{\vec{u}(t)}  \bra{-}_{\vec{u}(t)} \nn \\
    &=\frac{\om_0}{2} [(\cos^2{\frac{\theta(t)}{2}} - \sin^2{\frac{\theta(t)}{2}}) \ket{+}_{\vec{u}(t)} \bra{+}_{\vec{u}(t)} - (\cos^2{\frac{\theta}{2}} - \sin^2{\frac{\theta(t)}{2}})\ket{-}_{\vec{u}(t)} \bra{-}_{\vec{u}(t)})] \nn \\
    &= \om_0 \cos{\theta(t)}\left(\ket{+}_{\vec{u}(t)} \bra{+}_{\vec{u}(t)}-\ket{-}_{\vec{u}(t)} \bra{-}_{\vec{u}(t)})]\right).
\end{align}
such that
\be
    \delta {\cal W}_{t,t+dt} = \text{Tr}\left[\rho_t\left(H_{\vec{u}(t+dt)}-H_{\vec{u}(t)}\right)\right]  
\ee
As $\rho_t = C_\ii[\rho_t]$ and $\rho_{t+dt}=C_\ff[\rho_{t}]$, by definition of the representative of the Hamiltonian (see Eq.~\eqref{d:representative_comm}), we have $\text{Tr}\{H\rho_{t}\} = \text{Tr}\{H_{{\vec u}(t)}\rho_{t}\}$ and $\text{Tr}\{H\rho_{t+dt}\} = \text{Tr}\{H_{{\vec u}(t+dt)} \rho_{t+dt}]\}$ such that $\delta W_{t,t+dt} = \delta W_\text{ext}$. The work increment defined in \eqref{d:dW} therefore matches the energy change due to the Zeno rotation.

In addition, since the map ${\cal N}_{t,t+dt}$ can be considered the identity channel and is therefore unital (and so is the map $\Lambda_{t,t+dt}$), one simply has $K_t=0$, that is, $\delta {\cal Q}_{t,t+dt} = 0$ and $\Delta E_{t,t+dt} = \delta W_{t,t+dt}$.

\subsection{Open qubit}

We now move to the case where the qubit evolves between the two measurement according to a GKLS master equation generated by an environment at thermal equilibrium at temperature $T=1/\beta$ \cite{Alicki79}, namely:
\begin{align} \label{lindbladref}
    {\cal N}_{t,t+dt}[\rho(t)] &= \rho(t) - i [H, \rho_t] dt + \g_+ (\s_+ \rho(t) \s_- - \f12 \{\s_-\s_+,\rho(t)\}) dt \nn \\
    &+ \g_- (\s_- \rho(t) \s_+ - \f12 \{\s_+\s_-,\rho(t)\}) dt, \qquad \g_+ = e^{- \beta \om_0} \g_{-}, 
\end{align}
with $H = (\omega_0/2)\s_z$ the qubit Hamiltonian. Moreover, to look at the simplest setting first, we prevent the measurements $C_{\ii,\ff}$ from modifying the dynamics induced by the Lindbladian $\mathcal{N}_{t,t+dt}$ and actively inject energy into the system  
by assuming that they are both performed in the instantaneous eigenbases diagonalizing the density matrix at $t$ and $t+dt$, respectively, as presented in Section \ref{sec: heat and work}. As the projectors associated with the measurements $C_{\ii,\ff}$ are of rank $1$, the associated Petz recovery maps $P_{C_{\ii,\ff}, u}$ correspond to the identity channel.

Imagine that at $t$ the qubit is diagonal in the basis  $\{ \ket{+}_{\vec{u}(t)},  \ket{-}_{\vec{u}(t)}\}$ of the eigenstates of $\s_{\vec{u}(t)}$, where as before $\vec{u}(t) = (\sin{\theta(t)} \cos{\phi(t)}, \sin{\theta(t)} \sin{\phi(t)}, \cos{\theta(t)})$ being a unit vector on the Bloch sphere. 
Since $V_J = 1$, the representative of the density operator $\rho_t$ is once again the state itself and reads
\be 
    \rho_t = p_+ \ket{+}_{\vec{u}(t)} \bra{+}_{\vec{u}(t)} + p_- \ket{-}_{\vec{u}(t)} \bra{-}_{\vec{u}(t)}.
\ee
Note we can connect the ladder and spin operator in this basis to the Hamiltonian eigenbasis (corresponding to $\vec u = (0,0,1)$) via
\bea
    \s_{\pm,\vec{u}(t)} &=& \cos^2{\frac{\theta}{2}} \s_{\pm} - e^{\pm 2 i \varphi} \sin^2{\frac{\theta}{2}} \s_{\mp} - \f12 e^{\pm i \varphi} \sin{\theta} \s_z\nonumber\\
    \s_{\vec{u}(t)} &=& \cos{\theta} \s_z + \sin{\theta} (\s_+ e^{- i \varphi} + \s_- e^{i \varphi}). 
\eea

The evolution of the qubit state between time $t$ and $t + dt$ is given by the Lindblad equation
\begin{align} \label{lindbladref2}
    \rho(t+dt) &= \rho_t - i [H, \rho_t] dt + \g_+ (\s_+ \rho_t\s_- - \f14 [(\Id_2 - \s_z)\rho_t + \rho_t (\Id_2 - \s_z)] ) dt \nn \\
    &+ \g_- (\s_- \rho_t \s_+ - \f14 [(\Id_2 + \s_z)\rho_t+ \rho_t (\Id_2 + \s_z)]) dt, \qquad \g_+ = e^{- \beta \om_0} \g_{-}.
\end{align}

Given the measurement bases and the absence of coarse-graining, the map $\tilde\Lambda_{t,t+dt}=\text{A}_{t,t+dt}\circ\Lambda_{t,t+dt}$ we need to consider simply amounts to an evolution during $dt$ according to Eq.~\eqref{lindbladref2}, followed by a projection back onto the basis $\{ \ket{+}_{\vec{u}(t)},  \ket{-}_{\vec{u}(t)}\}$ diagonalizing $\rho(t) = \rho_t$.

However, as the rotation between the eigenbases of $\rho(t+dt)$ and $\rho(t)$ is of order $O(dt^2)$, the change of the population in the initial and final bases are the same. Namely, calling $\{ \ket{+}_{\vec{u}(t+dt)}, \ket{+}_{\vec{u}(t+dt)} \}$ the eigenbasis of $\rho(t + dt)$, one has
\be
    \ket{\pm}_{\vec{u}(t+dt)} = \ket{\pm}_{\vec{u}(t)} + \a_{\pm} dt \ket{\mp}_{\vec{u}(t)} + O(dt^2)
\ee
such that
$${}_{\vec{u}(t)}\! \bra{+}\rho(t + dt) \ket{+}_{\vec{u}(t)} = {}_{\vec{u}(t+dt)}\bra{+} \rho(t + dt) \ket{+}_{\vec{u}(t+dt)}+O(dt^2),$$
and $\rho_{t+dt}$ has a form similar to $\rho_t$ with updated populations $p_\pm(t+dt)$. The map $\tilde\Lambda_{t,t+dt}$ then simply leads to rate equations:
\begin{align} \label{probdis}
   \frac{d p_+}{dt} &= \frac{ d \text{\hspace{0.1 cm}}  {}_{\vec{u}(t)}\! \bra{+}\rho(t + dt) \ket{+}_{\vec{u}(t)}}{dt} = (\g_+ \cos^4{\frac{\theta}{2}} + \g_- \sin^4{\frac{\theta}{2}}) p_- - (\g_+ \sin^4{\frac{\theta}{2}} + \g_- \cos^4{\frac{\theta}{2}}) p_+ \nn \\
   \frac{d p_-}{dt} &= \frac{ d \text{\hspace{0.1 cm}} {}_{\vec{u}(t)}\! \bra{-}\rho(t + dt) \ket{-}_{\vec{u}(t)}}{dt} = (\g_+ \sin^4{\frac{\theta}{2}} + \g_- \cos^4{\frac{\theta}{2}}) p_+ - (\g_+ \cos^4{\frac{\theta}{2}} + \g_- \sin^4{\frac{\theta}{2}}) p_-
\end{align}
Solving for its instantaneous steady state $\sigma_t$ yields:
\begin{align}
    \frac{{}_{\vec{u}(t)}\bra{+}\s_t\ket{+}_{\vec{u}(t)}}{{}_{\vec{u}(t)}\bra{-}\s_t\ket{-}_{\vec{u}(t)}} &= \frac{\g_+ \cos^4{\frac{\theta}{2}} + \g_- \sin^4{\frac{\theta}{2}}}{\g_- \cos^4{\frac{\theta}{2}} + \g_+ \sin^4{\frac{\theta}{2}}} = \frac{ \cos^4{\frac{\theta}{2}} e^{-\beta \om_0} + \sin^4{\frac{\theta}{2}}}{ \cos^4{\frac{\theta}{2}} + e^{-\beta \om_0} \sin^4{\frac{\theta}{2}}} = \frac{ e^{-\beta \om_0} + \tan^4{\frac{\theta}{2}}}{ 1 + e^{-\beta \om_0} \tan^4{\frac{\theta}{2}}} \nn \\ 
    &= e^{- \beta \om_0} \frac{1 + e^{\beta \om_0} \tan^4{\frac{\theta}{2}}}{1 + e^{-\beta \om_0} \tan^4{\frac{\theta}{2}}}
\end{align}
so that 
\begin{align}
    {}_{\vec{u}(t)}\bra{+}\s_t\ket{+}_{\vec{u}(t)} = \frac{e^{- \beta \om_0} \frac{1 + e^{\beta \om_0} \tan^4{\frac{\theta}{2}}}{1 + e^{-\beta \om_0} \tan^4{\frac{\theta}{2}}}}{Z},  \qquad {}_{\vec{u}(t)}\bra{-}\s_t\ket{-}_{\vec{u}(t)} = \frac{1}{Z},
\end{align}
with $Z=1 + e^{- \beta \om_0} \frac{1 + e^{\beta \om_0} \tan^4{\frac{\theta}{2}}}{1 + e^{-\beta \om_0} \tan^4{\frac{\theta}{2}}}$.
We deduce 
\be 
K_t = \beta\left(\omega_0-T\ln{\frac{1 + e^{\beta \om_0} \tan^4{\frac{\theta}{2}}}{1 + e^{-\beta \om_0} \tan^4{\frac{\theta}{2}}}}\right)\ket{+}_{\vec u(t)}\bra{+} + \ln Z \Id.
\ee
Now, the Hamiltonian of the qubit being given by 
\be
    H = \frac{\om_0}{2} (\ket{+} \bra{+} - \ket{-} \bra{-}),
\ee
its representative in the commutative algebra associated with projectors onto states $\{ \ket{+}_{\vec{u}(t)}, \ket{-}_{\vec{u}(t)} \}$ is:
\begin{align}\label{eq:Htqubit}
    H_t&=
    {}_{\vec{u}(t)}\bra{+} H \ket{+}_{\vec{u}(t)} \ket{+}_{\vec{u}(t)}\!\bra{+}  + {}_{\vec{u}(t)}\bra{-} H \ket{-}_{\vec{u}(t)} \ket{-}_{\vec{u}(t)}\!  \bra{-} \nn \\
    &=\frac{\om_0}{2} [(\cos^2{\frac{\theta}{2}} - \sin^2{\frac{\theta}{2}}) \ket{+}_{\vec{u}(t)} \bra{+}_{\vec{u}(t)} - (\cos^2{\frac{\theta}{2}} - \sin^2{\frac{\theta}{2}})\ket{-}_{\vec{u}(t)} \bra{-}_{\vec{u}(t)})] \nn \\
    &= \om_0 \cos{\theta(t)}\ket{+}_{\vec{u}(t)} \bra{+}_{\vec{u}(t)}  - \frac{1}{2}\om_0 \cos{\theta(t)} \Id = \frac{\om_0}{2} (\ket{+}_{\vec{u}(t)} \bra{+}_{\vec{u}(t)} - \ket{-}_{\vec{u}(t)} \bra{-}_{\vec{u}(t)}) .
\end{align} 
Hence we see that the effective temperature appearing in Eq.~\eqref{eq: Kt in op basis} must be taken to be:
\be
    \frac{1}{\beta(t)}= T(t) = T \frac{\om_0 \cos{\theta}}{\om_0 - T \ln{\frac{1 + e^{\beta \om_0} \tan^4{\frac{\theta}{2}}}{1 + e^{-\beta \om_0} \tan^4{\frac{\theta}{2}}}}} \geq 0,
\ee
so that we have $K_t=\beta(t) H_{t} + \text{cste}$. This renormalization of the apparent temperature in the eigenbasis of $\rho_t$ is a consequence of the instantaneous fixed point being a projection of the actual thermal equilibrium state (stationary state of the Lindblad equation, lying on the $z$-axis of the Bloch sphere) onto the axis $\vec u(t)$. 

We can now compute 
\begin{align}\label{eq:Ktdiff}
    \langle K_t \rangle_{\rho(t+dt)} - \langle K_t \rangle_{\rho(t)} &= \beta \big( \om_0 - T \ln{\frac{1 + e^{\beta \om_0} \tan^4{\frac{\theta}{2}}}{1 + e^{-\beta \om_0} \tan^4{\frac{\theta}{2}}}} \big) (p_+(t+dt) - p_+(t) \big) \nn \\
    &= \beta dt \big( \om_0 - T \ln{\frac{1 + e^{\beta \om_0} \tan^4{\frac{\theta}{2}}}{1 + e^{-\beta \om_0} \tan^4{\frac{\theta}{2}}}} \big)\nn\\
    &\qquad\times [(p_- \g_+ - p_+ \g_-) \cos^4{\frac{\theta}{2}} - (p_+ \g_+ - p_- \g_-) \sin^4{\frac{\theta}{2}}] + O(dt^2)
\end{align}
which plays the role of the generalized heat (divided by the temperature) appearing in the entropy production rate Eq.\eqref{eq: second law loc t t+dt 3}.
Writing
\begin{align} \label{Hamiltoniandiff}
    \langle H_{t} \rangle_{\rho(t+dt)} - \langle H_{t} \rangle_{\rho(t)} &= \frac{\om_0}{2} \cos{\theta} \left[p_+(t+ dt) - p_+(t) - p_-(t+dt) + p_-(t)\right] + O(dt^2) \nn \\
    &= \om_0  dt\cos{\theta} \left[(p_- \g_+ - p_+ \g_-) \cos^4{\frac{\theta}{2}} - (p_+ \g_+ - p_- \g_-) \sin^4{\frac{\theta}{2}} \right] + O(dt^2),
\end{align}
we see that we have 
\be
    \frac{\Delta \langle H_{t} \rangle}{T(t)} = \Delta \langle K_t \rangle.
\ee
To compete the energy balance and write the first law as seen from those accessible subalgebras, we must take into account that the  projections of the Hamiltonian depends on time via $u(t)$ and $\theta(t)$ (see Eq.~\eqref{eq:Htqubit})
We then have
\begin{align} \label{firstlaw}
    d E &:= \langle H_{t + dt} \rangle_{\rho(t + dt)} - \langle H_{t} \rangle_{\rho(t)} = \tr{\big( \rho(t + dt) H_{t+dt} \big)} - \tr{\big( \rho(t) H_{t} \big)} \nn \\
    &= \tr{(\rho(t + dt) - \rho(t)) H_{t + dt} \big)} + \tr{\big( \rho(t) (H_{t+dt} - H_{t} \big)} \nn \\
    &= \tr{(\rho(t + dt) - \rho(t)) H_{t} \big)} + \tr{\big( \rho(t) (H_{t + dt} - H_{t}) \big)} + O(dt^2) \nn \\
    &= \d \mathcal{Q}_{t,t+dt} + \d \mathcal{W}_{t,t+dt}.
\end{align}
We have identified the heat contribution deduced from the second law via Eqs.~\eqref{eq:Ktdiff}-\eqref{Hamiltoniandiff}, and the work contribution from Eq.~\eqref{d:workincrementopen}. Of course, the natural definitions of heat and work we provided can be extended to any measurement along an axis $\vec{u}(t)$ at $t$, and $\vec{u}(t+dt)$ at $t+dt$, as long as $\vec{u}(t+dt) = \vec{u}(t) + O(dt)$ (i.e. not necessarily in the instantaneous eigenbasis). A geometrical interpretation of heat and work in the case of the qubit is depicted in Fig.~\ref{fig:BlochSphere} in the main text, where the heat (work) increment corresponds to the radial (radius-preserving) component of the evolution.

It may seem paradoxical that a nonzero work contribution arises in this example, since the measurement is designed not to inject energy and the qubit is otherwise coupled only to a heat bath. However, one should first note that the work contribution is always negative \emph{as long as the measurement is performed in the instantaneous energy eigenbasis of the qubit}. Indeed, one has
\be
    \d {\cal W}_{t,t+dt} = \tr{\left(\frac{d H_{\vec{u}(t)}}{d t} \rho(t)\right)} = - \frac{d \theta}{dt} \tan{\theta} \tr{(H_{\vec{u}(t)} \rho(t))} \leq 0 
\ee
since the qubit monotonously converges to the stationary thermal state, situated in the south hemisphere of the Bloch sphere, leading to $\frac{d \theta}{dt} \geq 0$ at any time. In the north hemisphere, $\tan{\theta} \geq 0$ but $\tr{\rho \hat{H}} \geq 0$, so $\d {\cal W}_{t,t+dt} \leq 0$, while in the south hemisphere $\tan{\theta} \leq 0$ but $\tr{\rho \hat{H}} \leq 0$, also leading to $\d {\cal W}_{t,t+dt} \leq 0$. Consequently, this term represents work performed by the qubit on its thermal environment, and is not in contradiction with the second law of thermodynamics. Moreover, the work-like nature of $\d {\cal W}_{t+dt}$ is visible in the fact that this amount of energy can be controllably and reversibly re-injected into the qubit by simply measuring the qubit again in the initial eigenbais in the direction $\vec u(t)$. 

More generally, the energy variation associated with the (controlled) measurement basis change is identified as work, while the energy variation related to the evolution along the $\vec u(t)$ axis is related to heat. The latter is uncontrolled and irreversible as it is not reverted simply by bringing the system back to the basis $\vec u(t)$.

\section{Example: Quantum emitter weakly coupled to 1D electromagnetic environment}

\label{app: collisional model}

In this appendix, we apply our framework to the case of a two-level quantum emitter (hereafter called qubit) and the one-dimensional electromagnetic field constituting its environment. The qubit as an open system has been treated in Appendix~\ref{app: qubit} and here we rather aim at treating the qubit and its environment using the autonomous coarse-grained approach outlined in the main text, exploring the consequences of choosing different subalgebras for the field. When considering a 1D, linear-dispersed, field as an environment, and under the condition of weak coupling to the qubit, the total dynamics of the system and environment can be efficiently mapped on a collision model, where discrete temporal modes of the field interact sequentially with the qubit, thereby allowing to access easily the outcome of various measurements on the qubit and/or the field  \cite{Prasad2026} (see Fig.~\ref{fig:collmodel} of main text). We first review this approach and discuss how it naturally singles out coarse-grained subalgebra of observables for the electromagnetic field. Then, we analyze the different contributions to entropy production appearing when applying our formalism to that case, for different choices of subalgebras (i.e. different measurement capabilities), as well at the exchanges of heat and work as defined in Secs.~\ref{sec: heat and work} and \ref{sec:algebraic_heat_and_work}.

\subsection{Introduction to the model}

Here we briefly review the model of Ref.~\cite{Prasad2026}. We consider an emitter modeled by a qubit (system A) interacting with a 1D electromagnetic field (environment B). The total Hamiltonian is given by 
\be
    H = H_A + H_I + H_B, \qquad H_A = \frac{\om_0}{2} \s_z, \quad H_{B} = \sum_k \om_k a_k^\dag a_k, \quad H_I =  i \sum_k g_k (a_k \s_+ - a_k^\dag \s_-),
\ee
in terms of environmental modes with frequency $\omega_k$ (and wavenumber labeled by $k$). The system is assumed to sit at position $x=0$ and the rotating wave approximation was performed ($g_k$ is the interaction strength between the system and the mode $a_k$). Moreover, $\omega_k = k \Delta\omega$, that is, the field has a linear dispersion relation for the sake of simplicity. In the interaction picture, one gets
\be
    \frac{d \rho_{AB}}{dt} = - i[V(t), \rho_{AB}(t)]
\ee
with 
\be \label{def: Vt}
    V(t) = i (B(0,t) \s_+ - \s_- B(0,t)^\dag)
\ee
where 
\be\label{eq:Bxt}
    B(x,t) = \sum_k g_k e^{- i \Delta \om_k (t - \frac{x}{c})} a_k, \qquad \Delta \om_k = \om_k - \om_0
\ee
represents the field operator at position $x$. We introduce the correlation function in the vacuum of the electromagnetic field 
\be
    C(t,t') = \bra{0} B(0,t) B(0,t')^\dag \ket{0} = C(t - t') =  \sum_k \lvert g_k \lvert^2 e^{i \Delta \om_k (t - t')}
\ee
that vanishes for $\lvert t - t' \lvert \gg \tau_c \sim \frac{1}{\Delta_\text{bw}} $ where $\Delta_\text{bw}$ is the range on which the coupling constants $g_k$ take non negligible values. In addition, one defines the spontaneous emission rate of the qubit:
\be \label{eq: gamma correl}
    \g = \int_{- \infty}^{+ \infty} C(\tau = t - t') d \tau = 2 \pi\sum_k \lvert g_k \lvert^2 \d (\Delta \om_k).
\ee

To map the problem onto a collision model, we split the dynamics into discrete time steps of duration $\Delta t$. The total system evolution operator between times $t_n = n \Delta t$ and $t_{n+1} = t_n + \Delta t$ is given by:
\be \label{eq: unitaryn}
    \mathcal{U}_n = \mathcal{T} \big( e^{-i \int_{t_n}^{t_{n+1}} V(t) dt} \big)
\ee
Assuming weak coupling $\g \t_c \ll 1$, a Magnus expansion allows us to get rid of the time-ordering operator in \eqref{eq: unitaryn} at first order and get
\be \label{eq: Un2}
    \mathcal{U}_n = e^{-i \int_{t_n}^{t_{n+1}} V(t) dt} := e^{- i V_n},
\ee
where
\be \label{eq: Vncollisionalmodel}
    V_n = i \sqrt{\g \Delta t} (b_n \s_+ - b_n^\dag \s_-).
\ee
We have identified the discrete temporal mode operators
\be \label{def: bn}
    b_n = \frac{1}{\sqrt{\gamma \Delta t}} \int_{t_n}^{t_{n+1}} dt B(0,t)
\ee
which satisfy the commutation relations
\begin{align}
    [b_n, b_m^\dag] &= \frac{1}{\gamma \Delta t} \int_{t_n}^{t_{n+1}} dt \int_{t_m}^{t_{m+1}} dt' (B(0,t) B^\dag(0,t') - B^\dag(0,t') B(0,t)) \nn \\
    &= \frac{1}{\gamma \Delta t}  \int_{t_n}^{t_{n+1}} dt \int_{t_m}^{t_{m+1}} dt' \sum_k \lvert g_k \lvert^2 e^{i \Delta \om_k (t - t')} [a_k, a_k^\dag] = \frac{1}{\gamma \Delta t} \int_{t_n}^{t_{n+1}} dt \int_{t_m}^{t_{m+1}} dt' C(t,t'),
\end{align}
reducing to $[b_n, b_m^\dag] = \d_{nm}$ (i.e. independent temporal modes) when $\tau_c \ll \Delta t \ll \gamma^{-1}$. Eq.~\eqref{eq: Un2}-\eqref{eq: Vncollisionalmodel} allow us to solve the dynamics by simply considering two-body short interactions between the system and a new oscillator $b_n$ at each time-step. 

From Eqs.~\eqref{eq:Bxt} and \eqref{def: bn} we deduce
\bea\label{eq:akbn}
 g_k \sinc \left(\tfrac{1}{2}\Delta\omega_k\Delta t\right)\frac{2\pi}{\Delta\omega} a_k = \sqrt{\gamma\Delta t}\sum_n e^{i\Delta\omega_k t_n}b_n.
\eea
Note that this formula only applies to modes $a_k$ at frequencies $\omega_k$ within a few $1/\Delta t$ of $\omega_0$. This is a consequence of considering only discrete temporal modes. A complete description of the field space including frequencies far from $\omega_0$ would be obtained by including higher harmonic modes in each time-interval \cite{Prasad2026}.

\vspace{0.3 cm}

We finally expand the equation \eqref{eq: Un2} up to first order $O(\g \Delta t)$ and get the variation $\Delta_n \rho_{AB} = \rho_{AB}(t_{n+1}) - \rho_{AB}(t_n)$ of the joint (qubit + electromagnetic field) state during time step $[t_n,t_{n+1}]$
\be \label{eq: deltanrhocollisions}
    \Delta_n \rho_{AB} = - i[V_n, \rho_{AB}(t_n)] - \frac{1}{2} [V_n, [V_n, \rho_{AB}(t_n)]].
\ee
Remarkably, it only involves temporal mode $b_n$, allowing its interpretation as a collision unit. We have neglected a Lamb-shift like term which can be absorbed in renormalized free qubit and reservoir Hamiltonians. Notice that in general $\Delta_n \rho_{AB} = O(\sqrt{\gamma \Delta t})$ because of \eqref{eq: Vncollisionalmodel}. For instance, the two first moments of the field evolution during a collision read
\bea\label{eq:fieldcollision}
\mean{b_n}_{t_{n+1}}&=& \mean{b_n}_{t_{n}}-\sqrt{\gamma\Delta t}\mean{\sigma_-}_{t_n}\nonumber\\
\mean{b_n^\dagger b_n}_{t_{n+1}}&=& \mean{b_n^\dagger b_n}_{t_{n}}-\sqrt{\gamma\Delta t}\left(\mean{\sigma_+}_{t_n}\mean{b_n}_{t_n}+\mean{\sigma_-}_{t_n}\mean{b_n^\dagger}_{t_n}\right)\nonumber\\&& + \gamma\Delta t\left[\left(\mean{b_n^\dagger b_n}_{t_{n}}+\tfrac{1}{2}\right)\mean{\sigma_z}_{t_{n}}+\tfrac{1}{2}\right].
\eea

\subsection{Algebraic thermodynamics for the collisional model: the electromagnetic field as a reservoir}

Let us assume that the qubit is described by its full algebra $\mathcal{A}_{\text{A}} =\mathcal{B} (\mathcal{H}_{\text{A}})$ while the electromagnetic field is described by a subalgebra $\mathcal{A}_{\text{B}} \subset \mathcal{B}(\mathcal{H}_{\text{B}})$ that, for example, may be generated by some of the local operators $\{ b_n, b_n^\dag \}_{n\in {\cal I}}$ for some interval ${\cal I}$.
Once some initial and final algebras are fixed, one can study the evolution of the joint system $\text{A} \cup \text{B}$ within the autonomous viewpoint and write the second law \eqref{eq : gen bipartite second law} between $t_n$ and $t_{n+1}$
\bea \label{eq: second law gen alg app bipartite}
    \Delta S_{\mathcal{A}_{\text{A}}} - \tr{((\rho_{\mathcal{A}_{\text{B}, n+1}}(t_{n+1}) - \rho_{\mathcal{A}_{\text{B}, n}}(t_{n}))\ln{\rho_{\mathcal{A}_{\text{B}, n}}(t_{n}))})} \nonumber\\\geq \Delta S_{\mathcal{A}_{\text{A}}\otimes \mathcal{A}_{\text{B}}} + \Delta I_{\mathcal{A}_{\text{A}} \otimes \mathcal{A}_{\text{B}}} + S(\rho_{\mathcal{A}_{\text{B}, n+1}}(t_{n+1}) \lvert \lvert \rho_{\mathcal{A}_{\text{B}, n}}(t_{n})).
\eea
The second term in the first line of \eqref{eq: second law gen alg app bipartite} can be written as $\beta_B(t_n) \delta Q_{\mathcal{A}_B}$ where $\beta_B(t_n)$ is the inverse temperature of the environment as defined in Appendix~\ref{app: negativetemperature}, and $\delta Q_{\mathcal{A}_B}$ the heat flux measured through the algebra $\mathcal{A}_B$ between $t_n$ and $t_{n+1}$. Of course, the heat and work fluxes, as well as the entropy production, depend on the specific choice we make for the field subalgebra, as we explore below.

Notice that, in the main text, when considering continuous quasi-static paths, we neglected the relative entropy contribution
$S(\rho_{\mathcal{A}_{\text{B}, n+1}}(t_{n+1}) \lvert \lvert \rho_{\mathcal{A}_{\text{B}, n}}(t_{n}))$ for the entropy production since it was of order $O(dt^2)$. However, in the collisional model mapping we analyze here, the continuous limit $\Delta t \to dt$ cannot be taken. Namely, the algebra used to describe the field is based on field operators averaged over a time interval $\Delta t$ 
satisfying $\tau_c \leq \Delta t \leq \g^{-1}$, which forbids to take $\Delta t \to dt$ arbitrarily small. The requirement $\Delta t > \tau_c$ is tightly connected to the possibility to obtain an emerging Markovian dynamics despite the correlation time of the field. The condition $\gamma\Delta t \ll 1$ still ensures a smooth dynamics despite this time-averaging.
One consequence is that $\Delta_n \rho_B = O(\sqrt{\gamma \Delta t})$ rather than $O(\gamma \Delta t)$, so that the relative entropy $S(\rho_{\mathcal{A}_{\text{B}, n+1}}(t_{n+1}) \lvert \lvert \rho_{\mathcal{A}_{\text{B}, n}}(t_{n}))$ is of first order in $\Delta t$, rather than of second order. Nevertheless, we will see that, in some cases, this contribution remains negligible at first order, although this is no longer systematic, in contrast with the rigorous continuous limit of section \ref{subsec: autonomous viewpoint gen}.

\subsubsection{Algebra of a single collision mode}

During the time interval $[t_n,t_{n+1}]$, the qubit interacts with a local field mode, referred to as the $n$-th collision mode, with annihilation and creation operators $b_n$ and $b_n^\dagger$. The corresponding Hilbert space $\mathcal{H}_{B_n}$ is the Fock space spanned by successive applications of $b_n^\dagger$ to the vacuum state $\ket{0}$. Since these operators act irreducibly on $\mathcal{H}_{B_n}$, the von Neumann algebra $\mathcal{A}_{B_n}$ they generate coincides with the algebra of all bounded operators on this space, $\mathcal{A}_{B_n}=\mathcal{B}(\mathcal{H}_{B_n})$.

We assume an initial thermal state for the field at inverse temperature $\beta$, which implies for the  n-th collision mode an initial state which is also thermal-like \cite{Prasad2026}:
\be
    \rho_{\mathcal{A}_{B_n}}(t_n) = \frac{e^{-\beta \om_0 b_n^\dag b_n }}{\tr{e^{-\beta \om_0 b_n^\dag b_n }}} = \frac{e^{-\beta H_n }}{\tr{e^{-\beta H_n }}}, \qquad H_n := \om_0 b_n^\dag b_n.
\ee
since $H_n$ is the representative of the Hamiltonian $H_B$ on $\mathcal{A}_{B_n}$. Then, the second term in the first line of \eqref{eq: second law gen alg app bipartite} becomes 
\be \label{eq: heat flux}
     - \tr{\left\{\left[\rho_{\mathcal{A}_{B_{n+1}}}(t_{n+1}) - \rho_{\mathcal{A}_{B_n}}(t_{n})\right]\ln{\rho_{\mathcal{A}_{B_{n}}}(t_{n})}\right\}} = \beta \delta Q_{\mathcal{A}_{B_n}}  
\ee
so that $\delta Q_{\mathcal{A}_{B_n}} = \Delta \langle H_n \rangle_{\rho_{\mathcal{A}_{B_n}}}$ is the heat transferred to the environment. Therefore, in this situation, the effective temperature of B is simply the initial temperature of the field, and the heat flux equals the energy variation of the collisional mode, 
implying a vanishing work increment. This remains true whenever the operators in the subalgebra of $B$ are in a thermal state, i.e., their expectation values satisfy the Kubo-Martin-Schwinger (KMS) condition at some positive temperature.

\begin{remark}
    This does not prevent the system from providing work to the environment, as it would be captured by the reverse description where the roles of A and B are swapped. In that case, we would find a non-zero work increment whenever the qubit starts out of equilibrium at $t_n$, i.e. has coherences or population inversion in its energy eigenbasis, see section \ref{subsec: qubit as reservoir}.
\end{remark}

What about the entropy production term? The entropy production term on the second line of \eqref{eq: second law gen alg app bipartite} consists of three contributions. The first one is the variation of the joint algebraic entropy. As we are describing the qubit  via its full algebra $\mathcal{A}_\text{A} = {\cal B}({\cal H}_\text{A})$ at both $t_n$ and $t_{n+1}$, its algebraic entropy coincides with its von Neumann entropy. Moreover, the algebra $\mathcal{A}_{B_n}$ taken for the n-th mode also corresponds to the full set of operators over the Hilbert space of this mode, such that the algebraic entropy of the latter, as well as the joint algebraic entropy of the emitter and the n-th collision mode, coincide with the corresponding von Neumann entropies. As in addition, the joint evolution of the qubit and the n-th mode is unitary over the interval $[t_n,t_{n+1}]$, their joint von Neumann entropy is conserved, and the first contribution to the entropy production vanishes. 

Regarding the relative entropy term $S(\rho_{\mathcal{A}_{B_n}}(t_{n+1}) \lvert \lvert \rho_{\mathcal{A}_{B_n}}(t_{n}))$, one needs to distinguish two cases. If the qubit state has no coherence at time $t_n$, one can show that $\rho_{\mathcal{A}_\text{B}}(t_{n+1}) - \rho_{\mathcal{A}_\text{B}}(t_{n}) = \tr_{\text{A}}{\Delta_n \rho_{\mathcal{A}_{\text{A}} \otimes \mathcal{A}_{B_n}}} = O(\g \Delta t)$ and it implies that $S(\rho_{\mathcal{A}_{B_n}}(t_{n+1}) \lvert \lvert \rho_{\mathcal{A}_{B_n}}(t_{n})) = O((\gamma \Delta t)^2)$. On the other hand, if the qubit carries coherence, the dominant term in the n-th collision mode evolution is given by 
\be \label{eq: first order state rhon coll model}
    \tr_{\text{A}}{[V_n, \rho_{\mathcal{A}_{\text{A}} \otimes \mathcal{A}_{B_n}}]} = i \sqrt{\g \Delta t} Z^{-1}(\beta)(1 - e^{\beta \om_0}) (\bra{g} \rho_{\mathcal{A}_{\text{A}}} \ket{e} b_n e^{- \beta H_n} + \text{c.c})
\ee
where $\ket{g}$ and $\ket{e}$ are respectively the ground and excited states of the qubit, and $Z(\beta) = \tr{e^{- \beta \om_0 b_n^\dag b_n}}$ the n-th mode partition function, such that $S(\rho_{\mathcal{A}_{B_n}}(t_{n+1}) \lvert \lvert \rho_{\mathcal{A}_{B_n}}(t_{n})) \sim O(\gamma \Delta t)$. 

Finally, the variation of mutual information $\Delta I_{\mathcal{A}_{\text{A}} \otimes \mathcal{A}_{\text{B}_n}}$ is giving a strictly positive contribution, since the initial mutual information between the qubit and the n-th mode vanishes. The latter is therefore the only contribution to entropy production when describing the field with this single collision mode algebra in the absence of qubit coherences (in presence of coherence, the relative entropy term also contributes, leading to a higher total entropy production).  

\subsubsection{The wave-train algebra: a coarse-grained Hamiltonian}
\label{sec: section wave train no coherences}

Now, instead of considering the algebra of a single collision mode at time $t_n$, we consider a wave train made of $N \gg \frac{1}{\gamma \Delta t} \gg 1$ collision modes. The length of the wave train is chosen to be much longer than the relaxation time $t_R = \g^{-1}$ of the qubit toward its ground state.
Therefore, we are interested in a von Neumann (sub)algebra $\bigotimes_{n = 1}^N \mathcal{A}_{B_n}$ generated by $\{ b_n, b_n^\dag \quad \lvert n \in (1, N) \}$. As we now follow the collisional modes after their interaction with the qubit, the state of the electromagnetic field at the beginning of each time step $t_n$ will change from $n=1$ to $n=N$. In particular, it is very likely that at least one photon will be emitted into the wave train at some point during the interval. 
This evolution of the field state yields different behavior for the work and heat increments compared with the description based on the single collision subalgebra.

The first point to notice is that if the qubit starts in a state with no coherences in the energy eigenbasis,  e.g. in the excited state, and if the wave train at $t = 0$ (before the first collision) is in a thermal state, then the wave train will not build any coherences in the energy eigenbasis as well. Focusing on this case for the rest of this subsection, we can therefore without loss of generality consider a subalgebra $\mathcal{A}_H$ generated by the Hamiltonian of the collision modes composing the wave train
\be \label{eq: train wave hamiltonian}
    H = \sum_{n \in [1,N]} H_n = \sum_{n \in [1,N]} \om_0 b_n^\dag b_n,
\ee
which captures frequency modes within the interval $\left(\omega_0-\frac{2\pi}{\Delta t},\omega_0+\frac{2\pi}{\Delta t}\right)$. Its eigenstates include, in particular, the vacuum state of the electromagnetic field $\ket{0}$ and the one photon states $\ket{1_k}=a_k^\dagger \ket{0}$, with $k \in K = \{ \om_0 - \frac{\Gamma (N - j)}{N} \lvert j \in (1, 2N) \}$ where $\Gamma$ is an frequency scale so that $\frac{1}{\Delta t} \gg \Gamma \gg \g$.

To further simplify the analysis, we assume that the field is initially prepared in a low temperature thermal state satisfying $\beta\omega_0\gg1$. In the relevant frequency range, this state is initially well approximated by the vacuum $\ket{0}$, and the wave-train can be safely assumed to contain at most one photon after interacting with the qubit.
We then define the projectors
\be
    \Pi_0 = \ket{0} \bra{0}, \qquad \Pi_1 = \sum_{k \in K} \ket{1_k} \bra{1_k},
\ee
on the zero and one photon spaces, respectively, associated with ranks $V_0 = \tr{\Pi_0} = 1$, and $V_1 = \tr{\Pi_1} = \lvert K \rvert \gg 1$, which could describe a photon-counting measurement setup insensitive to the photon frequency. The subalgebra generated by the operators $\{\Pi_0, \Pi_1 \}$ is a commutative von Neumann subalgebra $\mathcal{A}_\Pi \subset \mathcal{B}(\mathcal{H}_{0,1}^K)$ and $\mathcal{H}_{0,1}^K$ is a finite-dimensional Hilbert space generated by $\{ \ket{0}, \ket{1_k} \lvert k \in K \}$. The representative $H_{\mathcal{A}_\Pi}$ of the field Hamiltonian is 
\be \label{eq: cg Hamiltonian APi}
H_{\mathcal{A}_\Pi} = \frac{\tr{(H \Pi_0)}}{V_0} \Pi_0 + \frac{\tr{(H \Pi_1)}}{V_1} \Pi_1 = 0  + \om_0 \Pi_1.
\ee 
This setting is closely related to the spontaneous-emission problem studied in Appendix~\ref{app: spontaneous emission}. There, however, we considered only the entropy production term between the initial time $t=0$, when the interaction with the qubit is switched on, and the asymptotic regime $t\gg\gamma^{-1}$. In contrast, Eq.~\eqref{eq: second law gen alg app bipartite} allows us to study the entropy production, work, and heat associated with each individual interval $(t_n,t_{n+1})$, for arbitrary $n\in(1,N-1)$. 

Following the procedure of Section~\ref{sec: heat and work} and Appendix.~\ref{app: negativetemperature}, we parametrize the representative $\rho_{\mathcal{A}_\Pi}$ of the field state as
\be \label{eq: rhoapitqubiteff}
    \rho_{\mathcal{A}_\Pi}(t) = \frac{e^{- \beta_0(t)(H_{\mathcal{A}_\Pi} - \mu(t) \Pi_1)}}{\tr{e^{- \beta_0(t)(H_{\mathcal{A}_\Pi} - \mu(t) \Pi_1)}}}
\ee
where $\beta_0(t)=\beta(t)$ and $\mu(t) = 0$ as long as $\beta(t) \geq 0$ (of course, $\beta(0) = \beta$). However, as soon as $\beta(t) \leq 0$, $\beta_0(t)$ is identified from the thermal state $\om_{\mathcal{A}_\Pi}(t)$ with the same algebraic entropy as $\rho_{\mathcal{A}_\Pi}(t)$. 
In the present example, it simply leads to $\beta_0(t) = \lvert \beta(t) \lvert$ so that $\mu = 2 \om_0$ when $\beta(t)$ becomes negative. 
This implies that work may be provided by the field \text{B}. Its increment between $t_n$ and $t_{n+1}$ is given by 
\be \label{eq: work projector}
    \delta {\cal W}_{\text{B},t_n, t_{n+1}} = \mu(t)  \Delta \langle \Pi_1 \rangle_{ \rho_{\mathcal{A}_\Pi}}
\ee
and is non-vanishing only if the population of the wave train is inverted. Physically, this population inversion means that the wave train is in an active state able to drive another system out of equilibrium. 
The heat increment $\d Q_{\text{B}, t_n, t_{n+1}}$ is obtained via subtraction of \eqref{eq: work projector} to the total energy variation. Therefore, by considering the whole wave train instead of the collision mode one by one, one is able to account for new resources (here, population inversions witnessing the nonequilibrium nature of the qubit's radiation), and therefore use the field for more tasks.

On the other hand, we can quantify the entropy produced between $t_n$ and $t_{n+1}$ evaluated from this subalgebra. Coming back to the three terms in the right hand side of \eqref{eq: second law gen alg app bipartite}, we first notice that since the qubit has no initial coherences, the relative entropy $S(\rho_{\mathcal{A}_{\Pi}}(t_{n+1}) \lvert \lvert \rho_{\mathcal{A}_{\Pi}}(t_{n}))$ is still of order $O((\g \Delta t)^2)$. Second, the variation of mutual information is negligible here and can even be negative. Indeed, the temperature $T(t) = \frac{1}{\beta(t)}$ being very small, the mutual information at late time between the qubit and the wave train vanishes (since for $t \gg \gamma^{-1}$, qubit A is in its ground state). Therefore, all (global, i.e. integrated over the whole wave train) entropy production comes from the variation of the joint algebraic entropy which is dominated by the variation of observational entropy of the algebra $\mathcal{A}_\Pi$
\be
    \Delta S_{\mathcal{A}_{\text{A}}\otimes \mathcal{A}_{\text{B}}} \sim \Delta S_{\mathcal{A}_\Pi} = \sum_{J \in \{0,1\} } \Delta p_J (\ln{V_J} - \ln{p_J}) > 0.
\ee
Indeed, $V_1$ is very large and most of the entropy production goes into the Boltzmann entropy. This behavior is not only consistent with the analysis led in the Appendix \ref{app: spontaneous emission}, but it also shows that the entropy production term can have genuinely different origins depending on the algebra under consideration. While it mainly comes from the creation of correlation when collision modes are regarded one by one, it is instead attributed to the transfer of population from a low-dimensional space (the photon vacuum) to a high-dimensional one (the one photon space of the wave-train).

\subsubsection{The wave-train algebra: projection onto the zero- and one-particle sectors}

In this paragraph, one assumes now that the qubit is not initially prepared in an excited state, but instead in a quantum superposition of the energy eigenstates
\be \label{eq: qubit coherent state}
    \ket{\psi_A(t = 0)} = \frac{\ket{e} + \ket{g}}{\sqrt{2}} 
\ee
Then, the wave train made out of $N \gg \frac{1}{\gamma \Delta t}$ collision modes will also acquire some coherence in the Hamiltonian \eqref{eq: train wave hamiltonian} eigenbasis. Since $\beta \om_0 \gg 1$, one can then focus on the algebra of bounded operators $\mathcal{B}(\mathcal{H}_{0,1})$ belonging to the zero and one particle sectors of the Hilbert space $\mathcal{H}_{0,1}$ generated by the eigenstates $\ket{0}$ and $\{ \ket{1_k}\}$, respectively. Then, at any time $t$, a possible ansatz for the state of the wave train is
\be\label{eq:rho_B01}
    \rho_{\mathcal{B}(\mathcal{H}_{0,1})} = \frac{e^{-K_B(t)}}{\tr{e^{-K_B(t)}}}, \qquad K_B(t) = \beta(t) \left(H_{\mathcal{B}_{\mathcal{H}_{0,1}}} + \sum_k \l_k(t) \ket{1_k} \bra{0} + \l_k^\ast(t) \ket{0} \bra{1_k}\right)
\ee
where $H_{\mathcal{B}_{\mathcal{H}_{0,1}}} = \Pi_{0,1} H_B \Pi_{0,1}$, with $\Pi_{0,1}$ the projector onto the zero or one particle subspaces of the Fock space, is the representative of the Hamiltonian on the subalgebra $\mathcal{B}_{\mathcal{H}_{0,1}}$ and where $\beta(t) \geq 0$. The operators $C_k = \ket{0} \bra{1_k} = \Pi_{0,1} a_k \Pi_{0,1}$ are the restrictions of the ladder operators $a_k$ on ${\cal H}_{0,1}$, such that the parameters $\lambda_k(t)$ are directly connected to the coherent displacement of the field modes.
From Eq.~\eqref{eq: deltanrhocollisions}-\eqref{eq:fieldcollision} and Eq.~\eqref{eq:akbn}, the coherent displacement of the field $\alpha_k(t) = \mean{a_k(t)}$  obeys for $\omega_k$ in the vicinity of $\omega_0$:
\bea
  \alpha_k(t_{n+1})= \alpha_k(t_{n})+\frac{\gamma\Delta\omega}{2\pi g_k}\Delta t\mean{\sigma_-}_{t_n}e^{i\Delta\omega_kt_n}\simeq \alpha_k(t_{n})+\gamma\Delta t\mean{\sigma_-}_{t_n}e^{i\Delta\omega_kt_n}, 
\eea
such that in the continuous limit $\gamma\Delta t \to 0$, we have $\alpha_k(t) \simeq \gamma\int_0^t dt' \mean{\sigma_-}_{t'}e^{i\Delta\omega_k t'}$.
We have assumed that, before interacting with qubit A, the wave train was in a thermal state at the inverse temperature $\beta$ so that $\alpha_k(t = 0) = 0$. Therefore, for sufficiently early times $t$, one has $\vert\alpha_k(t)\vert \ll 1$.
Moreover, we can relate $\alpha_k$ to $\lambda_k$ using 
\be
    \l_k(t) = \tr \left(C_k \frac{e^{-K_B(t)}}{\tr{e^{-K_B(t)}}}\right) = \tr \left(a_k \frac{e^{-K_B(t)}}{\tr{e^{-K_B(t)}}}\right) = \a_k(t)
\ee
while
\bea\label{eq:relatingdispandcoh}
\tr \left(a_k \frac{e^{-K_B(t)}}{\tr{e^{-K_B(t)}}}\right) = \tr \left(a_k \Pi_{0,1} D_{\alpha_k} \frac{e^{-\beta(t) H_B}}{\tr{e^{-\beta(t)H_B}}}D^\dagger_{\alpha_k} \Pi_{0,1}\right) + O(\vert \a_k \vert^2),
\eea
with 
$D_{\alpha_k} = e^{\alpha_k a_k^\dag-\alpha_k^* a_k}$ the displacement operator acting on mode $k$. From the second equality of \eqref{eq:relatingdispandcoh}, we deduce
\be \label{eq: mod ham wtrain}
    K_B(t) = \beta(t)(H_{\mathcal{B}_{\mathcal{H}_{0,1}}}- \sum_k \omega_k(\alpha_k^*C_k+\alpha_k C_k^\dag)) + O(\vert \a_k \vert^2),
\ee
such that
\be
    \lambda_k(t) = -\omega_k\alpha_k(t)
\ee
at first order. Notice also that unlike in the previous subsection, since the qubit starts with only half a quantum of excitation in state  \eqref{eq: qubit coherent state}, the field will receive at most half a photon, and therefore no population in the energy eigenbasis inversion may occur. 
As a consequence, all the work contribution that can be identified from \eqref{eq: second law gen alg app bipartite} for this choice of accessible algebra is associated to the variation of the coherent displacements $\alpha_k$. In particular, the increment during the n-th collision reads (where the mean values $\langle \cdot \rangle$ below are taken in the state $\rho_{\mathcal{B}(\mathcal{H}_{0,1})}$)
\bea
  \delta W_{t_n,t_{n+1}}= -\sum_k (\lambda_k(t)\Delta\!\mean{C_k}^*+\lambda_k^*(t)\Delta\!\mean{C_k}) = 2\,\text{Re}\sum_k \omega_k\alpha_k(t)\Delta\!\mean{C_k}^*,
\eea
which, remembering that $\alpha_k(t) = \langle a_k \rangle = \langle C_k \rangle$ and that $\Delta\!\mean{C_k}$ is here an infinitesimal variation, can be rewritten as
\bea \label{eq: work increment coll model}
  \delta W_{t_n,t_{n+1}}= \sum_k \omega_k\Delta|\!\mean{C_k}\!|^2.
\eea
That is, the work simply corresponds to the increment of the coherent part of the amplitudes of the energy modes of the waveguide (or at least, its restriction to ${\cal H}_{0,1}$). Note that if higher photon numbers were also introduced, more parameters would be needed to parametrize the state, leading to extra work contribution. One known example corresponds to the case of a squeezed state where the variation squeezing parameter would lead to a previously identified work contribution. Once an accessible algebra is defined, the present approach generalizes those cases to an arbitrary measured variation of the field parameters.

\subsubsection{The wave train algebra: a rotating subalgebra}

From a complementary perspective, one may consider a smaller commutative algebra $\mathcal{A}_{\Pi_t} \subset \mathcal{B}(\mathcal{H}_{0,1})$ which could be obtained by measuring a single observable at each time $t$: One natural choice would be the fixed algebra $\mathcal{A}_{\Pi}$ generated by the projectors $\{ \Pi_0, \Pi_1 \}$ introduced in the previous paragraph. However, since it corresponds to the (coarse-grained) energy eigenbasis, no information about quantum coherences would be accessible. We instead consider an algebra $\mathcal{A}_{\Pi_t}$ generated by projectors on a time-dependent basis rotating with the field state. As in the qubit example discussed in Appendix \ref{app: qubit}, we introduce a unitary operator such that 
\begin{align}
    \forall t, \qquad \ket{0(t)} &= U_{ t} \ket{0} \nn \\
    \ket{1_k(t)} &= U_{t} \ket{1_k}
\end{align}
so that $\ket{0(0)} = \ket{0}$ and $\ket{1_k(0)} = \ket{1_k}$. The set of time-dependent projectors
\begin{align} \label{pi0andpi1}
    \forall t, \qquad \Pi_0(t) &= \ket{0(t)} \bra{0(t)} = U_t \ket{0} \bra{0} U_t^\dag = U_t \Pi_0 U_t^\dag \nn \\
    \Pi_1(t) &= \Id - \Pi_0(t) = U_t (\Id - \Pi_0) U_t^\dag = U_t \Pi_1 U_t^\dag
\end{align}
define a coarse-grained measurement that we assume to be performed at time t. Note that 
\be
    \Pi_1(t) = U_t \Pi_1(0) U_t^\dag = U_t \sum_k \ket{1_k} \bra{1_k} U_t^\dag = \sum_k \ket{1_k(t)} \bra{1_k(t)}.
\ee
The representative of the Hamiltonian $H$ on the algebra $\mathcal{A}_{\Pi_t}$ is defined as
\be \label{eq: hamilonian apit}
    H_{\mathcal{A}_{\Pi_t}}(t) = \frac{\tr{(H \Pi_0(t))}}{V_0} \Pi_0(t) + \frac{\tr{(H \Pi_1(t))}}{V_1} \Pi_1(t)
\ee
where $V_0 = \tr{\Pi_0(t)} = \tr{\Pi_0} = 1$ and $V_1 = \tr{\Pi_1(t)} = \tr{\Pi_1}$, while the coarse-grained state is given by 
\be \label{cgstate}
    \rho_{\mathcal{A}_{\Pi_t}}(t) = p_0(t) \Pi_0(t) + p_1(t) \frac{\Pi_1(t)}{V_1}.
\ee
which can be expressed a thermal state with respect the Hamiltonian \eqref{eq: hamilonian apit} for some effective inverse temperature $\beta(t)$. Still focusing on small initial temperatures such that $\beta(0) \om_0 \gg 1$, one assumes as before that the wave train is in the ground state $\ket{0}$, i.e. $p_0(0) \approx 1$ and $p_1(0) \approx 0$, as we did before. We then choose the unitary operator $U_t$ to keep information about the evolution of the wave train state in its Hilbert space. However, since the resulting algebra is only two-dimensional as a vector space, a part of the information contained in the field state is inevitably discarded. Such a loss of information does not occur when one has access to the much larger algebra $\mathcal{B}(\mathcal{H}_{0,1})$. The optimal approach consists in solving the quantum dynamics induced by the qubit on the wave train and diagonalizing the resulting state in its instantaneous eigenbasis. 

To this end, we use the parametrization of the state of the wave train via Eqs.~\eqref{eq:rho_B01} and \eqref{eq: mod ham wtrain}. At early times, when $\alpha_k(t) \ll 1$, a Taylor expansion of the ground state of $K_B(t)$ yields
\be
    \ket{0(t)} \approx \left[\Id + \sum_k \frac{\om_k}{\om_0} (\alpha_k^\ast a_k +\alpha_k a_k^\dag)\right] \ket{0} = \ket{0} + \sum_k \frac{\om_k}{\om_0}\alpha_k(t) \ket{1_k}.
\ee
The projectors $\{ \Pi_0(t), \Pi_1(t) \}$ can then be identified by comparison with Eqs.~\eqref{pi0andpi1}. At first order, Eq.~\eqref{eq: hamilonian apit} leads to:
\be
    H_{\mathcal{A}_{\Pi_t}}(t) = \sum_{k} \om_k \left(\frac{\om_k}{\om_0}\right)^2 \lvert \alpha_k(t) \lvert^2 \Pi_0(t) + \om_0 \Pi_1(t) \approx \sum_k \om_k \lvert \alpha_k(t) \lvert^2 \Pi_0(t) + \om_0 \Pi_1(t)
\ee
so that the work transfer at time $t$ is given at first order by
\be
    \d W_{\mathcal{A}_{\Pi_t}}(t) = \tr{\left(\frac{d H_{\mathcal{A}_{\Pi_t}}(t)}{dt} \rho_{\mathcal{A}_{\Pi_t}}\right)} =  \sum_k \om_k \Delta \lvert \alpha_k(t) \lvert^2 = \sum_k \om_k \Delta|\!\mean{C_k}\!|^2
\ee
which coincides exactly with the work increment obtained in Eq. \eqref{eq: work increment coll model}. In the present description, however, the algebra is only two-dimensional, and the work contribution appears as the variation of a time-dependent representative Hamiltonian associated to a time-dependency of the accessible algebra (as in the qubit example of App.~\ref{app: qubit}; see also sec.~\ref{sec:alg_work}) rather than from the variation of nonequilibrium parameters in the state. The strict equivalent between the two pictures holds only in the regime $\alpha_k(t) \ll 1$, which corresponds to early times since $\alpha_k(t = 0) = 0$. 

Instead, at later times $t \gg \g^{-1}$ such that the qubit is completely de-excited, the field state becomes pure (see app.~\ref{app: spontaneous emission}) and we have  
\be \label{cgstateex}
    \rho_{\mathcal{A}_{\Pi_t}}(t) = \ket{0(t)} \bra{0(t)}
\ee
with
\be
    \ket{0(t)} = U_{t} \ket{0} = \frac{1}{\sqrt{2}}(\ket{0} + \sum_k \alpha_k(t) \ket{1_k})
\ee  
and where $\lvert \alpha_k(t) \lvert^2$ approaches the Lorentzian distribution Eq.~\eqref{eq: energy dist}. We assume that the unitary operator $U_t$ has been chosen to track the evolution of the wave train in Hilbert space as accurately as possible. 
The final coarse-grained state being pure, its observational entropy vanishes. It does not imply, however, that the entropy production vanishes throughout the evolution. In the present framework, the observational entropy first increases and subsequently decreases, reflecting the fact that the system starts in a pure state and ultimately returns to a pure state. Its behavior thus differs markedly from that of the observational entropy associated with the coarse-grained Hamiltonian algebra considered in Sec. \ref{sec: section wave train no coherences}, as well as from the more elementary treatment presented in Appendix \ref{app: spontaneous emission}, in which the observational entropy grows to a very large value for $t \gg \g^{-1}$. Nevertheless, in both descriptions, the entropy production remains non-negative at all times.

\vspace{0.5 cm}

\subsection{Algebraic thermodynamics for the collisional model: the qubit as a reservoir}
\label{subsec: qubit as reservoir}

In this section, we take the opposite viewpoint with respect to the previous section and consider that the qubit A is the environment powering the evolution of the electromagnetic field B, which is now understood as the system. Contrary to the open system viewpoint analysis of the qubit thermodynamics developed in Appendix \ref{app: qubit}, the algebra of the qubit A is here fixed once and for all to be its complete operator space such that its quantum density operator can be perfectly characterized.
In contrast, we allow different choices for the algebra of the electromagnetic field B. Between $t_n$ and $t_{n+1}$ the second law  \eqref{eq: second law gen alg app bipartite} associated to the field evolution simply becomes
\begin{align} \label{eq: second law gen alg app bipartite 2}
    \Delta S_{\mathcal{A}_{\text{B}}} - \tr{((\rho_{\mathcal{A}_{\text{A}}}(t_{n+1}) - \rho_{\mathcal{A}_{\text{A}}}(t_{n}))\ln{\rho_{\mathcal{A}_{\text{A}}}(t_{n}))})} \nn \\
    \geq \Delta S_{\mathcal{A}_{\text{A}}\otimes \mathcal{A}_{\text{B}}} + \Delta I_{\mathcal{A}_{\text{A}} \otimes \mathcal{A}_{\text{B}}} + S(\rho_{\mathcal{A}_{\text{A}}}(t_{n+1}) \lvert \lvert \rho_{\mathcal{A}_{\text{A}}}(t_{n})).
\end{align}
To go further, one needs to compute the modular Hamiltonian of the qubit A at time $t_n$. On the Bloch sphere, one can decompose the state of the qubit in the orthonormal basis $\f12 \{ \Id, \s_x, \s_y, \s_z \}$ so that
\be
    \rho_{\mathcal{A}_{\text{A}}}(t_n) = \f12 (\Id + \langle \s_z \rangle \s_z + \langle \s_x \rangle \s_x + \langle \s_y \rangle \s_y )
\ee
where we denoted $\langle \s_{i} \rangle \equiv \langle \s_{i} \rangle_{\rho_{\mathcal{A}_{\text{A}}}(t_n)}$, for $i = x,y,z$. Then, we compute
\begin{align}
    K_{\mathcal{A}_{\text{A}}}(t_n) &= - \ln{\rho_{\mathcal{A}_{\text{A}}}(t_n)} \nn \\
    &= (\ln{2} - \f12 \ln{(1 - r^2)}) \Id - \frac{\arctanh{r}}{r} ( \langle \s_z \rangle \s_z + \langle \s_x \rangle \s_x + \langle \s_y \rangle \s_y )
\end{align}
where 
\be
    r = \sqrt{\langle \s_x \rangle^2 + \langle \s_y \rangle^2 + \langle \s_z \rangle^2 }.
\ee
In addition, let us assume that the state of the qubit A is in the south hemisphere (i.e. $\langle \s_z \rangle < 0$) so that the effective temperature of the qubit reads 
\be \label{def:temperature}
    T_{\text{A}}(t_n) = \frac{1}{\beta_{\text{A}}(t_n)} = - \frac{\om_0}{2} \frac{1}{\langle \s_z \rangle} \frac{r}{\arctanh{r}} \geq 0
\ee
and is then positive. Then, following the procedure detailed in the main text, the heat and work are defined through 
\begin{align} \label{eq: workqubitA}
    \delta \mathcal{W}_{\text{A},t_n,t_{n+1}} &= \frac{\om_0}{2 \langle \s_z \rangle}(\langle \s_x \rangle\text{d} \langle \s_x \rangle + \langle \s_y \rangle \text{d} \langle \s_y \rangle) = \frac{\om_0}{ 2 \langle \s_z \rangle} \text{d} C^2 \nn \\
    \delta \mathcal{Q}_{\text{A},t_n, t_{n+1}} &= \text{d} E_{t_n, t_{n+1}} -  \d \mathcal{W}_{\text{A},t_n} = \frac{\om_0}{2} \text{d} \langle \s_z \rangle - \frac{\om_0}{ 2 \langle \s_z \rangle} \text{d} C^2
\end{align}
where $C = |\langle \sigma_+ \rangle| = |\langle \sigma_- \rangle|$ quantifies the qubit coherence in its energy eigenbasis. Notice that the resulting work is proportional to the variation of the squared coherence, $\text{d} C^2$, as we found in the previous section in the case where the electromagnetic field was treated as the environment.
\begin{remark}
We note that the separate heat and work contributions may become singular as
$\langle\sigma_z\rangle\to0$, where the effective temperature diverges,
although their sum remains finite. Such a divergence is familiar from ordinary
thermodynamics: for a reversible heat exchange, $\d Q=T \text{d}S$, so that
$T\to\infty$ leads to a divergent heat transfer whenever $\text{d}S$ remains finite
and nonzero. As expected for an isentropic trajectory $\text{d}r = 0$, this singularity disappears, since the first equation of
Eq.~\eqref{eq: workqubitA} reduces to
\be
\delta\mathcal{W}_{\mathrm{A},t_n,t_{n+1}}
=
-\frac{\omega_0}{2}\,\text{d}\langle\sigma_z\rangle,
\ee
which remains finite at the equator.
\end{remark}

\subsubsection{The algebra of one collisional mode}

The main advantage of the collisional model is that it allows us to track the reservoir dynamics from one collision to the next. We can therefore first examine the thermodynamic balance of a single collision. In this case, the system algebraic entropy is the von Neumann entropy of one collisional mode. One has at first order $O(\g \Delta t)$:
\be
    \Delta S_{\mathcal{A}_{B_n}} - \frac{\frac{\om_0}{2} \Delta \langle \s_z \rangle + \frac{\om_0}{ 2 \langle \s_z \rangle} \Delta C^2}{T_{\text{A}}(t_n)} = I_{A B_n}(t_{n+1}) + O((\g \Delta t)^2) \geq 0
\ee
since $I_{A B_n}(t_n) = 0$. The qubit, seen as a reservoir, transfers some coherence to the collisional mode, which modifies its entropy. From the computation led in \cite{Prasad2026}, one can show that
\be \label{eq: entropy variation first order A}
    \Delta S_{\mathcal{A}_{B_n}} := \Delta S_{\text{v.N}, B_n} = \beta \om_0 \g \Delta t [(\frac{1}{e^{\beta \om_0} - 1} + \f12) \langle \s_z \rangle + \f12 - C^2] + O((\g \Delta t)^\frac{3}{2}).
\ee
 In particular, if $\langle \s_z \rangle = -\tanh{\frac{\beta \om_0}{2}}$, that is the qubit population is equal to the thermal equilibrium distribution at the incoming field temperature, then
\be
    \Delta S_{\mathcal{A}_{B_n}} = - \beta \om_0 \g \Delta t C^2 \leq 0. 
\ee
This means that the coherences of the qubit at the beginning of the collision constitute a resource which allows to decrease the field's entropy. More generally, for any value of $\langle \s_z \rangle$, Eq.\eqref{eq: entropy variation first order A} implies that 
\begin{align} \label{eq: variation free eenrgy abn}
   \Delta F_{\mathcal{A}_{B_n}} &= \Delta E_n - T \Delta S_{\mathcal{A}_{B_n}} \nn \\
   &= \om_0 \g \Delta t C^2 \geq 0 
\end{align}
at first order. Therefore, the qubit transfers thermodynamic resources to the collision mode through its coherence. The increase of the field free energy during the collision reflects the coherences generated by the interaction with the atom, which bring the field in a state able to perform work, e.g. if it later interacts with another qubit.

\subsubsection{The algebra generated by the local Hamiltonian}

Being able to measure the electromagnetic field after each collision
provides access to a wide range of thermodynamic resources.
However, if the accessible algebra is restricted to the commutative algebra $\mathcal{A}_{H_n}$ generated
by the local Hamiltonian $H_n = \om_0 b_n^\dag b_n$ attached to the n-th mode,
the coherence between the local energy eigenvectors are no longer accessible. Consequently, the thermodynamic resources
associated with this coherence can no longer be exploited. In order to compare our results with those obtained above for the full algebra of a collisional mode, we analyze again the interaction with a single collisional unit using the coarse-grained algebra for the electromagnetic field. We still assume the extremely low-temperature regime, $\beta \omega_0 \gg 1$, such that only the vacuum and one-particle states of the collisional mode are appreciably populated. Then, one has to evaluate the probability variation $\Delta p_1$ for finding one photon in the electromagnetic field. But we have already established that for very small initial temperature of the incoming n-th mode 
\be
    \Delta p_1 = \frac{\Delta E_n}{\om_0} = \g \Delta t [(\frac{1}{e^{\beta \om_0} - 1} + \f12) \langle \s_z \rangle + \f12]
\ee
so that the variation of the entropy 
\be
    S_{\mathcal{A}_{H_n}} = - \sum_{J \in \{ 0, 1\}} p_J \ln{\frac{p_J}{V_J}}
\ee
attached to the commutative subalgebra  $\mathcal{A}_{H_n}$ (where we have neglected populations with $J \geq 2$) is given at first order by 
\be
    \Delta S_{\mathcal{A}_{H_n}} = \Delta p_1 \ln{V_1} - \Delta p_1 \ln{\frac{p_1}{p_0}} = \g \Delta t [(\frac{1}{e^{\beta \om_0} - 1} + \f12) \langle \s_z \rangle + \f12] \ln{\frac{p_0 V_1}{p_1}}
\ee
and it becomes clear that restricting the description to the algebra $\mathcal{A}_{H_n}$ eliminates the coherence contribution from the entropy variation. Before the interaction with the qubit, the collisional mode is in thermal equilibrium at temperature $T$, so that $\frac{p_0(t_n) V_1}{p_1(t_n)} = e^{- \beta \om_0}$, implying
\be
     \Delta S_{\mathcal{A}_{H_n}} - \Delta S_{\mathcal{A}_{B_n}} = \beta \om_0 \g \Delta t C^2 \geq 0.
\ee
Thus, by discarding the coherence, the coarse-grained description assigns a larger entropy variation attached to the n-th mode. Since the energy variation is unchanged by this coarse-graining, the corresponding variation of the non-equilibrium free energy is reduced
\be \label{eq: diff free energy var dif alg}
     \Delta F_{\mathcal{A}_{H_n}} - \Delta F_{\mathcal{A}_{B_n}} = - \om_0 \g \Delta t C^2 \leq 0.
\ee
In addition, from \eqref{eq: variation free eenrgy abn} and \eqref{eq: diff free energy var dif alg}, one can directly deduce that $\Delta F_{\mathcal{A}_{H_n}} = 0$. Indeed, once the coherence has been suppressed, the remaining energy transfer is purely heat-like and arises solely from changes in the populations of the energy levels, which leaves the free energy unchanged.

Restricting the description to the algebra $\mathcal{A}_{H_n}$ therefore lowers the non equilibrium free energy attributed to the field and, consequently, reduces the predictable upper bound on the work that can subsequently be extracted from it, potentially increasing the amount of violations of the apparent second law.

\subsubsection{The algebra generated by the coarse-grained Hamiltonian}

In this final paragraph, we assume that local observables are no longer accessible and that only global observables of the wave train can be measured. More specifically, we assume access only to the number of particles attached to the total Hamiltonian $H=\sum_n H_n$, and therefore restrict the description to the commutative subalgebra $\mathcal{A}_\Pi$ already introduced in Section~\ref{sec: section wave train no coherences}. Since the temperature remains very low ($\beta \om_0 \gg 1$), and the dynamics is restricted to the single-excitation sector, the population of the wave train is essentially confined to the zero- and one-particle sectors. The corresponding observational entropy is
\be
S_{\mathcal{A}_\Pi} = - \sum_{J} p_J \ln{\frac{p_J}{V_J}}.
\ee
After complete spontaneous emission, when $p_1\simeq1$, this reduces to
\be
S_{\mathcal{A}_\Pi} = \ln{V_1}
\ee
where $V_1$ is the dimension of the global one-particle sector. The successive inclusions of accessible algebras $\mathcal{A}_\Pi \subset \otimes_{n=1}^{N} \mathcal{A}_{H_n} \subset \otimes_{n=1}^{N} \mathcal{A}_{B_n}$ imply the corresponding ordering of algebraic entropies 
\be
S_{\mathcal{A}_\Pi}
\geq
S_{\otimes_n\mathcal{A}_{H_n}}
\geq S_{\otimes_n \mathcal{A}_{B_n}} =
S_{\mathrm{v.N}},
\ee
where $S_{\mathrm{v.N}}$ denotes the von Neumann entropy of the full wave-train state. Since the total energy is unchanged by these different choices of accessible algebra, the corresponding non equilibrium free energies satisfy the reverse ordering,
\be
F_{\otimes_n\mathcal{A}_{B_n}}
\geq
F_{\otimes_n\mathcal{A}_{H_n}}
\geq
F_{\mathcal{A}_\Pi},
\ee
and similarly for their variations when evaluated with respect to the same reference temperature $T = \frac{1}{\beta}$.
The successive restrictions of the accessible algebra therefore lead to a decrease of the available thermodynamic resources. While the algebras generated by the local Hamiltonians are insensitive to local coherences, the algebra generated solely by the total Hamiltonian is additionally insensitive to the localization of the excitation along the wave train. This further loss of information substantially increases the observational entropy and correspondingly reduces the non equilibrium free energy, thereby lowering the amount of work that can be extracted from the electromagnetic field based on those different levels of knowledge of system B.

\section{An intuitive introduction to the algebraic approach to quantum mechanics}
\label{app: algerba review}

In this appendix, we sum up basic information about the algebraic approach to quantum mechanics, insisting on physicist's intuition behind the mathematical toolbox (see for instance \cite{FewsterRejzner2019, Witten2018, Sorce:2023fdx} for excellent modern reviews on the subject).
Quantum theory is naturally formulated in terms of algebras of observables. Indeed, it is natural to require that the set of observables quantities should be closed under addition, multiplication, and functional calculus, so that if $a$ is an observable, then any sufficiently regular function $f(a)$ is also an observable. The quantum theory describing a specific \emph{system}, in a very general sense of a set of degrees of freedom which can be measured and manipulated independently of the rest of the world, can then be constructed starting with a choice of operator algebra. 

To model a physically meaningful system, some additional constraints have to be put on the algebras to obtain what is called a von Neumann algebra.
First, we need to be able to construct observables from self-adjoint operators, and to build probabilities from positive operators of the form $b^\ast b$, where $b$ is an operator, and we have denoted its adjoint $b^\ast$ to connect with mathematical literature. The algebra must therefore be equipped with an adjoint operation, making it a unital $\ast$-algebra.

Moreover, physical observables are represented by bounded operators, which naturally calls for a normed algebraic structure. We are thus led to consider $\mathbb{C}^\ast$-algebras, namely $\ast$-algebras equipped with a norm satisfying $ \lvert \lvert a^\ast a \lvert \lvert =\lvert \lvert a \lvert \lvert^2$. Once the algebra is faithfully represented on a Hilbert space $\mathcal{H}$, this norm coincides with the usual operator norm. 

However, from a purely empirical perspective, one is naturally led to enlarge the algebra, since physical experiments probe expectation values of observables in a given state and transition amplitudes between states. Imposing, in addition, closure with respect to the weak operator topology leads to the notion of a von Neumann algebra. More precisely, a von Neumann algebra is required to be closed under weak operator limits: whenever a sequence (or more generally a net) of operators $\{a_n\}_n$ in the algebra converges weakly to an operator $a$, that is, 
\be
 \forall (\ket{\psi_1}, \ket{\psi_2}) \in \mathcal{H}^2,  \bra{\psi_2} a_n \ket{\psi_1} \underset{n \rightarrow + \infty}{\longrightarrow} \bra{\psi_2} a \ket{\psi_1},
\ee
then the limit operator $a$ must also belong to the algebra. 
Physically, this means that whenever all matrix elements of a sequence of observables converge, the limiting observable already belongs to the algebra. As classes of algebras, one therefore has  
\be
    \text{von Neumann algebras} \subset \mathbb{C}^\ast \text{-algebras} \subset \text{unital} \ast \text{-algebras}. 
\ee
Note that for algebras acting on finite dimensional Hilbert spaces one has 
\be
    \text{von Neumann algebras} = \mathbb{C}^\ast \text{-algebras} = \text{unital} \ast \text{-algebras}.
\ee

We finish by mentioning that von Neumann algebras can be classified in three types associated to their factor decomposition \cite{MurrayvonNeumann1936, Sorce:2023fdx}. A fundamental result \cite{TakesakiI,  Dixmier1981, KadisonRingroseII} states that every von Neumann algebra is isomorphic to a direct integral of factors. \footnote{More precisely, one has 
\be
    \mathcal{A} \simeq \int_{\oplus} \mathcal{A}(X) \text{d} \mu(X)
\ee
where $\mu(X)$ is a measure on the space indexing the factors.
}. In particular, type I von Neumann algebras, which are the primary focus of this work, admit in many physically relevant situations a decomposition into a direct sum of factors, each of which is isomorphic to the algebra of all bounded operators on some Hilbert space, i.e.
\be \label{eq: direct sum}
    \mathcal{A} \simeq \bigoplus_J \mathcal{A}_J, \quad \mathcal{A}_J \simeq \mathcal{B}(\mathcal{H}_J) \otimes \Id_{\mathcal{H}_J'}.
\ee
For finite dimensional systems, all von Neumann algebra are of type I. One characteristic feature of type I von Neumann algebra is that states can be represented by density operators and that pure states exist in the usual sense. The situation becomes substantially different when infinitely many degrees of freedom are present, leading to the existence of other classes of von Neumann algebra factors called type II and type III. Although the Murray-von Neumann classification of algebras is not formally a classification of entanglement, we think that it is often useful to keep the following heuristic picture in mind:
\begin{align}
    \text{Type I} &= \text{Finite amount of entanglement} \nn \\
    \text{Type II} &= \text{Infinite amount of entanglement but renormalizable} \nn \\ 
    \text{Type III} &= \text{Infinite amount of entanglement but non-renormalizable}.
\end{align}
Type II algebra lead to state still admitting a trace, allowing one to define density operator with finite von Neumann entropy despite the presence of infinitely many degrees of freedom. By contrast, in type III von Neumann algebras states possess no non-trivial trace and can no longer be represented by density matrices; the usual notion of von Neumann entropy ceases to exist. 

Type-III algebras arise naturally in relativistic quantum field theory. Consider the algebra of bounded observables localized within a spatial region. Even in the vacuum state, the degrees of freedom inside and outside the region are entangled across the boundary. This entanglement becomes increasingly singular at short distances. For example, for a free scalar field one has
\be
    \langle \phi(\vec{x},0) \phi(\vec{y},0) \rangle \sim \frac{1}{\lvert \vec{x} - \vec{y} \lvert^2} 
\ee
which diverges as $\vec x\to\vec y$. Such ultraviolet correlations are a hallmark of local quantum field theories and are closely related to the emergence of type-III von Neumann algebras. Consequently, many concepts familiar from finite-dimensional quantum mechanics, such as density matrices and von Neumann entropy, are no longer available and must be replaced by more general algebraic notions.   

\section{POVM-based formulation}
\label{app: POVM}

Our formalism allows us to consider POVM instead of PVM in some restricted contexts. A POVM is a set of positive operators $M_J \geq 0$ so that 
\be
    \sum_J M_J = \Id_{\mathcal{H}}.
\ee
Of course, a complete set of orthogonal projectors ${\Pi_J}$ defines a projective-valued measure and therefore constitutes a particular type of POVM, whereas a generic POVM needs not to be composed of projectors. As already emphasized in the main text, the crucial difference is that the elements ${M_J}$ of a generic POVM do not, in general, themselves form an algebra closed under the operator products, unlike the algebra generated by a complete family of mutually orthogonal projectors. Consequently, the POVM outcomes cannot be endowed directly with the same reduced quantum-theoretical structure that we constructed from the projectors.

Nevertheless, we show below that it is possible to replace the final measurement of the two-point measurement protocol we consider in the main text, and still obtain  a plausible formulation of the second law in terms of observational entropies. The first measurement, however, must still be a PVM. To obtain this extension, we first emphasize that when the final measurement is projective (as considered throughout the main text) expression of the entropy production only requires the knowledge of the system's final macrostate $\bar{\rho}(t_\ff)$, not the coarse-grained state, since
\be \label{eq: equalite coarse-grained macro}
    S(P_{C_\ff, u} \circ C_\ff \circ \mathcal{N}_{t_\ii, t_\ff}(\rho_i) \lvert \lvert P_{C_\ff, u} \circ C_\ff \circ \mathcal{N}_{t_\ii, t_\ff}(\s_\ii)) = S(C_\ff \circ \mathcal{N}_{t_\ii, t_\ff}(\rho_i) \lvert \lvert  C_\ff \circ \mathcal{N}_{t_\ii, t_\ff}(\s_\ii)). 
\ee
In other words, instead of considering the monotonicity of the relative entropy under the map $\Lambda_{t_\ii, t_\ff} = P_{C_\ff, u} \circ C_{\ff} \circ \mathcal{N}_{t_\ii, t_\ff}$, one may equivalently consider its monotonicity under the CPTP map $\l_{t_\ii, t_\ff} = C_\ff \circ \mathcal{N}_{t_\ii, t_\ff}$, and this is sufficient to derive Eq. \eqref{eq: generic second law commalg opens}.

Now, let us assume that the coarse-graining map $C_\ff$ is defined in terms of a POVM rather than a family of orthogonal projectors, i.e.
\be
C_f(\rho) = \sum_F \tr{M_F\rho} \ket{F}\bra{F}.
\ee
In this case, one obtains
\be
    S(P_{C_\ff, u} \circ C_\ff \circ \mathcal{N}_{t_\ii, t_\ff}(\rho_i) \lvert \lvert P_{C_\ff, u} \circ C_\ff \circ \mathcal{N}_{t_\ii, t_\ff}(\s_\ii)) < S(C_\ff \circ \mathcal{N}_{t_\ii, t_\ff}(\rho_i) \lvert \lvert  C_\ff \circ \mathcal{N}_{t_\ii, t_\ff}(\s_\ii))
\ee
instead of Eq. \eqref{eq: equalite coarse-grained macro}. This statement can be interpreted as the second law if one extends the definition
of coarse-grained entropy to be the von Neumann entropy of the coarse-grained state
\be
    \rho^{\text{cg}} = \sum_F \frac{p_F}{V_F} M_F, \qquad p_F = \tr{(\rho M_F)}\quad V_F=\tr{M_F}.
\ee
This definition coincides with the observational entropy defined in terms of the $p_F$ and $V_F$ solely if the POVM reduces to a PVM, and otherwise leads to a different quantity which is always larger \cite{BuscemiSchindlerSafranek2023}. Monotonicity of relative entropy implies that
\be \label{eq: POVM montonicity}
    S(\rho(t_\ii) \lvert \lvert \s_\ii) \geq S(\l_{t_\ii, t_\ff}(\rho(t_\ii)) \lvert \lvert \l_{t_\ii, t_\ff}(\s_\ii)) = S\Big(\sum_F p_F \ket{F} \bra{F} \big\|\sum_F V_F e^{-K_F} \ket{F} \bra{F}\Big)
\ee
since $C_\ff(\rho) = \sum_F \tr{(M_F \rho)} \ket{F} \bra{F}$ and where $e^{- K_F} = \tr{\left[\frac{M_F}{V_F} \mathcal{N}_{t_\ii, t_\ff}(\s_\ii)\right]}$. Here, $\s_\ii$ is taken to be the fixed point of the map $\tilde{\Lambda}_{t_\ii, t_\ff}$ defined as in \eqref{eq: tildelambda}, but taking the micro-macro channel $C_\ff$ associated with to the POVM defined from the operators $M_F$, while $C_\ii$ continues to represent a projective measurement. 
Then, Eq. \eqref{eq: POVM montonicity}, which can be straightforwardly generalized to the case of non-vanishing initial mutual information exactly as in the main text, reads
\be \label{eq: second law povm}
    \Delta S^{\text{ob}}(\rho) - \Delta \langle \bar{K}_{\ii, \ff} \rangle_{\bar{\rho}} \geq - I_{\text{out}}(t_\ii) - I_{\text{int}}(t_\ii)
\ee
where 
\begin{align} \label{eq: povm second law}
    \bar{K}_\ii = - \sum_I \tr{\big( \frac{\Pi_I}{V_I} \ln{\s_i} \big)  } \ket{I} \bra{I}, \qquad \bar{K}_\ff = -\sum_F \ln{\tr{\big( \frac{M_F}{V_F} \mathcal{N}_{t_\ii, t_\ff}(\s_\ii)} \big)} \ket{F} \bra{F}
\end{align}
since the initial measurement carried out by the map $C_\ii$ is still assumed to be projective. Therefore, Eq.\eqref{eq: second law povm} is the generalization of Eq.\eqref{eq: generic second law commalg opens} to POVMs. Notice that if both initial and final measurements are projective measurements,  Eq.\eqref{eq: second law povm} is strictly equivalent to Eq.\eqref{eq: generic second law commalg opens}. 

\printbibliography

\end{document}